\documentclass[11pt]{article}% добавляется twoside для двусторонней печати

\usepackage[T1]{fontenc}
\usepackage[utf8]{inputenc}%cp1251
\usepackage{textcomp}
\usepackage[centertags]{amsmath}
\usepackage[mediummath]{nccmath}
\usepackage{amsfonts}
\usepackage{amssymb}
\usepackage[pdftex]{hyperref}%hypertex,pdftex
\usepackage{braket}% бра-кет обозначения
\usepackage{graphicx}
\usepackage{graphbox}
\usepackage[numbers,sort&compress]{natbib}% упорядочивает ссылки

\usepackage{paperinitial}% Установка параметров страницы

\paperinitialization{15mm}{15mm}{15mm}{15mm}{2pt}{10pt}

\DeclareMathOperator{\re}{Re}
\DeclareMathOperator{\im}{Im}

\DeclareMathOperator{\pr}{pr}

\DeclareMathOperator{\arsh}{arsh}

\DeclareMathOperator{\Sp}{Sp}

\DeclareMathOperator{\diag}{diag}

\newcommand{\lan}{\langle}
\newcommand{\ran}{\rangle}
\newcommand{\bs}{\boldsymbol}

\newcommand{\e}{\varepsilon}
\newcommand{\vf}{\varphi}
\newcommand{\s}{\sigma}

\newcommand{\al}{\alpha}
\newcommand{\be}{\beta}
\newcommand{\ga}{\gamma}

\newcommand{\de}{\delta}
\newcommand{\De}{\Delta}

\newcommand{\la}{\lambda}

\newcommand{\ups}{\upsilon}

\newcommand{\spx}{\mathbf{x}}

\newcommand{\spq}{\mathbf{q}}
\newcommand{\spk}{\mathbf{k}}

\newcommand{\spe}{\mathbf{e}}
\newcommand{\spn}{\mathbf{n}}
\newcommand{\spb}{\mathbf{b}}
\newcommand{\spw}{\mathbf{w}}

\begin{document}
\allowdisplaybreaks[4]% позволяет переносить многострочные формулы
\frenchspacing% уменьшение пробелов после запятых
\setlength{\unitlength}{1pt}% устанавливает единицу длины в окружении picture
%\selectlanguage{english}

%\sle

\title{{\Large\textbf{High-energy photon hologram of a photon gas}}}
%Susceptibility of a photon gas at high energies
%Hologram of a photon gas at high energies

\date{}

\author{%
P.O. Kazinski${}^{1)}$\thanks{E-mail: \texttt{kpo@phys.tsu.ru}}\;
and A.A. Sokolov${}^{1),2)}$\thanks{E-mail: \texttt{asokolov@tpu.ru}}\\[0.5em]
{\normalsize ${}^{1)}$ Physics Faculty, Tomsk State University, Tomsk 634050, Russia}\\
{\normalsize ${}^{2)}$ Tomsk Polytechnic University, Tomsk 634050, Russia}
}

\maketitle

\begin{abstract}

The photon hologram of a one-particle density matrix of a photon gas is derived including the case where the energy of a probe photon is above the electron-positron pair creation threshold. The explicit expressions for the holograms of a photon gas with one-particle density matrix in the form of a single Gaussian and of coherent and incoherent lattices of Gaussians are obtained. The conditions for resonant cones of coherent scattering by coherent and incoherent lattices are found. These conditions turn out to be different. The explicit expression for the dielectric susceptibility tensor of a photon gas and of a single photon prepared in arbitrary quantum states are derived on the probe photon mass-shell. It is established that a photon gas and a single photon behave in coherent photon scattering as a medium with linear and circular birefringence. This medium is transparent below the electron-positron creation threshold and is absorbing otherwise. In the high-energy limit, $\sqrt{s}\gg 2m$, it has the dielectric susceptibility tensor of a birefringent plasma except for a slow logarithmic scaling. It is shown that, for the probe photon energies of order $1$ GeV and higher, the energies of target photons of order $1$ eV and higher, and the photon gas density such that the classical intensity parameter is of order unity, the hologram of the photon gas can be measured with existing experimental facilities.

\end{abstract}

\section{Introduction}

Recently, it has been shown that even elementary particles can be endowed by some properties that usually are ascribed to macroscopic objects \cite{pra103,KazSol2022,KazSol2023,radet,KazinskiFr24,AKS2025,KazSokNeut}. One of such properties is a dielectric susceptibility that appears in describing coherent scattering of photons by elementary particles interacting with electromagnetic field. Due to light-by-light scattering \cite{KarpNeu50,KarpNeu51,Tollis64,Tollis65,ConTollPist71,Varf66}, the dielectric susceptibility tensor can be introduced even for a single photon prepared in a certain quantum state. This tensor was constructed in \cite{KazSol2023} for the energies of a probe photon lower than the electron-positron pair creation threshold. Such a susceptibility tensor determines the outcomes of coherent scattering \cite{KazSol2022,BednNaum2021,Toll1952} of probe photons on target ones that form the hologram of the quantum state of a target photon on the detector \cite{KazinskiFr24,KazSokNeut}. This hologram is a result of the interference between the free passed part of the probe photon wave function and its scattered part. In this sense, it is a realization of a simple Gabor scheme for construction of a hologram \cite{Gabor1949,SpBookMicrosc2019}. In the present paper, we generalize the results of \cite{KazSol2023} and construct the dielectric susceptibility tensor and the hologram of a photon gas and of a single photon prepared in an arbitrary quantum state in the framework of perturbative quantum electrodynamics. The probe photon energies are assumed to be arbitrary but much lower than the muon pair creation threshold. The quantum state of target photons is supposed to be such that the usual perturbation theory is applicable.

The holograms and the susceptibility tensor of target electromagnetic fields arise naturally in strong field quantum electrodynamics in studying the so-called vacuum diffraction and birefringence in the presence of strong laser fields
\cite{Piazza2006,FMST07,King2010,Tommasini2010,BatRiz13,DHIMT14,DHIMT141,KarbShai15,BMKP17,Karb18,KUMZ22,FIKKSTT,
Ahmadiniaz2023,Wang2024,Berezin2024,Smid2025,Rinderknecht2025,Ahmadiniaz2025,Matheron2026}. These studies are commonly conducted either in the formalism of the Heisenberg-Euler effective action \cite{BBBB70,MarShuk06,FMST07,KarbShai15,KarbMos20,KUMZ22,Ahmadiniaz2023,FIKKSTT, Piazza2006,Matheron2026,Ahmadiniaz2025,Rinderknecht2025,Varf66,Smid2025,Berezin2024,Wang2024, Karb18,DHIMT14,DHIMT141,BatRiz13,Tommasini2010,King2010,Valialshchikov2025,Bu2026} or by using the photon polarization operator found in the presence of a traveling plane electromagnetic wave \cite{BeckMit75,BaiMilStr75,BMKP17}. In both cases, the state of a target electromagnetic field is assumed to be coherent and so can be described by classical fields. Moreover, in the latter case this classical field has a very specific profile. We get rid of these limitations at the cost of treating light-by-light scattering perturbatively and show that the hologram and the dielectric susceptibility tensor of a photon gas are determined by its one-particle density matrix.

Light-by-light scattering is a tiny effect and has not been verified experimentally yet with all photons participating in the process on-shell \cite{BMKP17,KUMZ22,Ahmadiniaz2023,Berezin2024,Smid2025,Rinderknecht2025,Ahmadiniaz2025,Matheron2026,MarShuk06, Yu2023,NakHom17,Takahashi18,EMRY22,Budker22}. Therefore, one of our aims is to find such parameters of the target photon gas and of the probe photon that the hologram of the target can be observed with existing experimental facilities. It is known \cite{LandLifQED} that the probability of this process grows with the center-of-mass energy $\sqrt{s}$ up to the electron-positron pair creation threshold. Therefore, it is reasonable to expect that the holograms of a sufficiently dense photon gas constructed with the aid of high energy probe photons can be recorded by the existing detectors \cite{BMKP17,Yu2023,NakHom17,Takahashi18,EMRY22,Budker22}. We show that this is indeed the case and derive a simple analytic expression for the hologram of a photon gas with one-particle density matrix given by a narrow in momentum space Gaussian with the dispersion matrix of a general form.

Furthermore, we investigate the influence of coherence of the target quantum state on its hologram. To this end, we consider coherent scattering of a probe photon on the two types of lattices made of photons. The one-particle density matrix of the first lattice is pure and is described by a coherent superposition of identical Gaussians with centers located at the lattice sites in momentum space at the instant of time $t=0$ in the interaction picture. The one-particle density matrix of the second lattice is mixed and is obtained in the same way as in the case of a coherent lattice but with Gaussians multiplied by stochastic phases. It turns out that the holograms of these two lattices differ qualitatively even in the case when the corresponding photon number densities coincides in the coordinate space at the instant of time $t=0$. This demonstrates a strong dependence of the hologram on coherence properties of a target quantum state.

The paper is organized as follows. In Sec. \ref{Hologram}, we derive the general expression for the inclusive probability to record a probe photon in scattering by a gas of photons in the leading nontrivial order of perturbation theory. This leading nontrivial order in the coupling constant describes coherent scattering and is nothing but a hologram of a target photon gas. In Sec. \ref{Dens_Matr_Scat}, we particularize the general formulas of Sec. \ref{Hologram} and obtain the expression for the density matrix of a scattered probe photon that determines the hologram. Section \ref{Evolut_Stokes_Param_Probe} is devoted to evolution of the Stokes parameters of a probe photon in coherent scattering in the small momentum transfer limit. In this limit, we obtain the equations specifying a change of the probe photon density matrix in coherent scattering. In Sec. \ref{Examples}, we consider several examples of holograms of a photon gas prepared in different quantum states. Here we deduce the explicit expressions for the holograms and investigate an impact of the quantum state coherence of a photon gas on its hologram. We also provide the estimates for the relative magnitude of the effect of coherent light-by-light scattering. The tensor of dielectric susceptibility of a photon gas is derived in Sec. \ref{Susc_Gas_Phot}. We also give a simple physical explanation of a large magnitude of coherent scattering in an almost head-on configuration. In Conclusion, we summarize the results. Some tedious derivations and bulky formulas are removed to Appendices \ref{Polarization_Vectors_App}, \ref{Evaluation_Gaauss_Int_App}, and \ref{Time_Shift_App}. We use the system of units such that $c=\hbar=1$ and $e^2=4\pi\al$, where $\alpha$ is the fine structure constant. The summation over repeated indices is implied unless otherwise stated.

\section{Hologram}\label{Hologram}

Consider scattering of a probe photon on a target consisting of photons in the framework of perturbative quantum electrodynamics. The system is prepared at the instant of time $t_{in}\rightarrow-\infty$ and a single probe photon is measured in the state picked out by the projector $D_{\bar{\ga}\ga}$ at the final instant of time $t_{out}\rightarrow+\infty$. Hereinafter, we will always imply the interaction picture  (see the details of the general formalism in the papers \cite{KazSol2022,radet}). The initial state of the system is taken in the form
\begin{equation}\label{densMatrIn}
    \hat{R} =  \hat{R}_{ph} \otimes |0\rangle_{e^-}\langle 0|_{e^-} \otimes|0\rangle_{e^+}\langle 0|_{e^+},
\end{equation}
where $ \hat{R}_{ph}$ is the photon density matrix, $\ket{0}_{e^-}$ and $\ket{0}_{e^+}$ are the vacuum states of electrons and positrons. The photon density matrix is given by
\begin{equation}\label{R_ph}
    \hat{R}_{ph} = \rho^h_{\ga\bar{\ga}} \hat{c}^\dag_\ga \hat{R}^s \hat{c}_{\bar{\ga}},
\end{equation}
where $\hat{c}^\dag_\ga$ and $\hat{c}_{\bar{\ga}}$ are the creation and annihilation operators of photons, $\rho^h_{\ga\bar{\ga}}$ is the density matrix of a hard probe photon, $\hat{R}^s$ is the density matrix of soft target photons, and
\begin{equation}
\begin{split}
    \hat{R}^s=&\sum_{N,M=0}^\infty \frac{1}{\sqrt{N!M!}}\rho^s_{\ga_N\cdots\ga_1|\bar{\ga}_1\cdots\bar{\ga}_M} \hat{c}^\dag_{\ga_1}\cdots \hat{c}^\dag_{\ga_N}|0\ran\lan0| \hat{c}_{\bar{\ga}_M}\cdots \hat{c}_{\bar{\ga}_1},\\
    \hat{R}=&\sum_{N,M=0}^\infty \frac{1}{\sqrt{N!M!}}\rho_{\ga_N\cdots\ga_1|\bar{\ga}_1\cdots\bar{\ga}_M} \hat{c}^\dag_{\ga_1}\cdots \hat{c}^\dag_{\ga_N}|0\ran\lan0| \hat{c}_{\bar{\ga}_M}\cdots \hat{c}_{\bar{\ga}_1}.
\end{split}
\end{equation}
It is implied in expression \eqref{R_ph} that the state of a probe photon does not overlap with the state of target photons,
\begin{equation}\label{orthogon_1}
    \rho^h_{\ga\bar{\ga}} \rho^s_{\ga\ga_M\cdots\ga_1|\bar{\ga}_1\cdots\bar{\ga}_M\bar{\ga}}=0,\qquad \forall M=\overline{0,\infty}.
\end{equation}
Then the normalization condition for the density matrix \eqref{R_ph} becomes
\begin{equation}\label{norm_cond}
    \sum_\ga\rho^h_{\ga\ga}=1,\qquad \Sp\hat{R}^s=1.
\end{equation}
Also we need the $M$-particle density matrix (see for details Appendix C of \cite{radet})
\begin{equation}
    \rho^{(M)}_{\ga_1\cdots\ga_M|\bar{\ga}_M\cdots\bar{\ga}_1}=\Sp(\hat{R}_{ph}\hat{c}^\dag_{\bar{\ga}_M}\cdots \hat{c}^\dag_{\bar{\ga}_1}\hat{c}_{\ga_1}\cdots \hat{c}_{\ga_M}).
\end{equation}
Taking into account relations \eqref{orthogon_1}, the $M$-particle density matrix is written as
\begin{equation}\label{rho_M}
    \rho^{(M)}_{\ga_1\cdots\ga_M|\bar{\ga}_M\cdots\bar{\ga}_1}= \rho^{s(M)}_{\ga_1\cdots\ga_M|\bar{\ga}_M\cdots\bar{\ga}_1} +\sum_{k,l=1}^M \rho^h_{\ga_l\bar{\ga}_k}\rho^{s(M-1)}_{\ga_1\cdots\hat{\ga}_l\cdots\ga_M|\bar{\ga}_M\cdots\hat{\bar{\ga}}_k\cdots\bar{\ga}_1},
\end{equation}
where the hat over the index means that this index is absent.

The measurement at the final instant of time is specified by the projector
\begin{equation}\label{proj_msrmnt}
    \hat{\Pi}_D = (1 -  \hat{\tilde{\Pi}}_D)\otimes \hat{1}_{e^-}\otimes\hat{1}_{e^+}, \qquad \hat{\tilde{\Pi}}_D=:\exp(-\hat{c}^\dag D\hat{c}):,
\end{equation}
where the colons denote the normal ordering. This projector distinguishes the states in the Fock space where there is at least one photon in the one-particle state determined by the projector $D_{\bar{\ga}\ga}$. The inclusive probability to record a photon in the state picked out by the projector $D_{\bar{\ga}\ga}$ reads
\begin{equation}\label{inclus_probab}
    P_D=\Sp(\hat{\Pi}_D\hat{S}\hat{R}\hat{S}^\dag).
\end{equation}
We suppose that with good accuracy
\begin{equation}\label{orthogon_2}
    \rho^s_{\ga\ga_M\cdots\ga_1|\bar{\ga}_1\cdots\bar{\ga}_M\bar{\ga}'} D_{\bar{\ga}'\bar{\ga}}=0= D_{\ga\ga'} \rho^s_{\ga'\ga_M\cdots\ga_1|\bar{\ga}_1\cdots\bar{\ga}_M\bar{\ga}},
\end{equation}
i.e., the target photons are prepared in the state that does not overlap with the state of a recorded photon.

Let us write out the contributions to the $S$-matrix up to the fourth power in the creation-annihilation operators taking into account the conservation laws. In that case, keeping in mind the form of the initial state \eqref{densMatrIn}, we have
\begin{equation}\label{S_matrix_gen}
     \hat{S}=\hat{1} + \hat{F} + \hat{C} +\ldots,
\end{equation}
where
\begin{equation}
     \hat{F} = \hat{a}^{\dag}_{\alpha} \hat{b}^{\dag}_{\beta} F_{ \beta \alpha  \bar{\ga}_2 \bar{\ga}_1} \hat{c}_{\bar{\ga}_1} \hat{c}_{\bar{\ga}_2}, \qquad \hat{C} = \hat{c}^{\dagger}_{\ga_2} \hat{c}^{\dagger}_{\ga_1} C_{ \ga_1 \ga_2 \bar{\ga}_2\bar{\ga}_1}  \hat{c}_{\bar{\ga}_1} \hat{c}_{\bar{\ga}_2},
\end{equation}
and $\hat{a}^\dag_{\al}$ and $\hat{a}_{\bar{\al}}$ are the creation and annihilation operators of electrons, $\hat{b}^\dag_{\al}$ and $\hat{b}_{\bar{\al}}$ are the creation and annihilation operators of positrons. The operator $\hat{F}$ is of second order in the coupling constant and describes the creation of an electron-positron pair from a pair of photons. The operator $\hat{C}$ is of fourth order in the coupling constant and describes light-by-light scattering. The terms denoted by ellipsis in \eqref{S_matrix_gen} do not contribute to the inclusive probability \eqref{inclus_probab} in the leading nontrivial order in the coupling constant. Substituting expression \eqref{S_matrix_gen} into the unitarity relation,
\begin{equation}
    \hat{S}^\dag \hat{S}=\hat{1},
\end{equation}
and bringing the creation-annihilation operators to the normal order, we obtain in particular
\begin{equation}\label{unitar_rel}
    C_{\ga_1\ga_2\bar{\ga}_2\bar{\ga}_1}+C^*_{\bar{\ga}_1\bar{\ga}_2 \ga_2 \ga_1}+F^*_{\be'\al'\ga_1\ga_2} F_{\be'\al'\bar{\ga}_2\bar{\ga}_1}=0.
\end{equation}
This relation can also be derived with the aid of the Cutkosky rules applied to the diagrams of light-by-light scattering.

On substituting expression \eqref{S_matrix_gen} for the $S$-matrix to the inclusive probability \eqref{inclus_probab}, we arrive at
\begin{equation}\label{inclus_probab_1}
    P_D=\Sp(\hat{\Pi}_D\hat{R})+\big[\Sp(\hat{\Pi}_D\hat{C}\hat{R}) +c.c.\big] +\Sp(\hat{\Pi}_D\hat{F}\hat{R}\hat{F}^\dag),
\end{equation}
in the leading nontrivial order in the coupling constant. The arising traces of operators are reduced to
\begin{equation}
\begin{split}
    \Sp(\hat{\Pi}_D\hat{R}_{ph})=&\,1-\rho^{(0)}_{\tilde{D}},\\
    \Sp(\hat{c}^\dag_{\bar{\ga}_1} \hat{c}^\dag_{\bar{\ga}_2}\hat{\Pi}_D\hat{c}_{\ga_2}\hat{c}_{\ga_1}\hat{R}_{ph})=&\,\rho^{(2)}_{\ga_2\ga_1|\bar{\ga}_1\bar{\ga}_2} -(\rho^{(2)}_{\tilde{D}})_{\ga_2\ga_1|\bar{\ga}_1\bar{\ga}_2},\\
    \Sp(\hat{\Pi}_D \hat{c}^\dag_{\bar{\ga}_1} \hat{c}^\dag_{\bar{\ga}_2}  \hat{c}_{\ga_2}\hat{c}_{\ga_1}\hat{R}_{ph})=&\,\rho^{(2)}_{\ga_2\ga_1|\bar{\ga}_1\bar{\ga}_2} -(\rho^{(2)}_{\tilde{D}})_{\ga_2\ga_1|\bar{\ga}'_1\bar{\ga}'_2}\tilde{D}_{\bar{\ga}'_1\bar{\ga}_1} \tilde{D}_{\bar{\ga}'_2\bar{\ga}_2},
\end{split}
\end{equation}
where $\tilde{D}_{\bar{\ga}\ga}=\de_{\bar{\ga}\ga}-D_{\bar{\ga}\ga}$, relations (B.13) from Appendix B of \cite{radet} have been used, and the projected $M$-particle density matrix has been introduced (see Appendix C of \cite{radet})
\begin{equation}
\begin{split}
    (\rho^{(M)}_{\tilde{D}})_{\ga_1\cdots\ga_M|\bar{\ga}_M\cdots\bar{\ga}_1}&=\Sp(\hat{R}_{ph}\hat{c}^\dag_{\bar{\ga}_M}\cdots \hat{c}^\dag_{\bar{\ga}_1}\hat{\tilde{\Pi}}_D \hat{c}_{\ga_1}\cdots \hat{c}_{\ga_M})=\\
    &=\sum_{N=M}^\infty \frac{N!}{(N-M)!} \rho_{\ga_1\cdots\ga_M\ga_{M+1}\cdots\ga_N|\bar{\ga}_N\cdots\bar{\ga}_{M+1}\bar{\ga}_M\cdots\bar{\ga}_1} \tilde{D}_{\bar{\ga}_N\ga_N}\cdots \tilde{D}_{\bar{\ga}_{M+1}\ga_{M+1}}.
\end{split}
\end{equation}
The traces of operators acting in the subspace of electrons and positrons are evaluated in a trivial way. Then the inclusive probability \eqref{inclus_probab_1} turns into
\begin{equation}\label{inclus_probab_2}
    P_D=1-\rho^{(0)}_{\tilde{D}}+\Big\{\big[(\rho^{(2)}_{\tilde{D}})_{\bar{\ga}_1\bar{\ga}_2|\ga_2\ga_1} -(\rho^{(2)}_{\tilde{D}})_{\bar{\ga}_1\bar{\ga}_2|\ga'_2\ga'_1} \tilde{D}_{\ga'_2\ga_2} \tilde{D}_{\ga'_1\ga_1}\big]C_{\ga_1\ga_2\bar{\ga}_2\bar{\ga}_1} +c.c. \Big\}.
\end{equation}
The contribution of the last term in \eqref{inclus_probab_1} is canceled by the terms stemming from the expression standing in the square brackets in \eqref{inclus_probab_1} due to unitarity relation \eqref{unitar_rel}. Therefore, the amplitude $F_{\be'\al'\bar{\ga}_2\bar{\ga}_1}$ is absent in expression \eqref{inclus_probab_2}.

We assume in what follows that the projector $D$ picks out a small volume in the phase space that is the case, for example, when $D$ projects to the state with a definite momentum. Then the expression for the inclusive probability \eqref{inclus_probab_2} should be developed as a series in powers of $D$ and the leading contribution in $D$ ought to be retained. Substituting expression \eqref{rho_M} for the  $M$-particle density matrix into \eqref{inclus_probab_2} and using the orthogonality requirement  \eqref{orthogon_2}, we come to (cf. formula (22) of \cite{KazSol2023})
\begin{equation}\label{inclus_probab_3}
    P_D=D_{\bar{\ga}_4\ga_4}(\rho^h_{out})_{\ga_4\bar{\ga}_4},
\end{equation}
where the density matrix of a scattered probe photon,
\begin{equation}\label{rho_out_gen}
    (\rho^h_{out})_{\ga_4\bar{\ga}_4}=\rho^h_{\ga_4\bar{\ga}_4}+ (\Phi_{\ga_4\ga_2}\rho^h_{\ga_2\bar{\ga}_4} +H.c.),
\end{equation}
and the amplitude of coherent photon-by-photon scattering have been introduced (cf. formula (21) of \cite{KazSol2023} and formula (30) of \cite{KazSokNeut})
\begin{equation}\label{coherent_amplitude}
    \Phi_{\ga_4\ga_2}=4 C_{\ga_4\bar{\ga}_1\ga_1\ga_2}\rho^{s(1)}_{\ga_1\bar{\ga}_1}.
\end{equation}
As it was thoroughly discussed in \cite{KazinskiFr24,AKS2025,KazSokNeut}, the inclusive probability \eqref{inclus_probab_3} containing a coherent contribution of the form \eqref{coherent_amplitude} entering into the density matrix of a scattered probe particle as in \eqref{rho_out_gen} is a hologram of the one-particle density matrix of a target.

\section{Density matrix of a scattered probe photon}\label{Dens_Matr_Scat}

Let us find the explicit expression for the inclusive probability \eqref{inclus_probab_3} for a small momentum transfer in the photon-by-photon scattering. Denote as
\begin{equation}
\begin{gathered}
    \ga_n = (\lambda_n, \mathbf{k}_n), \quad n=\overline{1,4},\qquad
    \sum_{\ga} = \sum_{\lambda} \int \frac{V d \mathbf{k}}{(2\pi)^3}, \\
    \rho^h_{\la_2 \la'_2}(\spk_2,\spk_2')=\frac{V}{(2\pi)^3} \rho^h_{\ga_2 \ga'_2},\qquad \rho^{s(1)}_{\la_1 \la'_1}(\spk_1,\spk_1')=\frac{V}{(2\pi)^3} \rho^{s(1)}_{\ga_1 \ga'_1}.
\end{gathered}
\end{equation}
We suppose that the quantum numbers $\ga_1$ and $\ga_3$ refer to the target photons before and after scattering, respectively, whereas $\ga_2$ and $\ga_4$ characterize the initial and final states of the probe photon. The polarization vectors of photons are defined in Appendix \ref{Polarization_Vectors_App}. In particular, the quantum numbers $\la_n=\pm1$, $n=\overline{1,4}$, specify a circular polarization of the respective photon. The kinematics of the process is characterized by the Mandelstam variables
\begin{equation}
\begin{split}
    s &= (k_1 + k_2)^2 = (k_3 + k_4)^2 = 2 (k_1 k_2) = 2 (k_3 k_4),\\
    t &= (k_1 - k_3)^2 = (k_2 - k_4)^2 = -2 (k_1 k_3) = -2 (k_2 k_4),\\
    u &= (k_1 - k_4)^2 = (k_2 - k_3)^2 = -2 (k_1 k_4) = -2 (k_2 k_3).
\end{split}
\end{equation}
Notice that $s+t+u=0$ and
\begin{equation}
    s=  k_0^1 k_0^2 (\mathbf{n}_1 - \mathbf{n}_2)^2=k_0^3 k_0^4 (\mathbf{n}_3 - \mathbf{n}_4)^2,
\end{equation}
where $\mathbf{n}_{n}:= \mathbf{k}_{n}/ |\mathbf{k}_{n}|$. Introduce the transferred momentum
\begin{equation}\label{q_mu}
    q^\mu=k^\mu_4-k^\mu_2=k^\mu_1-k^\mu_3,\qquad t=q^2.
\end{equation}
We will also use below the notation
\begin{equation}\label{k_13_k_24}
    k_{13}=k_{31}=k_s=(k_1+k_3)/2,\qquad k_{24}=k_{42}=k_h=(k_2+k_4)/2.
\end{equation}
The normalization conditions for the one-particle density matrices are written as
\begin{equation}\label{norm_conds}
    \sum_{\la_2}\int d\spk_2 \rho^h_{\la_2 \la_2}(\spk_2,\spk_2)=1,\qquad \sum_{\la_1}\int d\spk_1 \rho^{s(1)}_{\la_1 \la_1}(\spk_1,\spk_1)=N_s,
\end{equation}
where $N_s$ is the average number of photons in the target.

In the paper \cite{KazSol2023}, the inclusive probability to record a probe photon in scattering by a beam of soft photons was found in the case when the state of soft photons is coherent and $s<4m^2$, i.e., for the energies of initial photons lower than the electron-positron pair creation threshold. Besides, in that paper, the basis of photon polarizations \eqref{polarization_vects_KN} was used. This basis was introduced in the papers \cite{KarpNeu50,KarpNeu51,Tollis64,Tollis65,ConTollPist71} and is inconvenient in describing scattering of photon wave packets of a general form. In the case when the density matrices of scattered photons are not very narrow in the momentum space, it is more suitable for description of polarizations of such states to employ the polarization vectors \eqref{polarization_vects}, \eqref{polarization_vects_circ} defined in the laboratory frame, where $N^\mu=(1,0,0,0)$, with respect to a fixed spacelike vector $d^\mu$. Comparing formula \eqref{coherent_amplitude} with formula (21) of \cite{KazSol2023} and using formula (26) of \cite{KazSol2023}, we see that the density matrix of a scattered probe photon \eqref{rho_out_gen} in the basis of circular polarizations (the chiral basis) is given by
\begin{multline}
    (\rho^h_{out})_{\la_4'\la_4}(\spk_4',\spk_4)=\rho^h_{\la_4'\la_4}(\spk_4',\spk_4)+ \frac{i\pi}{2}\sum_{\la_1,\la_2,\la_3} \int\frac{d\spk_1 d\spk_2 d\spk_3}{(2\pi)^3}\Big[\de(k_3+k_4'-k_1-k_2)  \times\\
    \times \frac{\rho^{s(1)}_{\la_1\la_3}(\spk_1,\spk_3) \tilde{M}_{\la_3\la_4'\la_1\la_2}|_{k_4\rightarrow k_4'} \rho^h_{\la_2\la_4}(\spk_2,\spk_4)}{\sqrt{k_0(\spk_1) k_0(\spk_2) k_0(\spk_3) k_0(\spk_4')}} -\de(k_3+k_4-k_1-k_2) \frac{\rho^{s(1)*}_{\la_1\la_3}(\spk_1,\spk_3) \tilde{M}^*_{\la_3\la_4\la_1\la_2} \rho^{h*}_{\la_2\la'_4}(\spk_2,\spk'_4)}{\sqrt{k_0(\spk_1) k_0(\spk_2) k_0(\spk_3) k_0(\spk_4)}} \Big],
\end{multline}
where, in the notation used in formulas \eqref{rho_out_gen} and \eqref{coherent_amplitude},
\begin{equation}
    \ga_4=(\la_4',\spk_4'),\qquad \bar{\ga}_4=(\la_4,\spk_4),\qquad \ga_2=(\la_2,\spk_2),\qquad \ga_1=(\la_1,\spk_1),\qquad \bar{\ga}_1=(\la_3,\spk_3).
\end{equation}
Furthermore, the invariant scattering amplitude in the new basis of polarization vectors has been introduced
\begin{equation}
    \tilde{M}_{\la_3\la_4\la_1\la_2}=M_{\la'_3\la'_4\la'_1\la'_2}U^{3*}_{\la'_3\la_3} U^{4*}_{\la'_4\la_4} U^{1}_{\la'_1\la_1} U^{2}_{\la'_2\la_2}.
\end{equation}
The quantity $M_{\la_3\la_4\la_1\la_2}$ is the invariant light-by-light scattering amplitude found in \cite{KarpNeu50,KarpNeu51,Tollis64,Tollis65,ConTollPist71} (see also \cite{LandLifQED}) in the basis of polarization vectors \eqref{polarization_vects_KN}, \eqref{polarization_vects_circ}. It is the function of the variables $s$, $t$, and $u$. The transition matrices to the new basis of polarization vectors, $U^n$, are presented in formula \eqref{U_n_O_n_matrices}. Notice that if the initial density matrix of a probe photon describes a pure state, then the density matrix of a scattered probe photon also corresponds to a pure state in the order of perturbation theory we consider \cite{KazSokNeut,KazSol2023}.

If the detector records the momentum and the polarization of the probe photon, then the inclusive probability \eqref{inclus_probab_3} is determined only by the diagonal of the scattered probe photon density matrix in the momentum space
\begin{equation}\label{dens_matr_out1}
\begin{split}
    (\rho^h_{out})_{\la_4'\la_4}(\spk_4,\spk_4)=&\,\rho^h_{\la_4'\la_4}(\spk_4,\spk_4)+ \frac{i\pi}{2}\sum_{\la_1,\la_2,\la_3} \int\frac{d\spk_1 d\spk_2 d\spk_3}{(2\pi)^3} \de(k_3+k_4-k_1-k_2)   \times\\
    &\times \Big[\frac{\rho^{s(1)}_{\la_1\la_3}(\spk_1,\spk_3) \tilde{M}_{\la_3\la_4'\la_1\la_2} \rho^h_{\la_2\la_4}(\spk_2,\spk_4)}{\sqrt{k_0(\spk_1) k_0(\spk_2) k_0(\spk_3) k_0(\spk_4)}} - H.c. \Big].
\end{split}
\end{equation}
It is useful to expand the density matrices and the invariant scattering amplitude in terms of the $\s$-matrices. Then
\begin{equation}\label{dens_matr_repr}
\begin{gathered}
    \rho^{s(1)}_{\la_1\la_3}(\spk_1,\spk_3)=\frac{\rho^{s(1)}(\spk_1,\spk_3)}2[1+(\bs\xi^s(\spk_1,\spk_3)\bs\s)]_{\la_1\la_3},\qquad \rho^{h}_{\la_2\la_4}(\spk_2,\spk_4)=\frac{\rho^{h}(\spk_2,\spk_4)}2[1+(\bs\xi^h(\spk_2,\spk_4)\bs\s)]_{\la_2\la_4},\\
    (\rho^{h}_{out})_{\la_4'\la_4}(\spk_4,\spk_4)=\frac{\rho^{h}_{out}(\spk_4,\spk_4)}2[1+(\bs\xi^h_{out}(\spk_4,\spk_4)\bs\s)]_{\la_4'\la_4},
\end{gathered}
\end{equation}
and
\begin{equation}\label{tM_in_sigma}
    \tilde{M}_{\la_3\la_4\la_1\la_2}=\tilde{M}^{(0)} \de_{\la_3\la_1}\de_{\la_4\la_2} +\tilde{M}^{(1)}_a (\s_a)_{\la_3\la_1}\de_{\la_4\la_2} +\tilde{M}^{(2)}_a \de_{\la_3\la_1}(\s_a)_{\la_4\la_2} +\tilde{M}^{(3)}_{ab} (\s_a)_{\la_3\la_1}(\s_b)_{\la_4\la_2}.
\end{equation}
Consequently,
\begin{equation}
    \sum_{\la_1,\la_2,\la_3} \rho^{s(1)}_{\la_1\la_3} \tilde{M}_{\la_3\la_4'\la_1\la_2} \rho^h_{\la_2\la_4}-H.c.=i\de_{\la_4'\la_4} A+i(\s_a)_{\la_4'\la_4} B_a,
\end{equation}
where
\begin{equation}\label{A_B_defn}
\begin{split}
    A=&\im\big[\rho^{s(1)}\rho^h\big(\tilde{M}^{(0)} +\xi^s_a\tilde{M}^{(1)}_a +\tilde{M}^{(2)}_a\xi^h_a +\xi^s_a\tilde{M}^{(3)}_{ab}  \xi^h_b\big)\big],\\
    B_a=&\im\big\{ \rho^{s(1)}\rho^h \big[(\tilde{M}^{(0)} +\xi^s_b\tilde{M}^{(1)}_b)\xi^h_a +\tilde{M}^{(2)}_a +\xi^s_b \tilde{M}^{(3)}_{ba} +i\e_{abc} ( \tilde{M}^{(2)}_{b} +\xi^s_d \tilde{M}^{(3)}_{db}) \xi^h_c \big]\big\}.
\end{split}
\end{equation}
Substituting these expressions into the density matrix \eqref{dens_matr_out1}, moving from the integration variables $(\spk_1,\spk_3)$ to the new ones
\begin{equation}\label{spk_31_spq}
    \spk_{s}=(\spk_1+\spk_3)/2,\quad\spq=\spk_1-\spk_3;\qquad\spk_1=\spk_{s}+\spq/2,\quad \spk_3=\spk_{s}-\spq/2,
\end{equation}
and evaluating the integral over $\spk_2$, we arrive at
\begin{equation}\label{rho_out_0}
    (\rho^h_{out})_{\la_4'\la_4}(\spk_4,\spk_4)=\rho^h_{\la_4'\la_4}(\spk_4,\spk_4) -\frac{\pi}{2} \int\frac{d\spq d\spk_{s}}{(2\pi)^3} \frac{\de(k_3^0+k_4^0-k_1^0-k_2^0)}{\sqrt{k_1^0 k_2^0 k_3^0 k_4^0}} \big[\de_{\la_4'\la_4}A +(\s_a)_{\la_4'\la_4} B_a \big]\Big|_{\spk_2=\spk_4-\spq},
\end{equation}
where it is understood that $\spk_1$ and $\spk_3$ are expressed through $\spk_{s}$ and $\spq$ as in \eqref{spk_31_spq}. Using the representation \eqref{dens_matr_repr} for the density matrix of a scattered probe photon, we obtain
\begin{equation}\label{rho_out_1}
    \rho^h_{out}=\rho^h+a,\qquad\bs\xi^h_{out}=\frac{\rho^h\bs\xi^h+\mathbf{b}}{\rho^h+a}\approx\big(1-\frac{a}{\rho_h}\big)\bs\xi^h +\frac{\mathbf{b}}{\rho_h},
\end{equation}
where
\begin{equation}
    a=-\pi \int\frac{d\spq d\spk_{s}}{(2\pi)^3} \frac{\de(k_3^0+k_4^0-k_1^0-k_2^0)}{\sqrt{k_1^0 k_2^0 k_3^0 k_4^0}} A\Big|_{\spk_2=\spk_4-\spq},\qquad b_a=-\pi \int\frac{d\spq d\spk_{s}}{(2\pi)^3} \frac{\de(k_3^0+k_4^0-k_1^0-k_2^0)}{\sqrt{k_1^0 k_2^0 k_3^0 k_4^0}} B_a\Big|_{\spk_2=\spk_4-\spq}.
\end{equation}
It is seen from \eqref{A_B_defn} that if $\bs\xi^h(\spk_2,\spk_4)$ is real-valued, what is the case, for example, when $\bs\xi^h$ does not depend on $\spk_2$ and $\spk_4$, then the following relation holds,
\begin{equation}
    A=B_a \xi^h_a,
\end{equation}
for $(\bs\xi^h)^2=1$. As it was discussed in \cite{KazSokNeut}, this relation reflects the fact that, in scattering of a probe photon prepared initially in the pure state with respect to spin, its final state stays pure with respect to spin in the order of perturbation theory we consider.

\section{Evolution of the Stokes parameters of a probe photon}\label{Evolut_Stokes_Param_Probe}

Below we will consider the case of a small recoil $q^\mu\rightarrow0$ and retain only the leading contributions in powers of  $q^\mu$. We assume that
\begin{equation}\label{small_recoil_cond}
    |\spq|\ll|\spk^s_0|,\qquad |\spq|\ll|\spk^h_0|,\qquad -q^2\ll m^2,\qquad -q^2\ll s,
\end{equation}
where $\spk^s_0$ and $\spk^h_0$ are the typical values of momenta in the state of target photons and in the state of a probe one, respectively, and $m$ is the electron mass. The invariant scattering amplitude obeys the symmetry relations \cite{LandLifQED}
\begin{equation}
    M_{\la_1\la_2\la_3\la_4}=M_{-\la_1,-\la_2,-\la_3,-\la_4},\qquad M_{\la_1\la_2\la_3\la_4}=M_{\la_3\la_4\la_1\la_2},\qquad M_{\la_1\la_2\la_3\la_4}=M_{\la_2\la_1\la_4\la_3}.
\end{equation}
In the small momentum transfer limit, there are only the three independent nonzero chiral amplitudes,
\begin{equation}
    m_+:=M_{++++}(s),\qquad m_-:=M_{+-+-}(s),\qquad n:=M_{++--}(s),
\end{equation}
where
\begin{equation}\label{M_t0}
    m_+=8\al^2 f(s),\qquad  m_-=8\al^2 f(-s),\qquad n=-8\al^2 g(s),
\end{equation}
and
\begin{equation}
\begin{split}
    f(s)&=-\Big[1+\Big(2-\frac{4}{s'}\Big)B(s') +\Big(-4+\frac{4}{s'}\Big)B(-s') +\Big(\frac{4}{s'}-\frac{8}{s'^2}\Big)T(s') +\Big(2-\frac{4}{s'}-\frac{8}{s'^2}\Big)T(-s') \Big]_{s'\rightarrow s/m^2},\\
    g(s)&=-\Big[1 +\frac{4}{s'}B(s') -\frac{4}{s'}B(-s') +\frac{8}{s'^2} T(s') +\frac{8}{s'^2}T(-s') \Big]_{s'\rightarrow s/m^2}.
\end{split}
\end{equation}
The functions
\begin{equation}
    B(s)=\sqrt{1-\frac{4}{s}}\arsh \frac{\sqrt{-s}}{2}-1=\sqrt{\frac{4}{s}-1}\arcsin \frac{\sqrt{s}}{2}-1,\qquad T(s)=\arsh^2\frac{\sqrt{-s}}{2}=-\arcsin^2\frac{\sqrt{s}}{2},
\end{equation}
where the principal branches of the multivalued functions are chosen and $s\rightarrow s+i0$.

In the low energy limit, where $s$, $|t|$, and $|u|$ are much less than $4m^2$, there are five independent chiral amplitudes in the large quantum recoil regime. These independent amplitudes are reduced to
\begin{equation}\label{M_low_energy}
\begin{gathered}
    M_{++++}=\frac{11\al^2}{45 m^4}s^2 +\frac{4\al^2}{315 m^6}s^3,\qquad M_{+-+-}=\frac{11\al^2}{45 m^4}u^2 +\frac{4\al^2}{315 m^6}u^3,\qquad
    M_{+--+}=\frac{11\al^2}{45 m^4}t^2 +\frac{4\al^2}{315 m^6}t^3,\\
    M_{++--}=-\frac{\al^2}{15m^4}(s^2+t^2+u^2) -\frac{2\al^2}{189m^6}(s^3+t^3+u^3),\qquad M_{+++-}=-\frac{\al^2}{945 m^6}(s^3+t^3+u^3).
\end{gathered}
\end{equation}
In the general case, the explicit expressions for $M_{\la_1\la_2\la_3\la_4}$ are given in \cite{KarpNeu50,KarpNeu51,Tollis64,Tollis65,ConTollPist71,LandLifQED}. Notice that the scattering amplitude and the effective dielectric susceptibility of a classical electromagnetic field found by means of the Heisenberg-Euler Lagrangian \cite{BBBB70,MarShuk06,FMST07,KarbShai15,KarbMos20,KUMZ22,Ahmadiniaz2023,FIKKSTT, Piazza2006,Matheron2026,Ahmadiniaz2025,Rinderknecht2025,Varf66,Smid2025,Berezin2024,Wang2024, Karb18,DHIMT14,DHIMT141,BatRiz13,Tommasini2010,King2010,Valialshchikov2025,Bu2026} contain the terms not higher than of the fourth power in the photon momenta. Therefore, they correspond to the terms standing at $1/m^4$ in the invariant scattering amplitude \eqref{M_low_energy}.

As long as relations \eqref{phi_s_phi_h} hold for the transition matrices $U^n$ in the small quantum recoil limit, the invariant scattering amplitude in the basis of polarization vectors $f^\mu_{(\la)}(\spk_n)$ becomes
\begin{equation}
    \tilde{M}_{\la_3\la_4\la_1\la_2}= M_{\la_3\la_4\la_1\la_2} e^{i(\la_1-\la_3)\vf_s} e^{i(\la_2-\la_4)\vf_h}.
\end{equation}
The nonzero amplitudes are written as
\begin{equation}
    \tilde{M}_{++++}=\tilde{M}_{----}=m_+,\qquad \tilde{M}_{+-+-}=\tilde{M}_{-+-+}=m_-,\qquad \tilde{M}_{--++}=n_+, \qquad \tilde{M}_{++--}=n_-,
\end{equation}
where
\begin{equation}\label{n_pm}
    n_\pm=ne^{\pm2i(\vf_s+\vf_h)}.
\end{equation}
Then the coefficients entering into expansion \eqref{tM_in_sigma} are given by
\begin{equation}\label{tM_elements}
\begin{gathered}
    \tilde{M}^{(0)}=(m_++m_-)/2,\qquad \tilde{M}^{(1)}_a=\tilde{M}^{(2)}_a=0,\qquad \tilde{M}^{(3)}_{13}=\tilde{M}^{(3)}_{31}=\tilde{M}^{(3)}_{23}=\tilde{M}^{(3)}_{32}=0,\\
    \tilde{M}^{(3)}_{11}=-\tilde{M}^{(3)}_{22}=(n_++n_-)/4,\qquad \tilde{M}^{(3)}_{33}=(m_+-m_-)/2,\qquad \tilde{M}^{(3)}_{12}=\tilde{M}^{(3)}_{21}=-i(n_+-n_-)/4.
\end{gathered}
\end{equation}
If $\bs\xi^h=const$, then a variation of the density matrix of a probe photon \eqref{rho_out_1} is governed by the equations
\begin{equation}\label{de_rho_de_xi}
    \de\rho^h=\eta +(\bs\eta''\bs\xi^h),\qquad \rho^h\de\bs\xi^h= \bs\eta'' -(\bs\eta''\bs\xi^h) \bs\xi^h +[\bs\eta',\bs\xi^h],
\end{equation}
where
\begin{equation}
    \eta:=\im\lan\tilde{M}^{(0)}\ran,\qquad \eta_a:=\lan\xi^s_b \tilde{M}^{(3)}_{ba}\ran,\qquad\eta'_a=\re\eta_a,\qquad \eta''_a=\im\eta_a,
\end{equation}
and, for brevity, we have denoted by angle brackets the integral
\begin{equation}\label{inclus_int}
    \lan F\ran:=-\pi \int\frac{d\spq d\spk_{s}}{(2\pi)^3} \frac{\de(k_3^0+k_4^0-k_1^0-k_2^0)}{\sqrt{k_1^0 k_2^0 k_3^0 k_4^0}} \rho^{s(1)}(\spk_1,\spk_3) \rho^h(\spk_2,\spk_4) F \bigg|_{\substack{\spk_1=\spk_s+\spq/2,\\ \spk_2=\spk_4-\spq, \\ \spk_3=\spk_s-\spq/2}}.
\end{equation}
Introducing the one-particle density matrices at the instant of time $x^0$,
\begin{equation}
    \rho^{s(1)}(x^0;\spk_1,\spk_3)=e^{-ix^0(k_1^0 -k_3^0)} \rho^{s(1)}(\spk_1,\spk_3),\qquad \rho^{h}(x^0;\spk_2,\spk_4)=e^{-ix^0(k_2^0 -k_4^0)} \rho^{s(1)}(\spk_2,\spk_4),
\end{equation}
the main integral \eqref{inclus_int} can be cast into the form
\begin{equation}\label{inclus_int_t}
    \lan F\ran:=-\pi \int\frac{dx^0d\spq d\spk_{s}}{(2\pi)^4} \frac{\rho^{s(1)}(x^0) \rho^h(x^0)}{\sqrt{k_1^0 k_2^0 k_3^0 k_4^0}}  F,
\end{equation}
where, for brevity, the arguments having the same form as in \eqref{inclus_int} are omitted. Notice that the contraction appearing in the definition of $\eta_a$ is reduced to
\begin{equation}\label{xi_tM}
    \xi^s_b \tilde{M}^{(3)}_{b1}=\frac14(n_-\xi^s_++n_+\xi^s_-),\qquad \xi^s_b \tilde{M}^{(3)}_{b2}=\frac{i}4(n_-\xi^s_+-n_+\xi^s_-), \qquad \xi^s_b \tilde{M}^{(3)}_{b3}=\frac{1}2(m_+-m_-)\xi^s_3,
\end{equation}
where $\xi^s_\pm:=\xi^s_1\pm i\xi^s_2$.

The latter relation in \eqref{de_rho_de_xi} implies that a change of degree of polarization of a probe photon obeys the equation
\begin{equation}
    \rho^h \de\sqrt{1-\bs\xi_h^2}=-(\bs\eta''\bs\xi^h) \sqrt{1-\bs\xi_h^2},
\end{equation}
in the given order of perturbation theory. This equation shows, in particular, that a completely polarized state of the probe photon, i.e., a pure state with respect to the spin degree of freedom, remains completely polarized on scattering by target photons. Furthermore, it follows from the second equation in \eqref{de_rho_de_xi} that if the quantum state of target photons is not polarized, $\bs\xi^s=0$, then $\eta_a=0$ and the polarization of a probe photon does not change, $\de\bs\xi^h=0$. In that case, only the probability to record a photon with given momentum varies. On the other hand, if the state of target photons is polarized and the probe photon is not polarized, $\bs\xi^h=0$, then the probe photon can become polarized on scattering.

In the case $\xi^s_1=\xi^s_2=0$, i.e., when the state of target photons possesses a circular polarization\footnote{Recall that we work in the chiral basis where the Stokes parameters are related to the Stokes paramters in the basis of linear polarizations \eqref{polarization_vects} as $\xi_1^l=\xi_2$, $\xi_2^l=\xi_3$, and $\xi_3^l=\xi_1$.}, equations \eqref{de_rho_de_xi} determining the evolution of the Stokes parameters turn into
\begin{equation}\label{xi_evol_gyro}
    \rho^h\de\xi^h_3=\eta''_3\big(1-(\xi^h_3)^2\big),\qquad \rho^h\de\xi^h_+=i(\eta_3'+i\eta_3''\xi^h_3)\xi^h_+,
\end{equation}
where $\xi^h_\pm:=\xi^h_1\pm i\xi^h_2$ and relations \eqref{xi_tM} have been taken into account. Equations \eqref{xi_evol_gyro} formally coincide with equations (74) of the paper \cite{KazSokNeut} for a change of the Stokes parameters of a probe photon in coherent scattering by neutrons (or a neutron), the quantum state of neutrons being spin polarized. In Sec. \ref{Susc_Gas_Phot}, we shall show that the quantum state of target photons with circular polarization has a gyrotropic susceptibility tensor just as the spin polarized quantum state of neutrons.

The integral \eqref{inclus_int} can be simplified when the quantum recoil is small \eqref{small_recoil_cond}. In this case, the argument of the delta function expressing the energy conservation law,
\begin{equation}
    E(\spk_1,\spk_3;\spk_4):=k_3^0+k_4^0-k_1^0-k_2^0=|\spk_3|+|\spk_4|-|\spk_1|-|\spk_4-\spk_1+\spk_3|,
\end{equation}
is written as
\begin{equation}\label{energy_cons_law_app}
    E(\spk_s+\spq/2,\spk_s-\spq/2;\spk_4)=(\spn_4-\spn_{s})\spq -\frac{1}{2|\spk_4|}\big[\spq^2-(\spq\spn_4)^2\big] +O(q^3/\spk_s^2,q^3/\spk_h^2),
\end{equation}
where $\spn_4=\spk_4/|\spk_4|$ and $\spn_{s}=\spk_{s}/|\spk_{s}|$. Let us decompose the vector $\spq$ as
\begin{equation}\label{q_par_perp_split}
    \spq=q_\parallel \frac{\De\spn}{|\De\spn|}+\spq_\perp,\qquad q_\parallel:=\frac{(\spq\De\spn)}{|\De\spn|},\qquad \De\spn:=\spn_4-\spn_s,
\end{equation}
where $(\spq_\perp \De\spn)=0$. We will study the nondegenerate case when in the region, where the integration in \eqref{inclus_int} is effectively carried out, the absolute value of the first term on the right-hand side of \eqref{energy_cons_law_app} is much larger than the absolute value of the second and third terms, viz., when
\begin{equation}
    |\De\spn|\gg |\spq|/|\spk_4|,\qquad |\De\spn|\gg \spq^2/\spk_s^2,
\end{equation}
in this domain. Notice that the degenerate case is much harder to realize in experiments than the nondegenerate one since the probe photons move in the same direction as the target photons. The region of their interaction is large and the target photons can hit the detector of probe photons.

Further simplification of the integral \eqref{inclus_int} can be achieved provided the additional assumptions are made about the behavior of the one-particle density matrices. We suppose that the integration in \eqref{inclus_int} is effectively performed in the domain of the variables $\spq$ and $\spk_s$ where
\begin{equation}\label{slow_depend_rho}
    \Big|\frac{\partial\ln(\rho^{s(1)}\rho^h F)}{\partial q^i}\Big| \frac{\spq^2}{|\spk_4|}\ll1,\qquad \Big|\frac{\partial\ln(\rho^{s(1)}\rho^h F)}{\partial q^i}\Big| \frac{|\spq|^3}{|\spk_s|}\ll1.
\end{equation}
Then neglecting the second term on the right-hand side of \eqref{energy_cons_law_app} and the terms of higher order, we transform the integral \eqref{inclus_int} into
\begin{equation}\label{inclus_int_nondeg}
    \lan F\ran=- \int\frac{d\spq_\perp d\spk_{s}}{8 \pi^2 |\spk_4||\spk_{s}||\De\spn|} \rho^{s(1)}\rho^hF \Big|_{\substack{\spq=\spq_\perp,\\\spk_2=\spk_4-\spq_\perp}},
\end{equation}
in the leading order in $q$. In the next section, we shall evaluate this integral for Gaussian one-particle density matrices.

Now we consider the particular case where the one-particle density matrices of the probe photon and of the target photons are concentrated near the momenta $\spk^h_0$ and $\spk^s_0$, respectively, and the vector $\spk^h_0$ is not collinear with $\spk^s_0$. We assume that  estimates \eqref{small_recoil_cond} are fulfilled, whereas conditions \eqref{slow_depend_rho} are not imposed yet. In that case, it is clear from \eqref{A_B_defn} that $|\spq|$ cannot be much larger than the minimum of the standard deviations of momenta in the one-particle density matrices of a probe photon and of a target photon gas. Furthermore, it is useful to take the vector $\mathbf{d}$ specifying the polarization vectors to be orthogonal to both the vectors $\spk^h_0$ and $\spk^s_0$ in the laboratory frame. In this case, we infer from \eqref{a_n_b_n_appr} that
\begin{equation}
    b_h\approx -b_s,
\end{equation}
in the leading order in $q$. For $\spn^h_0\neq \spn^s_0$, in virtue of the energy conservation law,
\begin{equation}
    E=0,
\end{equation}
it follows from \eqref{energy_cons_law_app} that
\begin{equation}
    (\spn^h_0\spq)=(\spn^s_0\spq),
\end{equation}
in the leading order in $q$. Using this relation, it is not difficult to ascertain that
\begin{equation}
    a_h\approx a_s,
\end{equation}
for $\mathbf{d}$ orthogonal to $\spn^h_0$ and $\spn^s_0$. Therefore, under the above conditions, $\vf_h=-\vf_s$ and
\begin{equation}
    n_+=n_-=n.
\end{equation}
The matrix $\tilde{M}^{(3)}_{ab}$ becomes diagonal and
\begin{equation}
    \tilde{M}^{(0)}=\frac{m_++m_-}{2}=8\al^2 f_s(s),\qquad \tilde{M}^{(3)}_{11}=-\tilde{M}^{(3)}_{22}=\frac{n}2 =-4\al^2 g(s),\qquad \tilde{M}^{(3)}_{33}=\frac{m_+-m_-}{2}= 8\al^2 f_a(s),
\end{equation}
where $f_s(s):=(f(s)+f(-s))/2$ and $f_a(s):=(f(s)-f(-s))/2$. It is this case that was investigated in the paper \cite{KazSol2023}. Then, in the leading order, we have for the parameters determining the dynamics of the Stokes parameters
\begin{equation}\label{eta_narrow_wp}
    \eta=8\al^2 \im\big(f_s(s_0)\lan 1\ran\big),\qquad \eta_1=-4\al^2 g(s_0)\lan \xi^s_1\ran,\qquad \eta_2=4\al^2g(s_0) \lan \xi^s_2\ran,\qquad \eta_3=8\al^2f_a(s_0) \lan \xi^s_3\ran,
\end{equation}
where $s_0=|\spk^0_s||\spk_4|(\De\spn_0)^2$ and $\De\spn_0=\spn_4-\spn_s^0$. The main integral appearing in these expressions is written as
\begin{equation}\label{F_int}
    \lan F\ran=- \int\frac{d\spq d\spk_{s}}{8\pi^2} \frac{\de(E)}{|\spk_s^0| |\spk_4|} \rho^{s(1)}(\spk_1,\spk_3) \rho^h(\spk_2,\spk_4) F \bigg|_{\substack{\spk_2=\spk_4-\spq,\\ \spk_1=\spk_s+\spq/2,\\ \spk_3=\spk_s-\spq/2}},
\end{equation}
to the same accuracy. If, additionally, conditions \eqref{slow_depend_rho} are met, then we have
\begin{equation}
    \lan F\ran=- \frac{|\De\spn_0|}{s_0} \int\frac{d\spq_\perp d\spk_{s}}{8\pi^2}  \rho^{s(1)}(\spk_s+\spq_\perp/2,\spk_s-\spq_\perp/2) \rho^h(\spk_4-\spq_\perp,\spk_4) F \Big|_{\substack{\spk_1=\spk_s+\spq_\perp/2,\\ \spk_3=\spk_s-\spq_\perp/2}},
\end{equation}
in the leading order in $q$. In the case $\bs\xi^s=const$, we arrive at
\begin{equation}\label{eta_through_I}
    \eta=-\frac{\al^2}{\pi^2} \im\Big(\frac{f_s(s_0)}{s_0} I\Big),\qquad \eta_1=\frac{\al^2}{2\pi^2} \frac{g(s_0)}{s_0} I \xi^s_1,\qquad \eta_2=-\frac{\al^2}{2\pi^2} \frac{g(s_0)}{s_0}I \xi^s_2,\qquad \eta_3=-\frac{\al^2}{\pi^2} \frac{f_a(s_0)}{s_0} I \xi^s_3,
\end{equation}
where
\begin{equation}\label{I_integral}
    I:= |\De\spn_0|\int d\spq_\perp d\spk_{s} \rho^{s(1)}(\spk_{s}+\spq_\perp/2,\spk_{s}-\spq_\perp/2) \rho^h(\spk_4-\spq_\perp,\spk_4).
\end{equation}
Substituting these expressions into equations \eqref{de_rho_de_xi}, we describe the evolution of the Stokes parameters of the density matrix of a probe photon in scattering by target photons.

The integrals appearing in \eqref{F_int} and \eqref{I_integral} can be written as a convolution of the probe photon density matrix with the Fourier transforms of the particle number and spin densities of a target photon gas in the coordinate space. To this end, we introduce the Wigner function -- the Weyl symbol of the one-particle density matrix of a photon gas -- and the Weyl symbol of the spin density matrix (for definitions and properties of the different symbols of operators see, e.g., \cite{BerezMSQ})
\begin{equation}\label{Wigner_funcs}
\begin{split}
    \rho_s(x,\spk_s)&:=\int\frac{d\spq}{(2\pi)^3}e^{i\spq\spx}\rho^{s(1)}(x^0;\spk_{s}+\spq/2,\spk_{s}-\spq/2),\\
    \varsigma^s_a(x,\spk_s)&:= \int\frac{d\spq}{(2\pi)^3}e^{i\spq\spx}\rho^{s(1)}(x^0;\spk_{s}+\spq/2,\spk_{s}-\spq/2) \xi^s_a(x^0;\spk_{s}+\spq/2,\spk_{s}-\spq/2).
\end{split}
\end{equation}
Then the functions,
\begin{equation}\label{phot_num_dens_spin_dens}
    \rho_s(x)=\int d\spk_s \rho_s(x,\spk_s), \qquad \varsigma^s_a(x) =\int d\spk_s \varsigma^s_a(x,\spk_s),
\end{equation}
are the photon number density and the spin density in the coordinate space, respectively. Their Fourier transforms with respect to spatial coordinates are defined as
\begin{equation}
    \tilde{\rho}_s(x^0;\spq)= \int d\spx e^{-i\spq\spx} \rho_s(x),\qquad \tilde{\varsigma}^s_a(x^0;\spq)=\int d\spx e^{-i\spq\spx} \varsigma^s_a(x).
\end{equation}
Therefore, representing the delta function expressing the energy conservation law in \eqref{F_int} as the Fourier transform, we obtain
\begin{equation}\label{1_ave_as_convol}
\begin{split}
    \lan1\ran&=-\frac{\pi}{|\spk^0_s||\spk_4|}\int\frac{dx^0 d\spq}{(2\pi)^4} \rho^h(x^0;\spk_4-\spq,\spk_4) \tilde{\rho}_s(x^0;\spq),\\
    \lan\xi^s_a\ran&=-\frac{\pi}{|\spk^0_s||\spk_4|}\int\frac{dx^0 d\spq}{(2\pi)^4} \rho^h(x^0;\spk_4-\spq,\spk_4) \tilde{\varsigma}^s_a(x^0;\spq).
\end{split}
\end{equation}
As for the integral \eqref{I_integral}, we have analogously
\begin{equation}\label{I_integral_as_convol}
    I=|\De\spn_0| \int d\spq_\perp \rho^h(\spk_4-\spq_\perp,\spk_4) \tilde{\rho}_s(0;\spq_\perp).
\end{equation}
These formulas allow one to restore the photon number and spin densities of the target photon gas by using the hologram.

In the low energy limit, the functions entering into \eqref{eta_narrow_wp} and \eqref{eta_through_I} are given by
\begin{equation}\label{fsfag_asymp}
\begin{gathered}
    f_s(s)=\frac{11s^2}{360m^4}+\frac{13s^4}{21600m^8}+\cdots,\qquad f_a(s)=\frac{s^3}{630 m^6}+\frac{s^5}{17325m^{10}}+\cdots,\\
    g(s)=\frac{s^2}{60m^4}+\frac{s^4}{1890m^8}+\cdots.
\end{gathered}
\end{equation}
The values of these functions at the electron-positron pair creation threshold are given in formula (49) of \cite{KazSol2023}. Their asymptotics for $s\gg 4m^2$ look as
\begin{equation}\label{fsfag_asymp_HE}
    f_s(s)\approx -\frac12\Big(5-\frac{\pi^2}{4} +\ln^2\frac{s}{iem^2}\Big),\qquad f_a(s)\approx -\frac{i\pi}{2}\ln\frac{s}{ie^3m^2},\qquad g(s)\approx-1.
\end{equation}
As it has been already noted above, the Heisenberg-Euler Lagrangian reproduces only the terms at $1/m^4$ in expansions \eqref{fsfag_asymp}. In particular, $f_a(s)=0$ for such an approximation. The second equation in \eqref{de_rho_de_xi} with the parameters $\eta$ and $\eta_a$ given in \eqref{eta_through_I} resembles Eq. (15) from the paper \cite{KotSerb97} (see also \cite{Maishev97,Sawyer04,BelMai12}). They, of course, do not coincide if only because that the integral $I$ is absent in Eq. (15) of \cite{KotSerb97}. However, if $I$ is real and one identifies $dt$ with $(\al/m)^2 I/(4\pi\rho^h)$, where $dt$ is the reduced optical thickness introduced in \cite{KotSerb97}, then the second equation in \eqref{de_rho_de_xi} agrees with Eq. (15) of \cite{KotSerb97} for head-on scattering.

\section{Examples}\label{Examples}

\subsection{Scattering by a Gaussian wave packet}\label{Scat_by_Gaussian}

Let us consider in detail the case where the one-particle density matrices of the probe photon and of the target photons have the form of Gaussians,
\begin{equation}\label{one_part_dens_matr}
\begin{split}
    \rho^{s(1)}(\spk_s,\spk'_s)&=c_se^{-\frac14\de \spk_s g_s \de \spk_s -\frac14\de \spk'_s g_s \de \spk'_s}e^{-i(\spk_s-\spk'_s)\mathbf{b}_s},\\
    \rho^{h}(\spk_h,\spk'_h)&=c_he^{-\frac14\de \spk_h g_h \de \spk_h -\frac14\de \spk'_h g_h \de \spk'_h}e^{-i(\spk_h-\spk'_h)\mathbf{b}_h},
\end{split}
\end{equation}
where the matrix notation has been used, $g^{-1}_s$ and $g^{-1}_h$ are the matrices of variances in the momentum space, $\de \spk_s:=\spk_s-\spk_s^0$, $\de \spk'_s:=\spk'_s-\spk_s^0$, $\de \spk_h:=\spk_h-\spk_h^0$, and $\de \spk'_h:=\spk'_h-\spk_h^0$, and $\mathbf{b}_{s,h}$ are the impact parameters. It is supposed that these one-particle density matrices are narrow in the momentum space, i.e., the standard deviations of momenta are much less than their average values $\spk_s^0$ and $\spk_h^0$. The normalization constants are found from the normalization conditions \eqref{norm_conds},
\begin{equation}
    c_s=\frac{\det^{1/2}(g_s)}{(2\pi)^{3/2}} N_s,\qquad c_h=\frac{\det^{1/2}(g_h)}{(2\pi)^{3/2}}.
\end{equation}
We also assume that $\bs\xi^s=const$ and conditions \eqref{slow_depend_rho} are satisfied. Let $\mathbf{b}:=\mathbf{b}_h-\mathbf{b}_s$, $\s^2_s$ be minimal of the dispersions $g^{-1}_s$, and $\s^2_h$ be minimal of the dispersions $g^{-1}_h$. Then conditions \eqref{slow_depend_rho} are fulfilled provided that
\begin{equation}\label{b_conds}
    |\mathbf{b}|\max\Big(\frac{\min(\s^2_s,\s^2_h)}{|\spk_h||\De\spn_0|}, \frac{\min(\s^3_s,\s^3_h)}{\spk_s^2|\De\spn_0|}\Big)\ll1.
\end{equation}
When all the above conditions are met, it is sufficient to evaluate the integral \eqref{I_integral} in order to find the evolution of the Stokes parameters of the density matrix of a probe photon.

This integral reads
\begin{equation}\label{I_int_Gauss}
    I=c_sc_h|\De\spn_0|\int d\spq_\perp d\spk_s e^{-S},
\end{equation}
where using the matrix notation
\begin{equation}\label{S_intI}
    S:=\frac12 \de \spk_hg_h\de \spk_h +\frac12 \de \spk_sg_s\de \spk_s -(i\mathbf{b}_\perp +\frac12\de \spk_hg_h)\spq_\perp +\frac18 \spq_\perp g_s\spq_\perp +\frac14 \spq_\perp g_h \spq_\perp.
\end{equation}
At first, we evaluate the integral over $\spq_\perp$ employing formula \eqref{I_q_int}. Define the antisymmetric matrix
\begin{equation}
    \De_{ij}:=\e_{ijk}\De n_k.
\end{equation}
Then the integral \eqref{I_int_Gauss} is reduced to
\begin{equation}\label{I_int_Gauss_1}
    I=c_sc_h |\De\spn_0|^2\int  \frac{d\spk_s2\sqrt{2}\pi}{\sqrt{-\Sp(\De g\De g)}}\exp\Big(-\frac12 \de \spk_hg_h\de \spk_h -\frac12 \de \spk_sg_s\de \spk_s +\frac{(i\mathbf{b}_\perp +\frac12\de \spk_hg_h)\De g\De (i\mathbf{b}_\perp +\frac12g_h \de \spk_h)}{\Sp(\De g\De g)} \Big),
\end{equation}
where, for brevity, the matrix of inverse dispersions of the recoil momentum $\spq$ has been introduced,
\begin{equation}
    g:=\frac12 g_h+\frac14 g_s.
\end{equation}
Under the fulfilment of conditions \eqref{b_conds}, one can replace,
\begin{equation}
    \De_{ij}\rightarrow \De^0_{ij}=\e_{ijk}\De n^0_k,
\end{equation}
in the integrand of \eqref{I_int_Gauss_1}. As a result, the integral over $\spk_s$ is readily evaluated giving rise to
\begin{equation}\label{I_int_Gauss_2}
\begin{split}
    I&=  \frac{2\sqrt{2}\pi N_s |\De\spn_0|^2}{\sqrt{-\Sp(\De^0 g\De^0 g)}}\exp\Big(\frac{(i\mathbf{b} +\frac12\de \spk_hg_h)\De^0 g\De^0 (i\mathbf{b} +\frac12g_h \de \spk_h)}{\Sp(\De^0 g\De^0 g)} \Big)\rho^h(\spk_4,\spk_4)=\\
    &=\frac{N_s \det^{1/2}(g_h) |\De\spn_0|^2}{\sqrt{-\pi\Sp(\De^0 g\De^0 g)}}\exp\Big(-\frac12 \de \spk_hg_h\de \spk_h +\frac{(i\mathbf{b} +\frac12\de \spk_hg_h)\De^0 g\De^0 (i\mathbf{b} +\frac12g_h \de \spk_h)}{\Sp(\De^0 g\De^0 g)} \Big).
\end{split}
\end{equation}
Substituting this expression into \eqref{eta_through_I}, we obtain the parameters $\eta$ and $\eta_a$ determining a variation of the Stokes parameters of the one-particle density matrix of a probe photon.

It is assumed in expression \eqref{I_int_Gauss_2} that the one-particle density matrices of target photons and of a probe photon have the form \eqref{one_part_dens_matr} at the instant of time $t=0$. It is not difficult to generalize expression \eqref{I_int_Gauss_2} to the case where the density matrix of a probe photon is of the form \eqref{one_part_dens_matr} at the instant of time $t=t_0$ for not too large $|t_0|$. Namely, under the additional assumption
\begin{equation}\label{t_0_cond}
    \frac{|t_0||\spq|^3}{\min(\spk_s^2,\spk_h^2)}\ll1,
\end{equation}
it is sufficient to make the replacement
\begin{equation}\label{t_0_b_g_replacement}
    \spb\rightarrow\spb -t_0\spn_4,\qquad g_{ij}\rightarrow g_{ij}-\frac{it_0}{2|\spk_4|} (\de_{ij} -n^4_i n^4_j),
\end{equation}
in expression \eqref{I_int_Gauss_2} (see Appendix \ref{Time_Shift_App} for a detailed derivation). The replacement \eqref{t_0_b_g_replacement} takes into account the shift and spreading of the density matrix of a probe photon during the time $t_0$ in the leading order in $q$.

%%%%%%%%%% figure here %%%%%%%%%%%%%
%%%%%%%%%% hologram. Gaussian head-on %%%%%

\begin{figure}[t!]
\centering
\includegraphics*[width=0.325\linewidth]{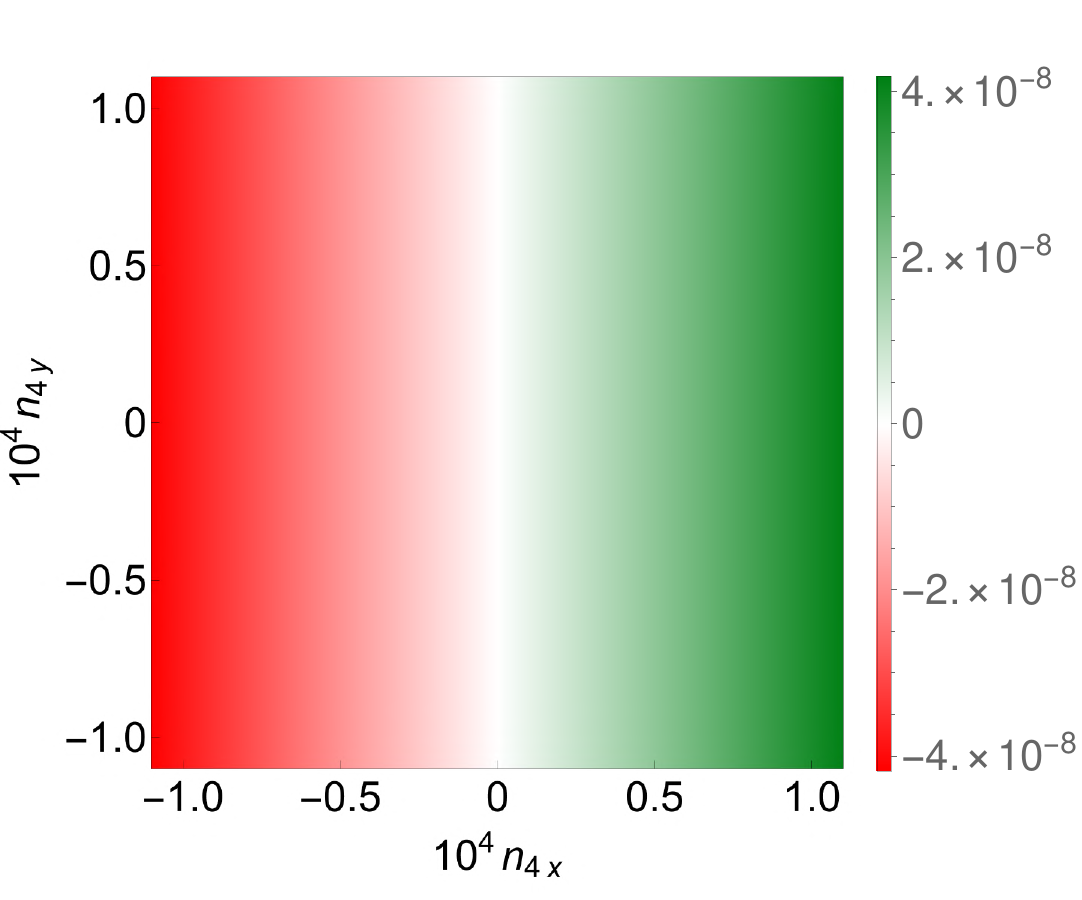}\,
\includegraphics*[width=0.307\linewidth]{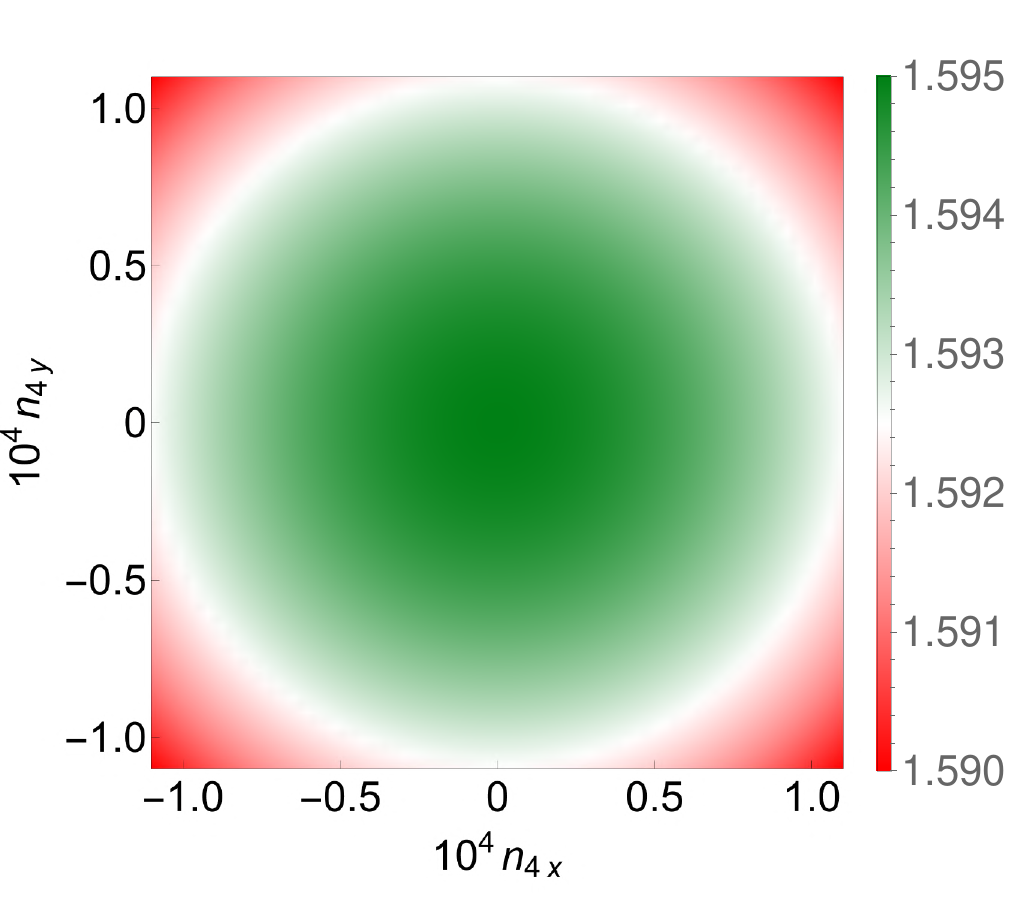}\,
\includegraphics*[width=0.337\linewidth]{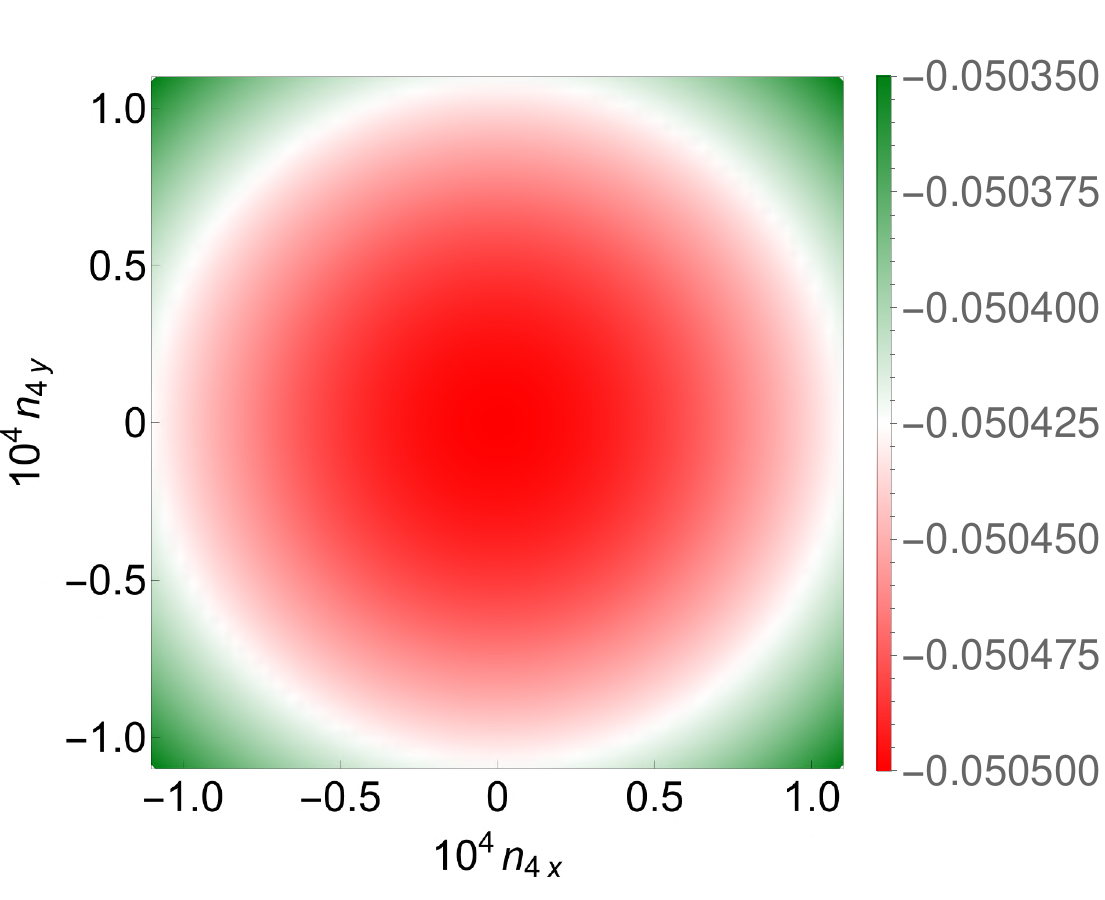}
\caption{{\footnotesize The parameters $\eta/\rho^h$ (left panel), $\eta'_1/\rho_h$ (middle panel), and $\eta'_3/\rho_h$ (right panel) determining a variation of the Stokes parameters of the density matrix of a probe photon in scattering by Gaussian in head-on configuration. The mean values of momenta: $\spk_s^0=(0,0,1.55)$ eV and $\spk_h^0=(0,0,-7)$ GeV. The matrices of inverse dispersions have the form \eqref{inverse_dispersions} with $\s_\perp^s/|\spk^0_s|=10^{-2}$, $\s_\parallel^s/|\spk^0_s|= 10^{-5}$, $\s_\perp^h/|\spk^0_h|= 7.3\times 10^{-5}$, and $\s_\perp^h/|\spk^0_h|=10^{-4}$. The impact parameter $\spb=(1,0,0)$ $\mu$m and the undulator strength parameter $K_u=1$. The domain of $(n_x,n_y)$ shown belongs to the region where $\de\spk_h g_h\de\spk_h<10$. We see that the parameter $\eta'_1/\rho_h$ defining the change of the probe photon polarization is of order of unity.}}
\label{Plot_Gaussian}
\end{figure}

%%%%%%%%%% figure here %%%%%%%%%%%%%

In order to estimate the magnitude of the effect of coherent light-by-light scattering, we consider a particular case of head-on scattering of a probe photon by a photon gas. In this case, $\spn_4\approx-\spn^0_s$ and we choose the $z$ axis so that $\spn^0_s=\spe_3$. Consequently, $\De\spn^0\approx-2\spe_3$ and $|\De\spn^0|\approx2$. Moreover, we will consider cylindrical beams with
\begin{equation}\label{inverse_dispersions}
\begin{gathered}
    g_s=\diag\big((\s^s_\perp)^{-2},(\s^s_\perp)^{-2},(\s^s_\parallel)^{-2}\big),\qquad g_h=\diag\big((\s^h_\perp)^{-2},(\s^h_\perp)^{-2},(\s^h_\parallel)^{-2}\big),\\
    g=\diag\big((\s_\perp)^{-2},(\s_\perp)^{-2},(\s_\parallel)^{-2}\big),
\end{gathered}
\end{equation}
where
\begin{equation}
    \frac{1}{\s_\perp^2}:=\frac{1}{2(\s^h_\perp)^2}+ \frac{1}{4(\s^s_\perp)^2},\qquad \frac{1}{\s_\parallel^2}:=\frac{1}{2(\s^h_\parallel)^2}+ \frac{1}{4(\s^s_\parallel)^2}.
\end{equation}
We suppose that the impact parameter, $\mathbf{b}$, and the momentum of recorded probe photon, $\spk_4$, are such that the exponent on the first line of formula \eqref{I_int_Gauss_2} is of order of unity. Then
\begin{equation}
    I/\rho^h\approx4\pi N_s\s_\perp^2.
\end{equation}
The average number of photons in the beam of target photons can be expressed through the intensity of radiation at the center of the Gaussian beam (see, e.g., formula (72) of \cite{rydet})
\begin{equation}
    I_m=\frac{4}{\pi(2\pi)^{1/2}}|\spk_s^0|N_s\s^s_\parallel(\s^s_\perp)^2,
\end{equation}
or through the undulator strength parameter \cite{BaKaStrbook,Bord.1}
\begin{equation}\label{K_u_defn}
    K_u^2:=\frac{e^2\mathbf{A}^2}{m^2}\approx 2\pi\al \frac{I_m}{m^2 |\spk_s^0|^2}.
\end{equation}
In the definition of the undulator strength parameter, the electromagnetic potential is taken in the Coulomb gauge. In physics of strong laser fields, this parameter is called the classical intensity parameter or the laser strength parameter. Then it follows from equations \eqref{de_rho_de_xi} describing dynamics of the Stokes parameters of a probe photon that the relative magnitude of the effect is  approximately
\begin{equation}\label{estimate_rel_mag_1}
    \Big|\frac{\de\rho^h}{\rho^h}\Big|\approx(2\pi)^{1/2} \al^2 \frac{|f_s(s_0)|}{s_0} \frac{I_m}{|\spk_s^0|\s^s_\parallel}\Big(\frac{\s_\perp}{\s^s_\perp}\Big)^2\approx \frac{\al}{(2\pi)^{1/2}} \frac{m^2|f_s(s_0)|}{s_0} K_u^2 \frac{|\spk_s^0|}{\s^s_\parallel} \Big(\frac{\s_\perp}{\s^s_\perp}\Big)^2.
\end{equation}
Taking $\s^s_\perp\ll\s^h_\perp$, we have $(\s_\perp/\s^s_\perp)^2\approx 4$. In that case, for the energy of photons in the target  $|\spk_s^0|=1.55$ eV and the energy of the probe photon $|\spk_4|=7$ GeV \cite{Hajima2016}, the variable $s_0\approx0.17 m^2$ and
\begin{equation}\label{effect_rel_val}
    \Big|\frac{\de\rho^h}{\rho^h}\Big|\approx6.0\times10^{-5} K_u^2 \frac{|\spk_s^0|}{\s^s_\parallel}.
\end{equation}
For the relative laser bandwidth of order $\s^s_\parallel/|\spk_s^0| \sim 10^{-5}$, we infer that the relative magnitude of the effect \eqref{effect_rel_val} is of order of unity for $K_u\sim 1$. Such values of the undulator strength parameter are achievable at the modern laser facilities \cite{FIKKSTT}. Furthermore, for $K_u\lesssim 1$, the perturbation theory that we use to describe light-by-light scattering is applicable. For $K_u\gg 1$, the external electromagnetic field has to be taken into account nonperturbatively. The polarization of GeV probe photons can be measured by the gamma-ray polarimetry \cite{NakHom17,BCKLP22,BMKP17}. The plots of the parameters $\eta/\rho^h$ and $\eta_a/\rho^h$ determining a relative magnitude of the effect are presented in Fig. \ref{Plot_Gaussian}.

\subsection{Scattering by a lattice of Gaussians}\label{Scatt_by_Lattice}

Let us consider scattering of a probe photon with the initial density matrix given on the second line of \eqref{one_part_dens_matr} by a photon gas being in the quantum state with the one-particle density matrix
\begin{equation}\label{dens_matr_N_Gauss_coh}
    \rho^{s(1)}(\spk_s,\spk'_s)=c_s^N e^{-\frac14\de \spk_s g_s \de \spk_s -\frac14\de \spk'_s g_s \de \spk'_s} \sum_{n,n'=1}^{N} \kappa_n\kappa^*_{n'} e^{-i\spk_s \mathbf{b}_n +i\spk'_s \mathbf{b}_{n'}},
\end{equation}
where the vectors $\mathbf{b}_n$ constitute a lattice with generators $\mathbf{b}_{(a)}$, $a=\overline{1,3}$, and
\begin{equation}
    \mathbf{b}_n=\sum_{a=1}^3 \mathbf{b}_{(a)} n_a,\qquad n_a=\overline{-N_a,N_a},\qquad N=\prod_{a=1}^3(2N_a+1).
\end{equation}
The index $n$ is uniquely defined by the set $\{n_a\}$. We suppose that the basis $\{\spb_{(a)}\}$ forms a right-handed triple. The normalization constant is equal to
\begin{equation}
    c_s^N=c_s/ \sum_{n,n'=1}^N \kappa_n\kappa^*_{n'} e^{-\frac12 \mathbf{b}_{nn'}g^{-1}_s \mathbf{b}_{nn'}-i \spk_s^0\mathbf{b}_{nn'}},
\end{equation}
where $\mathbf{b}_{nn'}:=\mathbf{b}_n-\mathbf{b}_{n'}$. The one-particle density matrix \eqref{dens_matr_N_Gauss_coh} describes $N$ Gaussians shifted relative to the origin by the vectors $\mathbf{b}_n$ in the coordinate space and multiplied by the complex factors $\kappa_n$. We assume that $N_a\gg1$ and the eigenvalues of the matrix $b_{(a)}g_s^{-1}b_{(b)}$ are of order of or larger than five. The last condition means that the distance between Gaussians is much larger than their typical sizes in the coordinate space. Then
\begin{equation}\label{c_s_N_appr}
    c_s^N\approx c_s/\sum_{n=1}^N |\kappa_n|^2.
\end{equation}
For $|\kappa_n|=1$ we have
\begin{equation}
    c_s^N\approx c_s/N.
\end{equation}
Below, we shall consider the two cases: $\kappa_n=1$ and $\kappa_n=e^{i\chi_n}$, where $\chi_n$ are stochastic phases. Introduce the photon number density in the coordinate space at the instant of time $t=0$ as
\begin{equation}\label{phot_num_dens_latt}
\begin{split}
    \rho_s(\spx)=&\int\frac{d\spk d\spk'}{(2\pi)^3} e^{i(\spk-\spk')\spx}\rho^{s(1)}(\spk,\spk')=\\
    =&\,\frac{8c_s^N}{\det(g_s)}\sum_{n,n'=1}^N \kappa_n\kappa^*_{n'} e^{-i\spk_s^0\spb_{nn'}} e^{-(\spx-\spb_n)g^{-1}_s(\spx-\spb_n) -(\spx-\spb_{n'})g^{-1}_s(\spx-\spb_{n'})}\approx\\
    \approx&\, \frac{8c_s^N}{\det(g_s)}\sum_{n=1}^N  |\kappa_n|^2 e^{-2(\spx-\spb_n)g^{-1}_s(\spx-\spb_n)},
\end{split}
\end{equation}
where in the last approximate equality we have made the same approximation as in \eqref{c_s_N_appr}, i.e., have neglected the overlap of  Gaussians in the coordinate space. Of course, formula \eqref{phot_num_dens_latt} is in concordance with the general definition \eqref{phot_num_dens_spin_dens}. We see that the photon number density in the coordinate space at the instant of time $t=0$ is approximately the same in the two cases we investigate.

In both cases, we assume that all the conditions guaranteeing that the parameters specifying the evolution of the Stokes parameters of a probe photon have the form \eqref{eta_narrow_wp} are satisfied. In other words, we assume that the small recoil approximation \eqref{small_recoil_cond} is valid and the one-particle density matrices of the target photons and of the probe photon are concentrated near the momenta $\spk^0_s$ and $\spk^0_h$, respectively. To simplify the calculations, we also suppose that $\bs\xi^s=const$, i.e., all Gaussians possess the same polarizations. Introduce standardly the basis vectors of the reciprocal lattice, $\mathbf{w}_{(a)}$, $a=\overline{1,3}$,
\begin{equation}
    (\spb_{(a)}\spw_{(b)})=2\pi\de_{ab}.
\end{equation}
We additionally require that
\begin{equation}
    |(\spb_h\spw_{(a)})|\ll2N_a+1,
\end{equation}
i.e., the impact parameter is much smaller than the size of the target photon gas. In fact, in order to describe the evolution of the Stokes parameters, it is sufficient to evaluate the integral \eqref{F_int} with $F=1$.

\subsubsection{Coherent lattice}\label{Scatt_by_Lattice_Coh}

Let us start with the case $\kappa_n=1$, i.e., with the case of a coherent sum of Gaussians. Then the structure factor in the one-particle density matrix \eqref{dens_matr_N_Gauss_coh} is equal to
\begin{equation}
    \sum_{n,n'=1}^{N} e^{-i\spk \mathbf{b}_n +i\spk' \mathbf{b}_{n'}}=\prod_{a=1}^3 \Big[\frac{\sin\big(\frac{\spk\spb_{(a)}}{2}(2N_a+1)\big)}{\sin\frac{\spk\spb_{(a)}}{2}} \frac{\sin\big(\frac{\spk'\spb_{(a)}}{2}(2N_a+1)\big)}{\sin\frac{\spk'\spb_{(a)}}{2}}\Big].
\end{equation}
This expression is concentrated near the momenta
\begin{equation}\label{resonant_momenta}
    \spk=\spk_r=\sum_{a=1}^3 \spw_{(a)} r_a,\qquad \spk'=\spk_{r'}=\sum_{a=1}^3 \spw_{(a)} r'_a,
\end{equation}
where the indices $r$ and $r'$ are uniquely defined by the sets $\{r_a\}$ and $\{r'_a\}$, respectively. The dimensions of the humps in the momentum space are of order
\begin{equation}
    |(\spk-\spk_r,\spb_{(a)})|\approx\frac{\pi}{N_a},\qquad |(\spk'-\spk_{r'},\spb_{(a)})|\approx\frac{\pi}{N_a}.
\end{equation}
Since under the assumptions made above these peaks are narrow on the scale of variations of the one-particle density matrix of a probe photon and of the envelope of the one-particle density matrix of target photons, the integral \eqref{F_int} can be approximated by
\begin{equation}\label{1_averaged_0}
    \lan 1\ran \approx-\sum_{\{r_a\},\{r'_a\}=-\infty}^\infty \frac{\rho^{s(1)}_0(\spk_r,\spk_{r'}) \rho^h(\spk_4-\spk_r+\spk_{r'},\spk_4)}{8\pi^2|\spk_s^0| |\spk_4|}F_{rr'}(\spk_4),
\end{equation}
where
\begin{equation}\label{F_int_cell}
\begin{split}
    F_{rr'}(\spk_4)&:=\int_{C_{rr'}} d\spk d\spk' \de \big(E(\spk,\spk';\spk_4)\big) \prod_{a=1}^3 \Big[\frac{\sin\big(\frac{\spk\spb_{(a)}}{2}(2N_a+1)\big)}{\sin\frac{\spk\spb_{(a)}}{2}} \frac{\sin\big(\frac{\spk'\spb_{(a)}}{2}(2N_a+1)\big)}{\sin\frac{\spk'\spb_{(a)}}{2}}\Big],\\
    \rho^{s(1)}_0(\spk_s,\spk'_{s})&:=c_s^N e^{-\frac14\de \spk_s g_s \de \spk_s -\frac14\de \spk'_s g_s \de \spk'_s},
\end{split}
\end{equation}
and $C_{rr'}$ is the unit cell in the momentum space $(\spk,\spk')$ constituted by the basis vectors of the reciprocal lattice with the center at the point \eqref{resonant_momenta}. It remains to evaluate the integral $F_{rr'}(\spk_4)$, which is the function of $\spk_4$ and the sets of integer numbers $\{r_a\}$ and $\{r_a'\}$ only. It does not depend on the form of the one-particle density matrices.

In order to find the estimate for this integral, we approximate the product of sine ratios in the integrand in the unit cell $C_{rr'}$ by Gaussian exponents
\begin{equation}\label{form_fact_coh}
    \prod_{a=1}^3 \Big[\frac{\sin\big(\frac{\spk\spb_{(a)}}{2}(2N_a+1)\big)}{\sin\frac{\spk\spb_{(a)}}{2}} \frac{\sin\big(\frac{\spk'\spb_{(a)}}{2}(2N_a+1)\big)}{\sin\frac{\spk'\spb_{(a)}}{2}}\Big]\approx \prod_{a=1}^3 (2 N_a+1)^2 e^{-\frac{N_a(N_a+1)}{\pi} \big[(\spk\spb_{(a)}-2\pi r_a)^2 + (\spk'\spb_{(a)}-2\pi r'_a)^2\big]}.
\end{equation}
This approximation gives the exact value of the approximated expression at its maximum and the integral of it over the unit cell $C_{rr'}$ coincides with the integral of the exact expression over this cell when $N_a\gg1$. If necessary, one can use more accurate approximations in the form of Gaussian exponents multiplied by some polynomials. All the below calculations can be easily generalized to the approximants of such a form.

%%%%%%%%%% figure here %%%%%%%%%%%%%
%%%%%%%%%% hologram. coh lattice. compar %%%%%

\begin{figure}[t!]
\centering
\includegraphics*[width=0.334\linewidth]{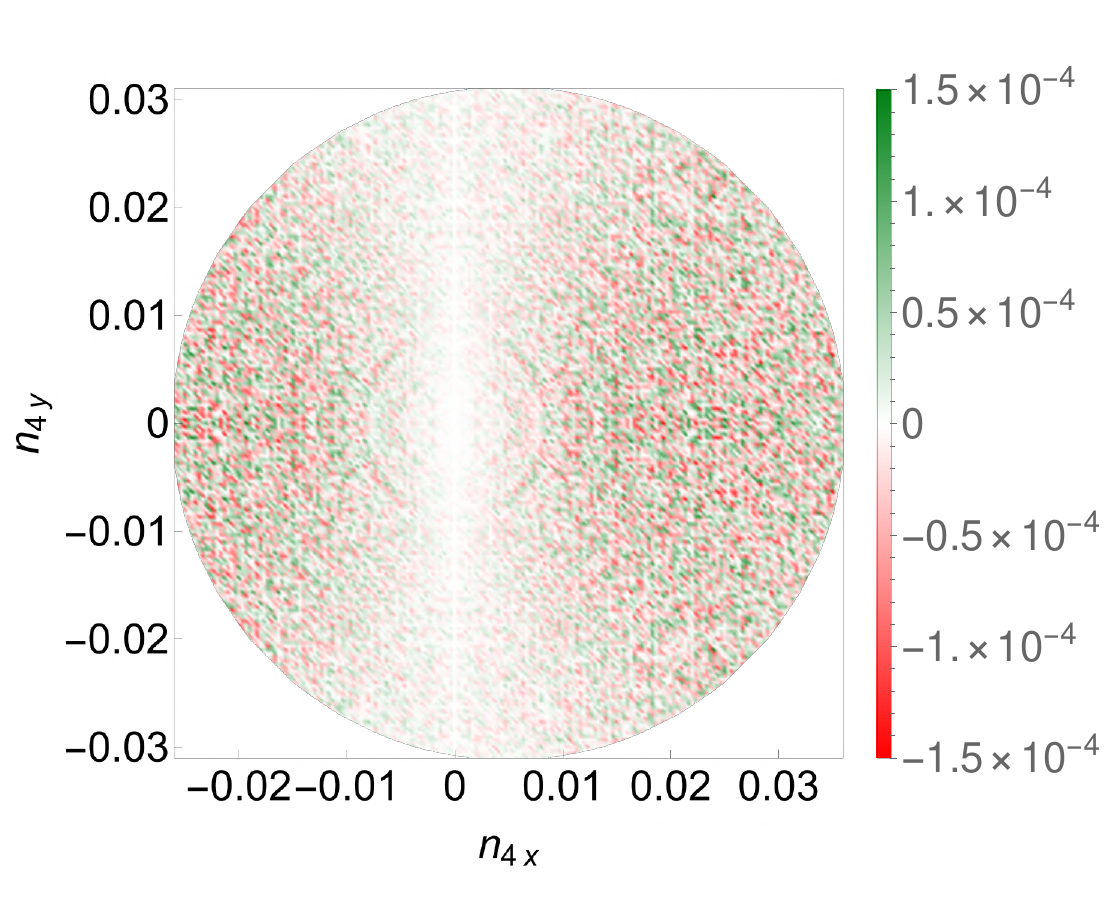}\,
\includegraphics*[width=0.307\linewidth]{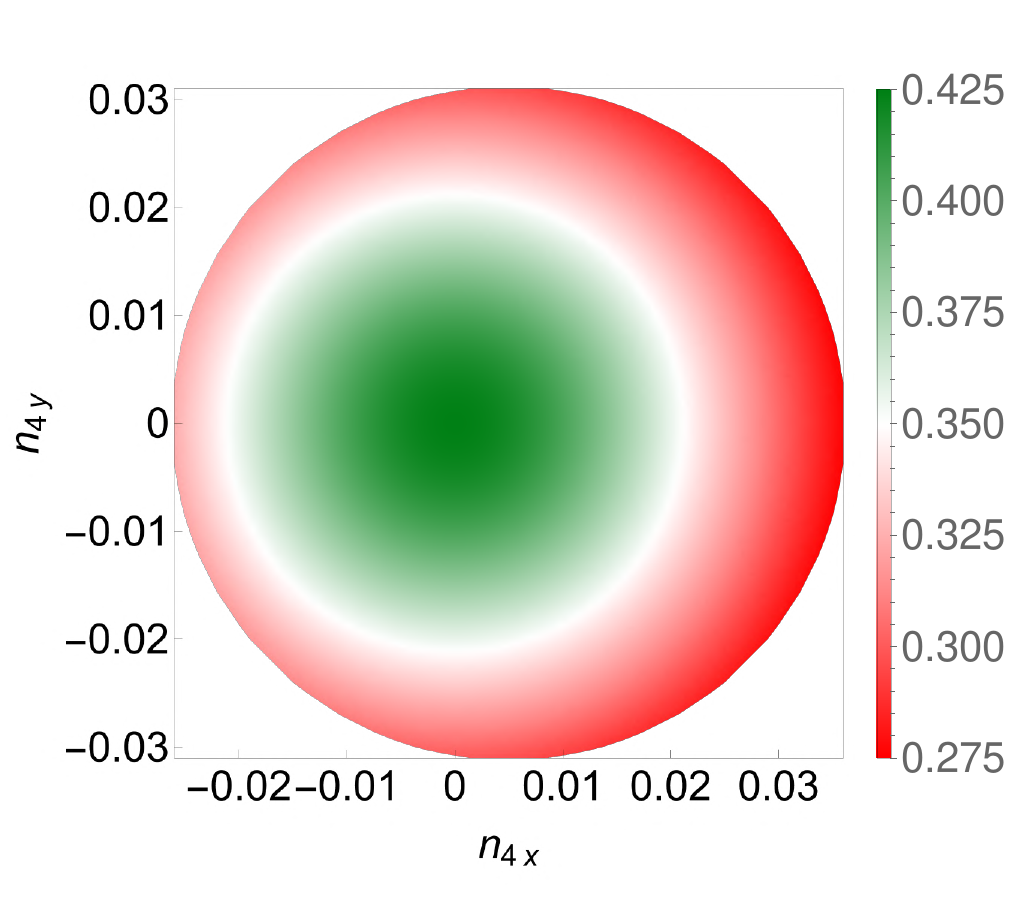}\,
\includegraphics*[width=0.316\linewidth]{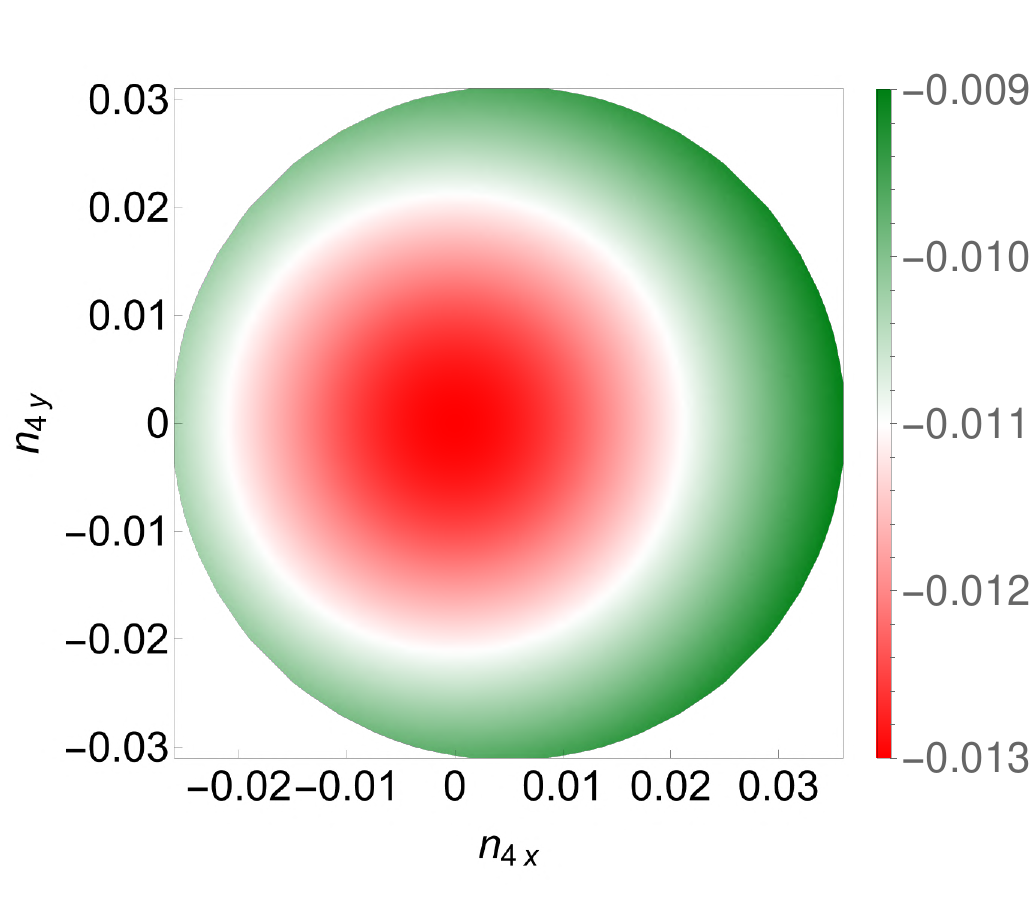}
\caption{{\footnotesize The parameters $\eta/\rho^h$ (left panel), $\eta'_1/\rho_h$ (middle panel), and $\eta'_3/\rho_h$ (right panel) determining a variation of the Stokes parameters of the density matrix of a probe photon in scattering by a coherent lattice. The mean values of momenta: $\spk_s^0=(0,0,1.55)$ eV and $\spk_h^0=(0,0,-7)$ GeV. The matrices of inverse dispersions have the form \eqref{inverse_dispersions} with $\s_\perp^s/|\spk^0_s|=10^{-3}$, $\s_\parallel^s/|\spk^0_s|=10^{-3}$, $\s_\perp^h/|\spk^0_h|=10^{-2}$, and $\s_\perp^h/|\spk^0_h|=10^{-3}$. The parameters of the lattice: $N_1=N_2=N_3=20$, $\spb_1=(3,0,0)/\s_\perp$, $\spb_2=(0,3,0)/\s_\perp$, and $\spb_3=(0,0,3)/\s_\parallel$. The density matrix of the probe photon is rotated by an angle of $5\times10^{-3}$ around the $y$ axis so that an almost head-on scattering is realized. The impact parameter $\spb=(1,0,0)$ $\mu$m and the undulator strength parameter $K_u=1$. The domain of $(n_x,n_y)$ is shown where $\de\spk_h g_h\de\spk_h<10$.}}
\label{Plot_Coh_Latt_1}
\end{figure}

%%%%%%%%%% figure here %%%%%%%%%%%%%

The argument of the delta function, $E$, is a slowly varying function of $\spk$ and $\spk'$ in the domain where expression \eqref{form_fact_coh} is essentially different from zero. Therefore, we can develop $E$ as a series in terms of deviation from the point  \eqref{resonant_momenta} up to the first order in $(\de\spk_r,\de\spk'_{r'})$,
\begin{equation}\label{E_expansion}
    E\approx E_0 +(\bs\ups_{rr'},\de\spk_r) -(\bs\ups'_{rr'},\de\spk'_{r'})+\cdots,
\end{equation}
where
\begin{equation}
\begin{gathered}
    E_0:=E(\spk_r,\spk_{r'};\spk_4),\qquad\bs\ups_{rr'}:=\spn_{2;rr'}-\spn_r,\qquad \bs\ups'_{rr'}:=\spn_{2;rr'}-\spn_{r'},\\ \spn_r:=\frac{\spk_r}{|\spk_r|},\qquad \spn_{r'}:=\frac{\spk_{r'}}{|\spk_{r'}|}, \qquad\spn_{2;rr'}:=\frac{\spk_4 -\spk_r +\spk_{r'}}{|\spk_4-\spk_r+\spk_{r'}|}.
\end{gathered}
\end{equation}
We assume that $\spk_4$ is taken so that $E_0=0$. If $E_0\neq0$ for given $\spk_4$, then $F_{rr'}(\spk_4)$ is small for sufficiently large $N_a$ in comparison with $F_{rr'}(\spk_4)$ with momentum $\spk_4$ satisfying the condition $E_0=0$ (see formula \eqref{1_averaged_1_E_0} below). To evaluate the integral \eqref{F_int_cell}, we employ the representation of the delta function as the limit
\begin{equation}\label{delta_epsilon}
    \de(x)=\lim_{\e\rightarrow0}\frac{1}{\e\sqrt{2\pi}} e^{-x^2/(2\e^2)}.
\end{equation}
Then replacing integration over the unit cell in the integral \eqref{F_int_cell} by integration over the whole  momentum space, we arrive at the Gaussian integral of the form \eqref{Gaussian_int_gen},
\begin{equation}\label{F_int_cell_appr}
    F_{rr'}(\spk_4)\approx \frac{\prod_{a=1}^3(2N_a+1)^2}{\e\sqrt{2\pi}}\int d\spk d\spk' e^{-\frac{1}{2\e^2}[(\bs\ups_{rr'}\de\spk_r)-(\bs\ups'_{rr'}\de\spk'_{r'})]^2} e^{-\sum_a \frac{N_a(N_a+1)}{\pi}[(\de\spk_r\spb_{(a)})^2 +(\de\spk'_{r'}\spb_{(a)})^2]}.
\end{equation}
The details of calculations of this integral are presented in Appendix \ref{Evaluation_Gaauss_Int_App}. As a result, we have
\begin{equation}\label{F_int_cell_appr1}
    F_{rr'}(\spk_4)\approx \frac{\det\nolimits^2(w^{(a)}_i)}{\sqrt{2\pi}\sqrt{\bs\ups_{rr'} b^{-1}\bs\ups_{rr'} +\bs\ups'_{rr'} b^{-1}\bs\ups'_{rr'}}},
\end{equation}
where it has been used that $\e\rightarrow0$ and $N_a\gg1$. The explicit expression for $b^{-1}_{ij}$ is given in formula \eqref{det_b_b_inv}. Hence
\begin{equation}\label{1_averaged}
\begin{split}
    \lan 1\ran &\approx-\frac{\det\nolimits^2(w^{(a)}_i)}{8\pi^2|\spk_s^0| |\spk_4|} \sum_{\{r_a\},\{r'_a\}=-\infty}^\infty \frac{\rho^{s(1)}_0(\spk_r,\spk_{r'}) \rho^h(\spk_4-\spk_r+\spk_{r'},\spk_4)}{\sqrt{2\pi}\sqrt{\bs\ups_{rr'} b^{-1}\bs\ups_{rr'} +\bs\ups'_{rr'} b^{-1}\bs\ups'_{rr'}}}\approx\\
    &\approx -\frac{\det\nolimits^2(w^{(a)}_i)}{8\pi^2|\spk_s^0| |\spk_4| } \sum_{\{r_a\},\{r'_a\}=-\infty}^\infty \frac{\rho^{s(1)}_0(\spk_r,\spk_{r'}) \rho^h(\spk_4-\spk_r+\spk_{r'},\spk_4)}{\sqrt{4\pi\De\spn_0 b^{-1}\De\spn_0}},
\end{split}
\end{equation}
where the summation is carried out over those $\{r_a\}$, $\{r'_a\}$ that obey the condition $E_0=0$ for given $\spk_4$. In the last approximate equality, the small recoil approximation \eqref{small_recoil_cond} has been taken into account. The condition $E_0=0$ is fulfilled at $r_a=r'_a$ for any $\spk_4$. Singling out this contribution, we obtain
\begin{equation}\label{1_averaged_1}
    \lan 1\ran = -\frac{\det\nolimits^2(w^{(a)}_i)}{8\pi^2|\spk_s^0| |\spk_4| }\frac{ \rho^h(\spk_4,\spk_4)}{\sqrt{4\pi\De\spn_0 b^{-1}\De\spn_0}} \Big[\sum_{\{r_a\}=-\infty}^\infty \rho^{s(1)}_0(\spk_r,\spk_{r}) +\sideset{}{'}\sum_{\{r_a\},\{r'_a\}=-\infty}^\infty \frac{\rho^{s(1)}_0(\spk_r,\spk_{r'}) \rho^h(\spk_4-\spk_r+\spk_{r'},\spk_4)}{\rho^h(\spk_4,\spk_4)}\Big],
\end{equation}
where the prime at the sum sign means that the terms with $r_a=r'_a$ are absent. Expression \eqref{1_averaged} or \eqref{1_averaged_1} ought to be substituted into formula \eqref{eta_narrow_wp} for the parameters $\eta$, $\eta_a$ that determine a variation of the Stokes parameters of the density matrix of a probe photon. It is seen that for a fixed average number of target photons proportional to $N_s/N$, expression \eqref{1_averaged} is proportional to $\max(N_a)$. This means that for a fixed energy density of the target electromagnetic field, a variation of the density matrix of a probe photon increases by a factor of $\max(N_a)$ due to modulation of the one-particle density matrix of target photons.

Expression \eqref{1_averaged_1} is readily generalized to the case where $E_0\neq0$. Substituting expansion \eqref{E_expansion} into \eqref{delta_epsilon} and evaluating the resulting Gaussian integral for $F_{rr'}(\spk_4)$ with the aid of formulas  \eqref{Gaussian_int_gen}, \eqref{S_albe}, \eqref{S_det}, and \eqref{S_inv}, we come to
\begin{equation}\label{1_averaged_1_E_0}
    \lan 1\ran = -\frac{\det\nolimits^2(w^{(a)}_i)}{8\pi^2|\spk_s^0| |\spk_4| }\frac{ \rho^h(\spk_4,\spk_4)}{\sqrt{4\pi\De\spn_0 b^{-1}\De\spn_0}} \sum_{\{r_a\},\{r'_a\}=-\infty}^\infty \frac{\rho^{s(1)}_0(\spk_r,\spk_{r'}) \rho^h(\spk_4-\spk_r+\spk_{r'},\spk_4)}{\rho^h(\spk_4,\spk_4)} e^{-\frac{E^2(\spk_r,\spk_{r'};\spk_4)}{2(\bs\ups_{rr'} b^{-1}\bs\ups_{rr'} +\bs\ups'_{rr'} b^{-1}\bs\ups'_{rr'})}}.
\end{equation}
The generalization of expression \eqref{1_averaged_1} to the case where the density matrix of a probe photon has the form \eqref{one_part_dens_matr} at the instant of time $t=t_0$ for not too large $|t_0|$ satisfying the estimate \eqref{t_0_appr_applic3} is given in formula \eqref{1_averaged_1_t_0}.

\begin{figure}[t!]
\centering
\includegraphics*[width=0.32\linewidth]{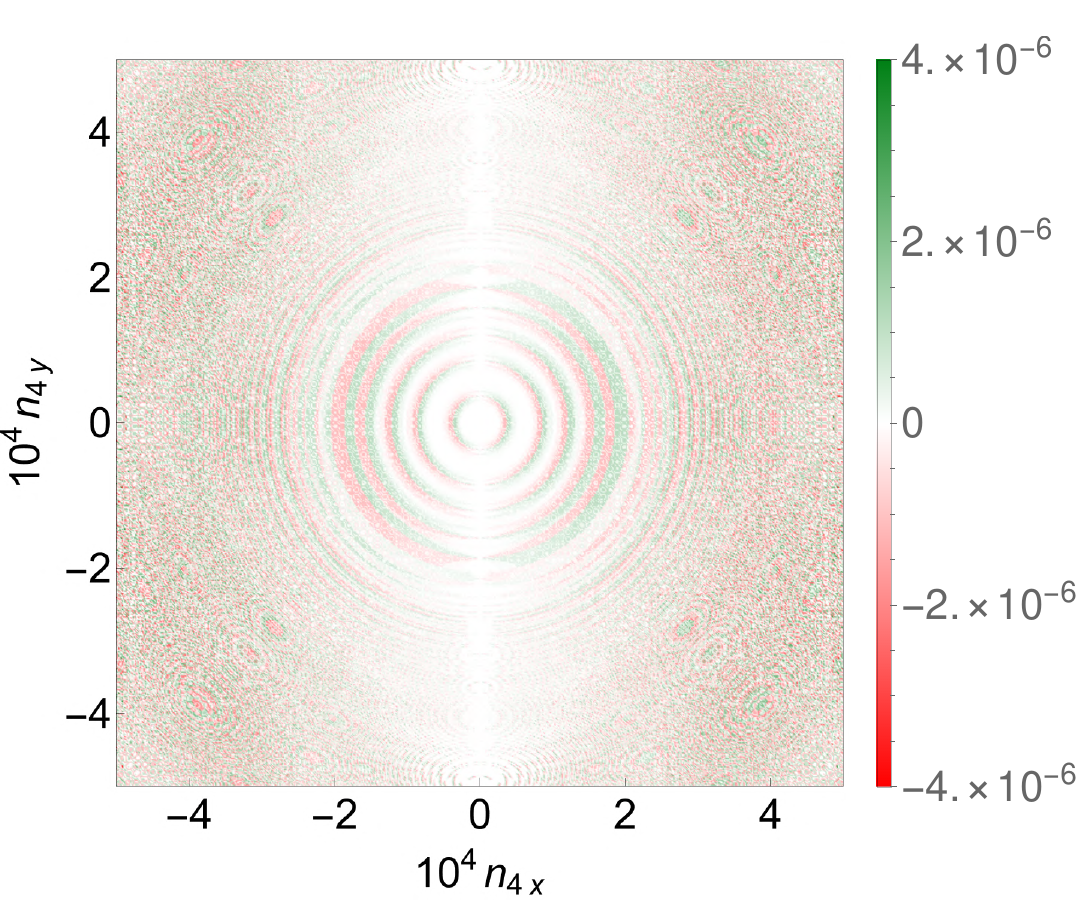}\,
\includegraphics*[width=0.315\linewidth]{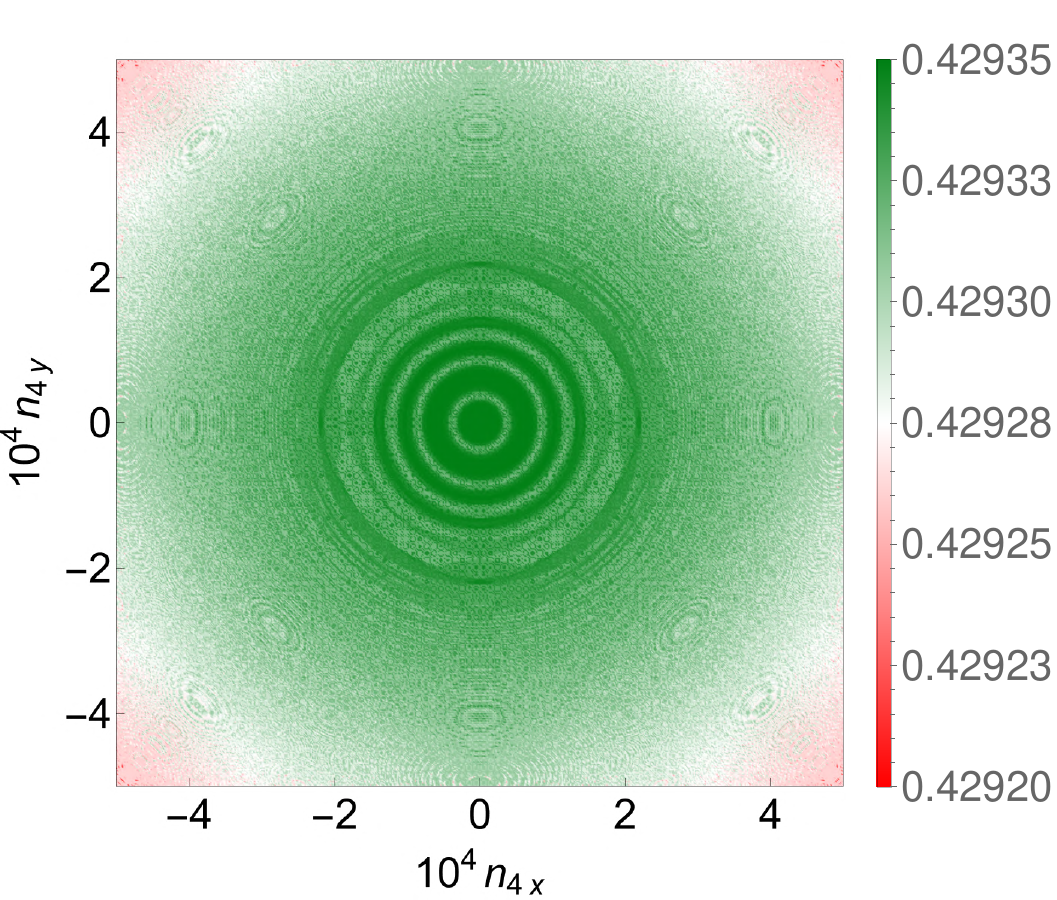}\,
\includegraphics*[width=0.326\linewidth]{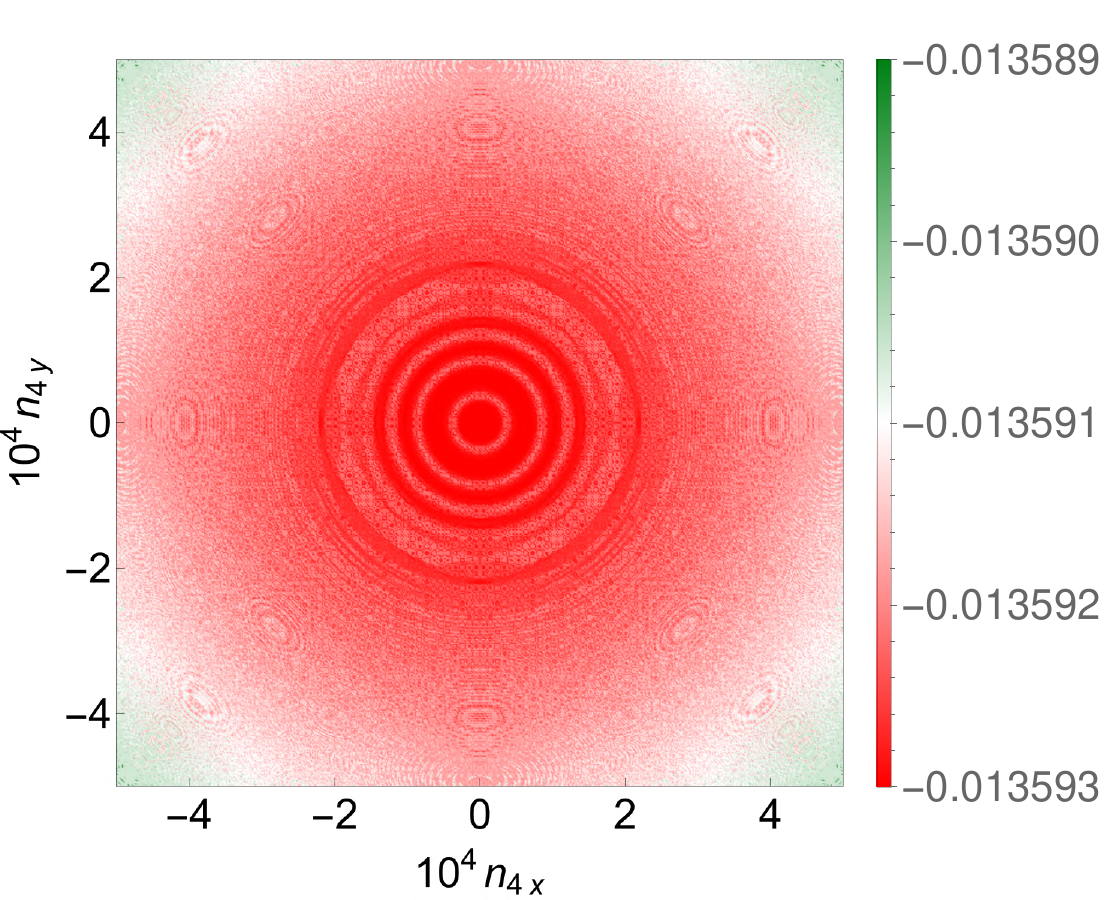}
\caption{{\footnotesize The same as in Fig. \ref{Plot_Coh_Latt_1} but for smaller values of $(n_x,n_y)$.}}
\label{Plot_Coh_Latt}
\end{figure}

In the small recoil limit, the condition $E_0=0$ means that $\spk_4$ is chosen so that
\begin{equation}\label{resonance_cond}
    \spn_4\spq_{rr'}=\frac{\spk_r+\spk_{r'}}{|\spk_r+\spk_{r'}|}\spq_{rr'}, \qquad\spq_{rr'}=\spk_r-\spk_{r'}.
\end{equation}
This condition can be written for $\spq_{rr'}\neq0$ as
\begin{equation}
    \cos(\widehat{\spn_4\spq_{rr'}})=\frac{|\spk_r|-|\spk_{r'}|}{|\spk_r-\spk_{r'}|},
\end{equation}
i.e., in this case, for given $\{r_a\}$ and $\{r'_a\}$, the condition $E_0=0$ is satisfied on the cone in the space of momenta $\spk_4$. The axis of this cone is directed along $\spq_{rr'}$ and one of its generatrices is co-directed with the vector $(\spk_r+\spk_{r'})/2$. If the vector $\spk_r$ is aligned along $\spk_{r'}$, then the cone degenerates to the ray directed along the vector $\spk_r$. At the points where these cones intersect, resonance coherent scattering is amplified in the case of constructive interference. The plots of the parameters $\eta/\rho^h$ and $\eta_a/\rho^h$ are presented in Figs. \ref{Plot_Coh_Latt_1}, \ref{Plot_Coh_Latt}.

%%%%%%%%%% figure here %%%%%%%%%%%%%
%%%%%%%%%% hologram. coh lattice. small interval %%%%%

%%%%%%%%%% figure here %%%%%%%%%%%%%

Expression \eqref{1_averaged_1} can be simplified in the case where the eigenvalues of the matrix $w_{(a)}g_sw_{(b)}$ are much less than unity, i.e., the lattice sites in the coordinate space are separated by distances much larger than the typical sizes of Gaussians in the lattice sites. In this case, using the Euler-Maclaurin formula, the first term in the square brackets \eqref{1_averaged_1} is approximately
\begin{equation}
     \sum_{\{r_a\}=-\infty}^\infty \rho^{s(1)}_0(\spk_r,\spk_{r})\approx  \frac{N_s}{N\det w^{(a)}_i}.
\end{equation}
If, additionally, $\rho^h(\spk_4-\spk_r+\spk_{r'},\spk_4)$ is a slowly varying function of $\spk_r$ and $\spk_r'$ on the scale of the basis vectors of reciprocal lattice, then the Euler-Maclaurin formula can be applied to the second term in the square brackets in \eqref{1_averaged_1}. This is realized when the eigenvalues of the matrices $w_{(a)}g_sw_{(b)}$ and $w_{(a)}g_hw_{(b)}$ are much less than unity and $|(\spw_{(a)}\spb_h)|\ll1$. From physical point of view, it means that, besides the restriction on the typical sizes of  Gaussians in the lattice sites, the typical sizes of the probe photon wave packet and its impact parameter must be much less than the distances between the lattice sites in the coordinate space. The Euler-Maclaurin formula implies
\begin{equation}
    \sideset{}{'}\sum_{\{r_a\},\{r'_a\}=-\infty}^\infty\approx \int \frac{d\spk d\spk'}{\det\nolimits^2(w^{(a)}_i)}=\int \frac{d\spk_s d\spq}{\det\nolimits^2(w^{(a)}_i)},
\end{equation}
where the condition $E_0=0$ should also be taken into account. In the small recoil limit, this condition takes the form
\begin{equation}
    (\spq\De\spn_0)=0 \;\;\Rightarrow\;\; q_\parallel=0,
\end{equation}
where the decomposition of the vector $\spq$ has been introduced as in \eqref{q_par_perp_split},
\begin{equation}
    \spq=q_\parallel\frac{\De\spn_0}{|\De\spn_0|} +\spq_\perp.
\end{equation}
We also need to take into account the integration measure in passing from the discrete summation indices to the integration over continuous ones
\begin{equation}
    dq_\parallel=\frac{\sqrt{\sum_a(\De\spn_0 \spw_{(a)})^2}}{|\De\spn_0|}dr.
\end{equation}
It leads to
\begin{equation}
    \sideset{}{'}\sum_{\{r_a\},\{r'_a\}=-\infty}^\infty\approx \int \frac{d\spk_s d\spq_\perp}{\det\nolimits^2(w^{(a)}_i)}\frac{\sqrt{\sum_a(\De\spn_0 \spw_{(a)})^2}}{|\De\spn_0|}.
\end{equation}
Then the second term in the square brackets in \eqref{1_averaged_1} becomes
\begin{multline}
    \sideset{}{'}\sum_{\{r_a\},\{r'_a\}=-\infty}^\infty \frac{\rho^{s(1)}_0(\spk_r,\spk_{r'}) \rho^h(\spk_4-\spk_r+\spk_{r'},\spk_4)}{\rho^h(\spk_4,\spk_4)}\approx \frac{|\spk^0_s||\spk_4|\sqrt{\sum_a(\De\spn_0 \spw_{(a)})^2}}{s_0\det\nolimits^2(w^{(a)}_i)\rho^h(\spk_4,\spk_4)}\times\\
    \times|\De\spn_0|\int d\spk_s d\spq_\perp \rho^{s(1)}_0(\spk+\spq_\perp/2,\spk-\spq_\perp/2) \rho^h(\spk_4-\spq_\perp,\spk_4).
\end{multline}
The integral on the last line coincides with \eqref{I_integral} and has been evaluated in formula \eqref{I_int_Gauss_2}, where one should bear in mind that $\rho^{s(1)}$ contains the factor $c_s^N=c_s/N$ instead of $c_s$ in the case we consider. As we see, apart from the factor
\begin{equation}
    \frac{1}{N}\sqrt{\frac{\sum_a(\De\spn_0 \spw_{(a)})^2}{4\pi\De\spn_0 b^{-1}\De\spn_0}}\sim\frac{\max(N_a)}{N},
\end{equation}
this contribution to expression \eqref{1_averaged_1} coincides with the contribution considered in Sec. \ref{Scat_by_Gaussian} under the conditions described above.

\subsubsection{Incoherent lattice}\label{Scatt_by_Lattice_Incoh}

Now we turn to the case of coherent scattering of a probe photon by a photon gas prepared in the quantum state with the one-particle density matrix that is an incoherent sum of Gaussians constituting a lattice. In this case, $\kappa_n=e^{i\chi_n}$, where $\chi_n$ are random phases with equiprobable distribution law. Then the one-particle density matrix of target photons \eqref{dens_matr_N_Gauss_coh} is reduced to
\begin{equation}\label{one_part_dens_matr_incoh}
\begin{split}
    \rho^{s(1)}(\spk_s,\spk'_s)&=c_s^N e^{-\frac14\de \spk_s g_s \de \spk_s -\frac14\de \spk'_s g_s \de \spk'_s} \sum_{n=1}^{N} e^{-i(\spk_s-\spk'_s) \mathbf{b}_n}=\\
    &= c_s^N e^{-\frac14\de \spk_s g_s \de \spk_s -\frac14\de \spk'_s g_s \de \spk'_s} \prod_{a=1}^3 \frac{\sin\big(\frac{\spq\spb_{(a)}}{2}(2N_a+1)\big)}{\sin\frac{\spq\spb_{(a)}}{2}}.
\end{split}
\end{equation}
The normalization constant is $c_s^N=c_s/N$. This density matrix describes a mixed state. Scattering of a photon by photons in such a state cannot be described with the aid of the dielectric susceptibility following from the Heisenberg-Euler Lagrangian even in the low energy limit as the Heisenberg-Euler Lagrangian is derived for background electromagnetic fields prepared in a coherent state. The product of sine ratios in \eqref{one_part_dens_matr_incoh} is concentrated near
\begin{equation}
    \spq=\spq_r=\sum_{a=1}^3 \spw_{(a)}r_a,
\end{equation}
where the index $r$ is uniquely defined by the set $\{r_a\}$. Expression \eqref{one_part_dens_matr_incoh} for the one-particle density matrix is to be substituted to the integral  \eqref{F_int} with $F=1$. In this integral, we divide the integration domain of the variable  $\spq$ into the cells $C_r$ and replace $\spq\rightarrow\spq_r$ in the integrand everywhere save the product of sine ratios. This approximation is justified for all the cells apart from the cell $C_0$ corresponding to $\spq_r=0$. Its contribution requires a separate consideration. Then we have approximately
\begin{equation}\label{1_averaged_incoh}
    \lan1\ran\approx -\frac{\rho^h(\spk_4,\spk_4)}{8\pi^2 N|\spk_s^0| |\spk_4|}\Big[ G^0(\spk_4)  +\sideset{}{'}\sum_{\{r_a\}=-\infty}^\infty \frac{\rho^h(\spk_4-\spq_r,\spk_4) }{\rho^h(\spk_4,\spk_4)} e^{-\frac18 \spq_rg_s\spq_r} G_{r}(\spk_4)\Big],
\end{equation}
where the prime at the sum sign means that the term with $\spq_r=0$ is absent and
\begin{equation}\label{G_r_ints}
\begin{split}
    G^0(\spk_4)&:=c_s\int d\spk_s e^{-\frac12\de \spk_sg_s\de \spk_s} \int_{C_0} d \spq \de\big(E(\spk_s+\spq/2,\spk_s-\spq/2;\spk_4) \big) \prod_{a=1}^3 \frac{\sin\big(\frac{\spq\spb_{(a)}}{2}(2N_a+1)\big)}{\sin\frac{\spq\spb_{(a)}}{2}},\\
    G_{r}(\spk_4)&:=\det(w^{(a)}_i) c_s\int d\spk_s e^{-\frac12\de \spk_sg_s\de \spk_s} \de\big(E(\spk_s+\spq_r/2,\spk_s-\spq_r/2;\spk_4) \big).
\end{split}
\end{equation}
The integral on the second line is, in fact, an average of the delta function, expressing the energy conservation law, with respect to the one-particle density matrix of target photons.

The integral over $\spq$ on the first line of \eqref{G_r_ints} can be evaluated in the same way as the integral \eqref{F_int_cell} has been done. On replacing the product of sine ratios by the Gaussian exponent and the delta function by the delta shaped sequence  \eqref{F_int_cell} with the argument
\begin{equation}
    E(\spk_s+\spq/2,\spk_s-\spq/2;\spk_4)\approx (\spq\De\spn),
\end{equation}
we obtain the Gaussian integral of the form \eqref{Gaussian_int_gen}. Applying the formulas from Appendix \ref{Evaluation_Gaauss_Int_App}, we deduce for $\e\rightarrow0$ and $N_a\gg1$ that
\begin{equation}\label{G_0_int}
    G^0(\spk_4)\approx c_s\int d\spk_s e^{-\frac12\de \spk_sg_s\de \spk_s} \frac{\det(w^{(a)}_i)}{\sqrt{2\pi\De \spn b^{-1}\De \spn}}\approx \frac{N_s\det(w^{(a)}_i)}{\sqrt{2\pi\De \spn_0b^{-1}\De \spn_0}},
\end{equation}
where it has been taken into account in the last approximate equality that the one-particle density matrix of target photons is narrow in the momentum space. Expression \eqref{G_0_int} is proportional to $N_a$ just as the contributions to the parameters $\eta$ and $\eta_a$ characterizing scattering by a lattice in the form of coherent superposition of Gaussian exponents \eqref{1_averaged}.

The integral on the second line \eqref{G_r_ints} can also be simplified. In the small recoil limit in the nondegenerate case, $\spn_s\neq\spn_4$, we have
\begin{equation}
    E(\spk_s+\spq_r/2,\spk_s-\spq_r/2;\spk_4)\approx (\spq_r\De\spn).
\end{equation}
Introducing the decomposition
\begin{equation}
    \spk_s=k^\parallel_s\bs\tau_r+\spk^\perp_s,\qquad\bs\tau_r:=\spq_r/|\spq_r|,
\end{equation}
and removing one integration with the help of the delta function, we arrive at
\begin{equation}\label{G_r_int}
    G_{r}(\spk_4)=\det(w^{(a)}_i)\frac{c_s}{|\spq_r|} \int d\bs\kappa_\perp |\bs\kappa_\perp| e^{-\frac12\de \spk_sg_s\de \spk_s}\Big|_{\substack{k_s^\parallel=|\bs\kappa_\perp|\cos\theta_r,\\ \spk_s^\perp=\bs\kappa_\perp \sin\theta_r}},
\end{equation}
where
\begin{equation}
    \cos\theta_r=(\spn_4\bs\tau_r),\qquad \sin\theta_r=\sqrt{1-(\spn_4\bs\tau_r)^2}.
\end{equation}
Expression \eqref{G_r_int} is exponentially suppressed except for the points in the space of momenta $\spk_4$ where
\begin{equation}\label{resonance_cond_incoh}
    (\spn_4\bs\tau_r)\approx (\spn^0_s\bs\tau_r).
\end{equation}
This equality distinguishes the set of resonant cones in the space of momenta $\spk_4$ with the axis directed along $\spq_{r}$ and one of the generatrices co-directed with the vector $\spn^0_s$. Notice that these cones differ from the cones distinguished by condition \eqref{resonance_cond}.

%%%%%%%%%% figure here %%%%%%%%%%%%%
%%%%%%%%%% hologram. incoh lattice %%%%%

\begin{figure}[t!]
\centering
\includegraphics*[width=0.352\linewidth]{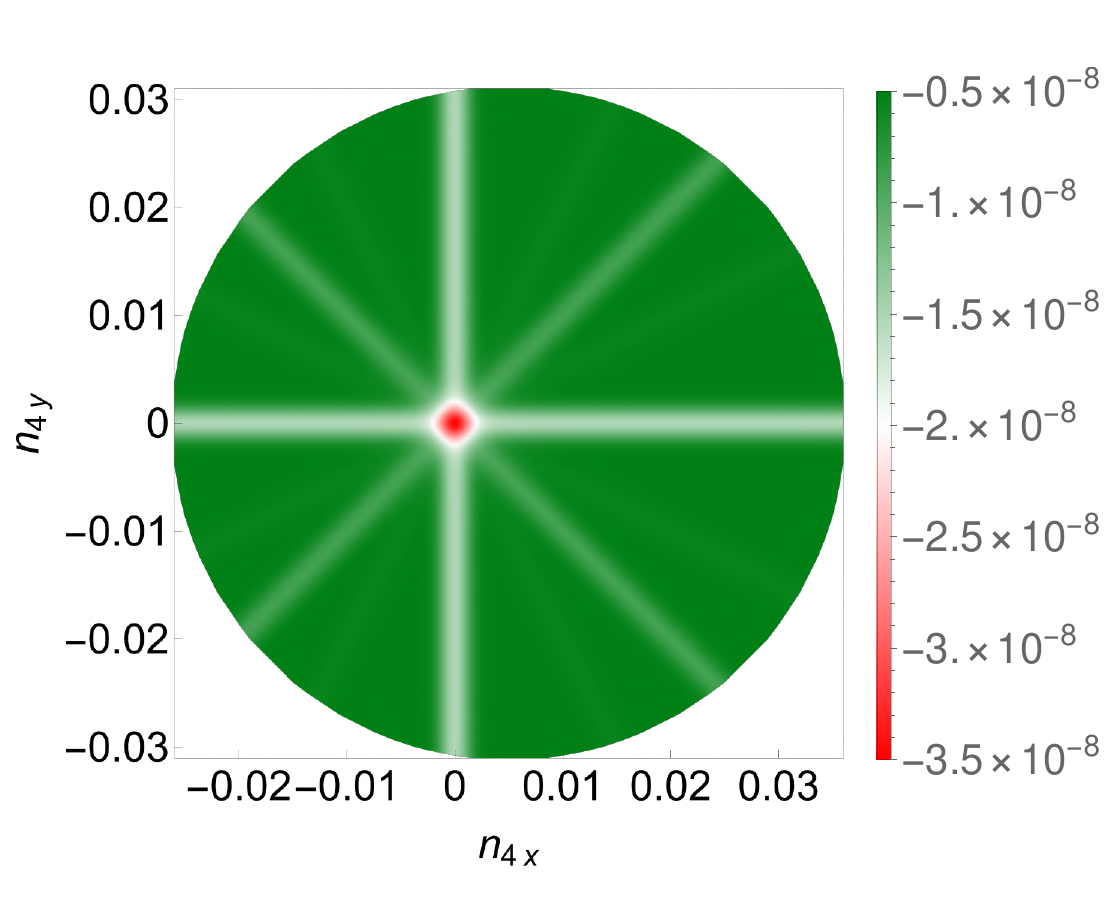}\,
\includegraphics*[width=0.298\linewidth]{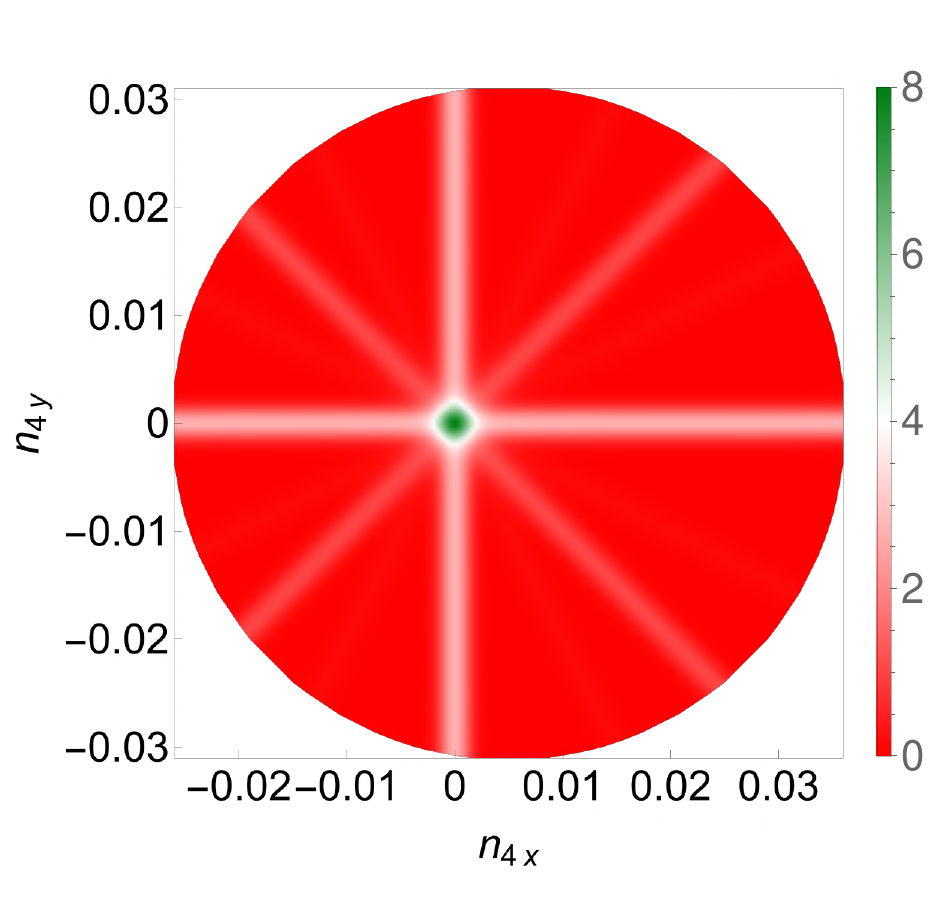}\,
\includegraphics*[width=0.325\linewidth]{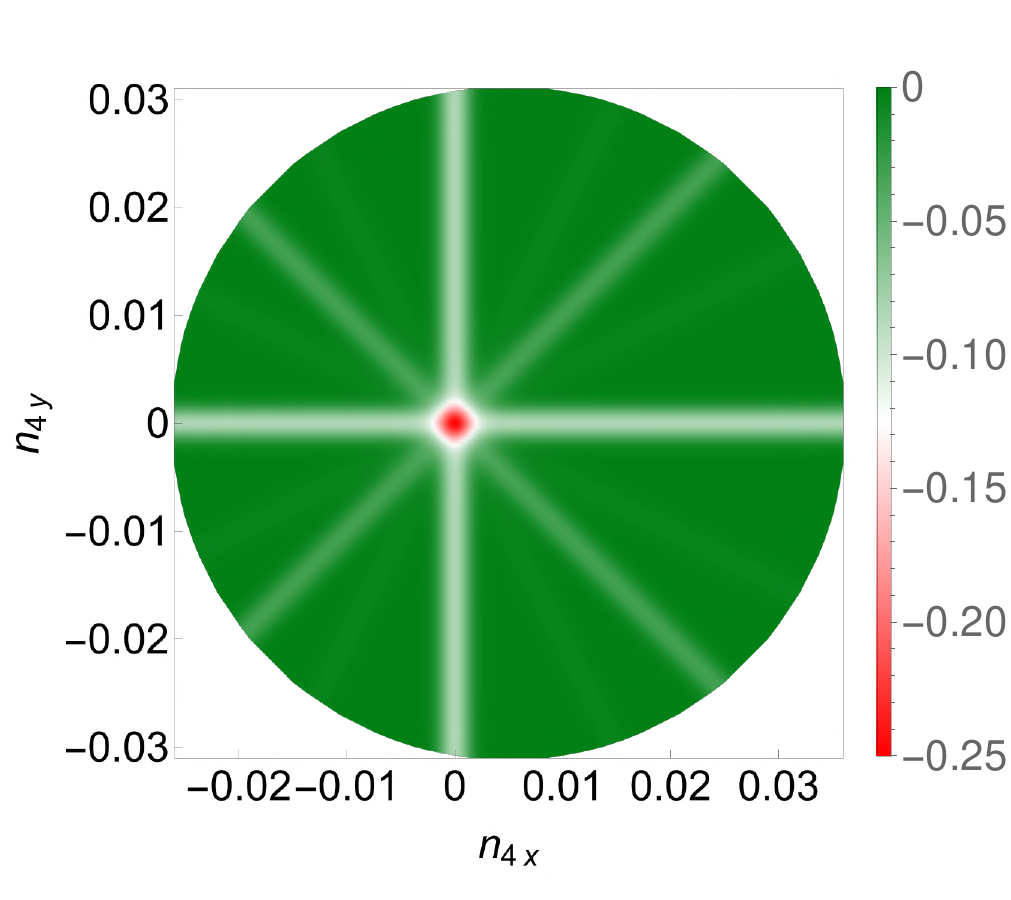}
\caption{{\footnotesize The parameters $\eta/\rho^h$ (left panel), $\eta'_1/\rho_h$ (middle panel), and $\eta'_3/\rho_h$ (right panel) determining a variation of the Stokes parameters of the density matrix of a probe photon in scattering by an incoherent lattice. The mean values of momenta: $\spk_s^0=(0,0,1.55)$ eV and $\spk_h^0=(0,0,-7)$ GeV. The matrices of inverse dispersions have the form \eqref{inverse_dispersions} with $\s_\perp^s/|\spk^0_s|=10^{-3}$, $\s_\parallel^s/|\spk^0_s|=10^{-3}$, $\s_\perp^h/|\spk^0_h|=10^{-2}$, and $\s_\perp^h/|\spk^0_h|=10^{-3}$. The parameters of the lattice: $N_1=N_2=N_3=20$, $\spb_1=(3,0,0)/\s_\perp$, $\spb_2=(0,3,0)/\s_\perp$, and $\spb_3=(0,0,3)/\s_\parallel$. The density matrix of the probe photon is rotated by an angle of $5\times10^{-3}$ around the $y$ axis so that an almost head-on scattering is realized. The impact parameter $\spb=(1,0,0)$ $\mu$m and the undulator strength parameter $K_u=1$. The domain of $(n_x,n_y)$ is shown where $\de\spk_h g_h\de\spk_h<10$.}}
\label{Plot_Incoh_Latt}
\end{figure}

%%%%%%%%%% figure here %%%%%%%%%%%%%

The two-dimensional integral \eqref{G_r_int} can be reduced to the one-dimensional one with respect to the azimuth angle. This last integral is readily evaluated numerically. Indeed, introduce the orthonormal vectors $\spe_1$, $\spe_2$ so that $\{\spe_1,\spe_2,\bs\tau_r\}$ constitute a right-handed orthonormal triple. Then representing
\begin{equation}
    \spk_s=|\bs\kappa_\perp|\tilde{\spk}_s,\qquad \tilde{\spk}_s:=\cos\theta_r\bs\tau_r +\sin\theta_r(\cos\vf\spe_1+\sin\vf\spe_2),
\end{equation}
where $\vf$ is the azimuth angle, we obtain
\begin{equation}
    \lan\de\spk_s,\de\spk_s\ran = |\bs\kappa_\perp|^2\lan\tilde{\spk}_s,\tilde{\spk}_s\ran -2|\bs\kappa_\perp| \lan\tilde{\spk}_s,\spk^0_s\ran +\lan \spk^0_s,\spk^0_s\ran,
\end{equation}
and
\begin{equation}
\begin{split}
    \lan\tilde{\spk}_s,\tilde{\spk}_s\ran=&\,\cos^2\theta_r \lan\bs\tau_r,\bs\tau_r\ran +\sin^2\theta_r \big(\cos^2\vf \lan\spe_1,\spe_1\ran +\sin^2\vf \lan\spe_2,\spe_2\ran +\sin(2\vf) \lan\spe_1,\spe_2\ran\big)+\\
    &+\sin(2\theta_r) \big(\cos\vf \lan\bs\tau_r,\spe_1\ran +\sin\vf \lan\bs\tau_r,\spe_2\ran\big),\\
    \lan\tilde{\spk}_s,\spk^0_s\ran=&\cos\theta_r \lan\bs\tau_r,\spk^0_s\ran +\sin\theta_r \big(\cos\vf\lan\spe_1,\spk^0_s\ran +\sin\vf\lan\spe_2,\spk^0_s\ran\big).
\end{split}
\end{equation}
Here, for brevity, we have denoted the inner product by the angle brackets,
\begin{equation}
    \lan\mathbf{a},\mathbf{b}\ran=a_i g_s^{ij} b_j.
\end{equation}
The function \eqref{G_r_int} takes the form
\begin{equation}
    G_{r}(\spk_4)=\det(w^{(a)}_i)\frac{2c_s}{|\spq_r|} e^{-\frac12\lan\spk^0_s,\spk^0_s\ran} \int_{-\pi}^\pi \frac{d\vf}{\lan\tilde{\spk}_s,\tilde{\spk}_s\ran^{3/2}} D_{-3}\Big(-\frac{\lan\tilde{\spk}_s,\spk^0_s\ran}{\lan\tilde{\spk}_s,\tilde{\spk}_s\ran^{1/2}}\Big) \exp\Big( \frac{\lan\tilde{\spk}_s,\spk^0_s\ran^2}{4\lan\tilde{\spk}_s,\tilde{\spk}_s\ran}\Big),
\end{equation}
where $D_p(z)$ is the parabolic cylinder function \cite{GrRy}.

The expression derived for $\lan1\ran$ in \eqref{1_averaged_incoh} is to be substituted into formula \eqref{eta_narrow_wp} for the parameters $\eta$ and $\eta_a$. The generalization of formulas \eqref{G_0_int} and \eqref{G_r_int} to the case where the density matrix of a probe photon has the form \eqref{one_part_dens_matr} at the instant of time $t=t_0$ for $|t_0|$ satisfying the estimate \eqref{t_0_appr_applic3} is given in formula \eqref{G_r_ints_t0}. The plots of the quantities $\eta/\rho^h$ and $\eta_a/\rho^h$ determining the evolution of the Stokes parameters in coherent scattering by the lattice constituted by an incoherent sum of Gaussians are presented in Fig. \ref{Plot_Incoh_Latt} for the same parameters of the probe photon and the lattice as in Figs. \ref{Plot_Coh_Latt_1}, \ref{Plot_Coh_Latt}. It is clearly seen that the holograms are qualitatively different.

\section{Dielectric susceptibility of a photon gas}\label{Susc_Gas_Phot}

As it was thoroughly discussed in \cite{KazSol2022,KazSol2023,radet,KazinskiFr24,AKS2025,KazSokNeut}, the tensor of dielectric susceptibility can be ascribed to the quantum state of photons and, in particular, to the quantum state of a single photon. This tensor describes coherent scattering of a probe photon by  target photons prepared in the given quantum state in the sense that coherent scattering of a probe photon results in the same scattering data as if it were scattered by a medium with this susceptibility tensor. It particular, it allows one to provide a physical interpretation to the coherent scattering data obtained in the previous section. In the paper \cite{KazSol2023}, this tensor was found for the energies of a probe photon below the electron-positron pair creation threshold. The analogous dielectric susceptibility tensors of quantum targets made of electrons and neutrons were calculated in the papers \cite{KazSol2022,AKS2025,KazSokNeut}.

The amplitude of coherent photon-by-photon scattering \eqref{coherent_amplitude} reads
\begin{equation}
    \Phi_{\ga_4\ga_2}=\frac{i\pi}{2V} \sum_{\la_1,\la_3}\int d\spk_1 d\spk_3\de(k_3+k_4-k_1-k_2) \rho^{s(1)}_{\la_1\la_3}(\spk_1,\spk_3) \frac{\tilde{M}_{\la_3\la_4\la_1\la_2}}{\sqrt{k_0^1k_0^2k_0^3k_0^4}},
\end{equation}
where $\ga_{2,4}=(\la_{2,4},\spk_{2,4})$. Representing the delta function in the form of the Fourier integral, we come to
\begin{equation}
\begin{split}
    \Phi_{\ga_4\ga_2}=&\,\frac{i\pi}{2V} \sum_{\substack{\la_1,\la_3,\\\la_2',\la_4'}}\int d^4x e^{i(k_4-k_2)x}\int \frac{d\spk_1 d\spk_3}{(2\pi)^4} e^{i(k_3-k_1)x} \rho^{s(1)}_{\la_1\la_3}(\spk_1,\spk_3) \frac{\tilde{M}_{\la_3\la_4'\la_1\la_2'}}{\sqrt{k_0^1k_0^2k_0^3k_0^4}}\times\\
    &\times f^{j*}_{(\la_2')}(\spk_2) f^i_{(\la_4')}(\spk_4) f^j_{(\la_2)}(\spk_2) f^{i*}_{(\la_4)}(\spk_4).
\end{split}
\end{equation}
Comparing this expression with formula (33) of \cite{KazSol2023} (see also \cite{KazKor22}), we find the Weyl symbol of the dielectric susceptibility tensor
\begin{equation}\label{chi_gen}
    \chi^{ij}(x,k_{h})=\pi \sum_{\substack{\la_1,\la_3,\\\la_2',\la_4'}}\int\frac{d\spk_{s}d\spq}{(2\pi)^4}e^{i\spq\spx} \frac{\rho^{s(1)}_{\la_1\la_3}(x^0;\spk_{s}+\spq/2,\spk_{s}-\spq/2)f^i_{(\la_4')}(\spk_{h}+\spq/2)  \tilde{M}_{\la_3\la_4'\la_1\la_2'} f^{j*}_{(\la_2')}(\spk_{h}-\spq/2)}{|\spk_{h}-\spq/2| |\spk_{h}+\spq/2|\sqrt{|\spk_{s}-\spq/2| |\spk_{s}+\spq/2|}},
\end{equation}
where the time dependent one-particle density matrix has been introduced,
\begin{equation}\label{dens_matr_t_dep}
    \rho^{s(1)}_{\la_1\la_3}(x^0;\spk_{1},\spk_{3})=e^{-i(k_0^1-k_0^3)x^0}  \rho^{s(1)}_{\la_1\la_3}(\spk_{1},\spk_{3}).
\end{equation}
Notice that the small quantum recoil is not understood in formula \eqref{chi_gen}. The dielectric susceptibility tensor is not determined only by the forward scattering amplitude as it is often assumed (see, e.g., \cite{Toll1952,Ahmadiniaz2025}). The dependence of the scattering amplitude on the transferred momentum, $q$, is relevant in \eqref{chi_gen}. Only in the case when the one-particle density matrix of the photon gas is concentrated near $\spq=0$ or, what is equivalent, the particle number density of this gas and its polarization are slowly varying functions of $\spx$, can the scattering amplitude in the integrand of \eqref{chi_gen} be replaced by the forward scattering amplitude.

In order to simplify the general expression \eqref{chi_gen} for the dielectric susceptibility of a photon gas and to provide a physical interpretation to it, we consider below a small recoil limit \eqref{small_recoil_cond}. Expanding the density matrix \eqref{dens_matr_t_dep} in terms of the $\s$-matrices as in formula \eqref{dens_matr_repr} and representing the invariant scattering amplitude in the form \eqref{tM_in_sigma}, we deduce with the use of relations \eqref{xi_tM} in the small recoil limit that
\begin{equation}\label{chi_sml_recoil}
\begin{split}
    \chi^{ij}(x,k_{h})=&\,\frac{1}{\spk_h^2} \int\frac{d\spk_{s}d\spq}{2|\spk_s|(2\pi)^3} e^{i\spq\spx} \rho^{s(1)}(x^0;\spk_{s}+\spq/2,\spk_{s}-\spq/2)\Big\{\frac12(m_++m_-) \de_\perp^{ij}-\\
    &-\frac{i}{2}(m_+-m_-)\xi^s_3\e_{ijk}n_h^k+\frac14(n_-\xi^s_++n_+\xi^s_-)(f^i_1f^j_1-f^i_2f^j_2) +\frac{i}4(n_-\xi^s_+-n_+\xi^s_-)f^i_{(1} f^j_{2)} \Big\},
\end{split}
\end{equation}
where $\de^{ij}_\perp=\de^{ij}-n^i_hn^j_h$, the round brackets at the pair of indices mean symmetrization without the factor $1/2$, and
\begin{equation}
    \xi^s_a\equiv\xi^s_a(x^0;\spk_{s}+\spq/2,\spk_{s}-\spq/2), \qquad f^i_l\equiv f^i_{l}(\spk_{h}).
\end{equation}
Notice that $n^i_hn^j_h$ in the expression for $\de^{ij}_\perp$ entering into the dielectric susceptibility tensor can be omitted because it is assumed that this tensor is contracted with the transverse polarization vectors.

The first term in the curly brackets in \eqref{chi_sml_recoil} is the isotropic contribution to the dielectric susceptibility. It is this term that determines coherent scattering of a probe photon by a photon gas in the case when the quantum state of this gas is not polarized. The second term in the curly brackets in \eqref{chi_sml_recoil} describes a gyrotropy of a photon gas \cite{KazSol2023}. If $\xi_1^s=\xi^s_2=0$ and $\xi^s_3\neq0$, viz., the target photons possess a circular polarization, then this target participates in coherent scattering as a gyrotropic medium and has only circular birefringence \cite{LandLifshECM}. The last two terms in the curly brackets in \eqref{chi_sml_recoil} responsible for linear birefringence vanish in this case. Since the gyration vector is proportional to $k_h^i$, a gas of photons possesses a natural optical activity. The magnitude of this effect is determined by $m_+-m_-\sim f_a(s)$ and in the low energy limit $f_a(s)\sim s^3$. Therefore, this effect cannot be reproduced by the use of the Heisenberg-Euler Lagrangian.

The third and fourth terms in the curly brackets \eqref{chi_sml_recoil} describe linear birefringence. In the general case, in order to evaluate these terms, it is necessary to employ expressions \eqref{M_t0}, \eqref{n_pm}, and \eqref{a_n_b_n_appr} and to take into account that
\begin{equation}
    e^{i\vf_{h,s}}=a_{h,s}+ib_{h,s}.
\end{equation}
The resulting expression is rather huge and we do not write it out here.

Instead, we will consider a special case where the one-particle density matrices of a probe photon and of target photons are concentrated near the momenta $\spk^h_0$ and $\spk^s_0$, whereas the vector $\mathbf{d}$ entering into the definition of the polarization vectors is taken to be orthogonal to both the vectors $\spk^h_0$ and $\spk^s_0$ in the laboratory frame. So, as it has been discussed in Sec. \ref{Evolut_Stokes_Param_Probe}, $\vf_h\approx-\vf_s$ and $n_+\approx n_-\approx n$. If
\begin{equation}
    \xi^s_a(x^0;\spk_{s}+\spq/2,\spk_{s}-\spq/2)\approx \xi^s_a(x^0;\spk_{s},\spk_{s}),
\end{equation}
then $\xi^s_a\in \mathbb{R}$ and the expressions responsible for linear birefringence in \eqref{chi_sml_recoil} become
\begin{equation}
    \frac14(n_-\xi^s_++n_+\xi^s_-)=\frac{n\xi^s_\perp}{2}\cos(2\psi_s),\qquad \frac{i}4(n_-\xi^s_+-n_+\xi^s_-) =- \frac{n\xi^s_\perp}{2}\sin(2\psi_s),
\end{equation}
where $\xi^s_\perp=|\xi^s_+|$ and $2\psi_s=\arg\xi^s_+$. As is well known (see, e.g., \cite{BaKaStrbook}), $\psi_s$ is the angle between the polarization vector $\mathbf{f}_1$ and the plane of predominant linear polarization. This angle is counted in the plane spanned by the vectors $\{\mathbf{f}_1,\mathbf{f}_2\}$ from the vector $\mathbf{f}_1$ counterclockwise as is seen from the detector measuring the radiation.

The third and fourth terms in the curly brackets in \eqref{chi_sml_recoil} can be written as
\begin{equation}
    \frac14(n_-\xi^s_++n_+\xi^s_-)(f^i_1f^j_1-f^i_2f^j_2) +\frac{i}4(n_-\xi^s_+-n_+\xi^s_-)f^i_{(1} f^j_{2)} =\frac{n\xi_\perp^s}{2} \big(f^i_{-\psi_s}f^j_{-\psi_s} -\frac12\de^{ij}_\perp \big),
\end{equation}
where the vector $\mathbf{f}_{-\psi_s}$ is obtained from the vector $\mathbf{f}_1$ by rotating it by an angle of $-\psi_s$ in the plane $\{\mathbf{f}_1,\mathbf{f}_2\}$ counterclockwise. Assuming that the one-particle density matrix of a target photon possesses a certain polarization, i.e., it is justified to replace
\begin{equation}
    \xi^s_a(x^0;\spk_{s},\spk_{s})\rightarrow \xi^s_a(x^0;\spk_{s}^0,\spk_{s}^0),
\end{equation}
in the integrand of \eqref{chi_sml_recoil}, the dielectric susceptibility tensor of a photon gas can be cast into the form
\begin{equation}\label{chi_app_1}
\begin{split}
    \chi^{ij}(x,k_{h})=&\,4\al^2\frac{\rho_s(x)}{|\spk_s^0|\spk_h^2} \Big[ \big(f_s+\frac{\xi^s_\perp}{4}g \big) \de_\perp^{ij}
    - i\xi^s_3 f_a\e_{ijk}n_h^k -\frac{\xi^s_\perp}{2}g f^i_{-\psi_s} f^j_{-\psi_s} \Big]=\\
    =&\,4\al^2\frac{\rho_s(x)|\spk_s^0|}{s_0^2}(\mathbf{n}_h-\mathbf{n}_s^0)^4 \Big[ \big(f_s+\frac{\xi^s_\perp}{4}g \big) \de_\perp^{ij}
    - i\xi^s_3 f_a\e_{ijk}n_h^k -\frac{\xi^s_\perp}{2}g f^i_{-\psi_s} f^j_{-\psi_s} \Big],
\end{split}
\end{equation}
where we are to put
\begin{equation}\label{s_0_variable}
    s=s_0=|\spk_h| |\spk_s^0| (\mathbf{n}_h-\mathbf{n}_s^0)^2,
\end{equation}
in the arguments of the functions $f_s$, $f_a$, and $g$. Besides, we have used the definition of the photon number density \eqref{phot_num_dens_spin_dens}. In particular, for head-on scattering of a probe photon by a photon gas, the vector $\mathbf{f}_{-\psi_s}$ lies in the plane of  predominant linear polarization of the photon gas state.

%%%%%%%%%% figure here %%%%%%%%%%%%%
%%%%%%%%%% susceptibility. eigenvalues %%%%%

\begin{figure}[t!]
\centering
\includegraphics*[width=0.47\linewidth]{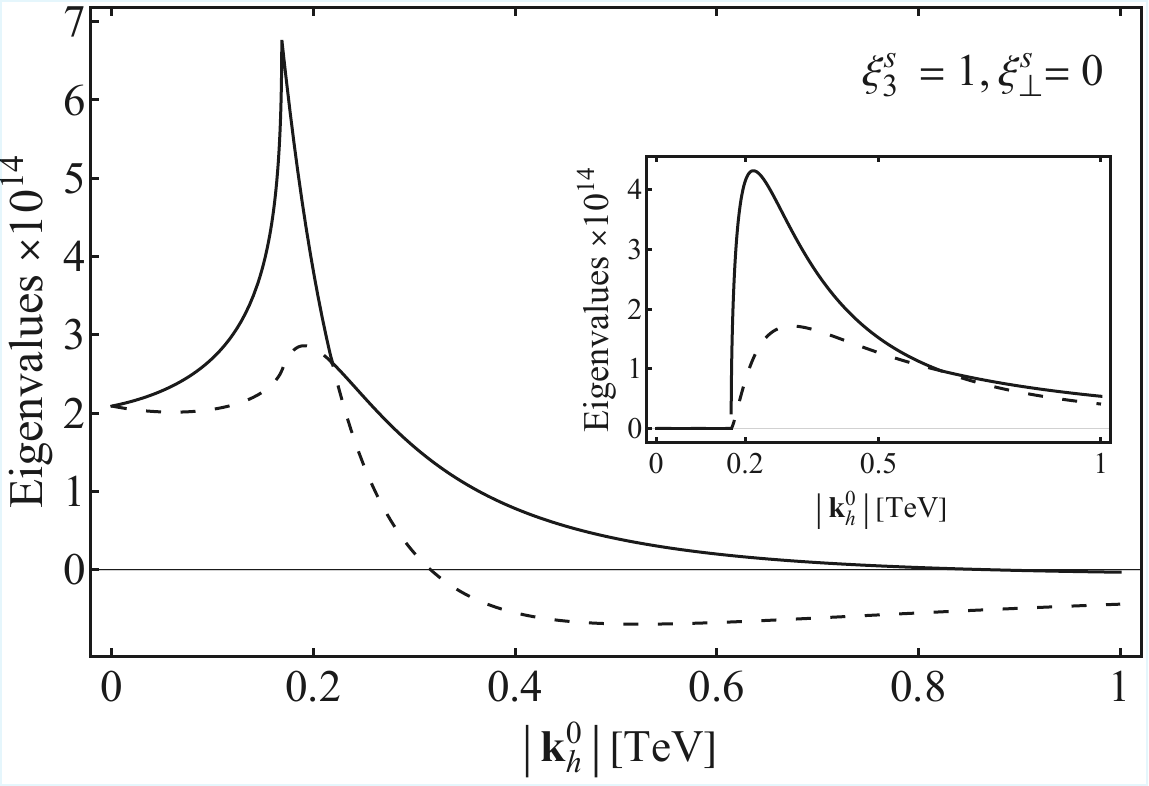}\,
\includegraphics*[width=0.47\linewidth]{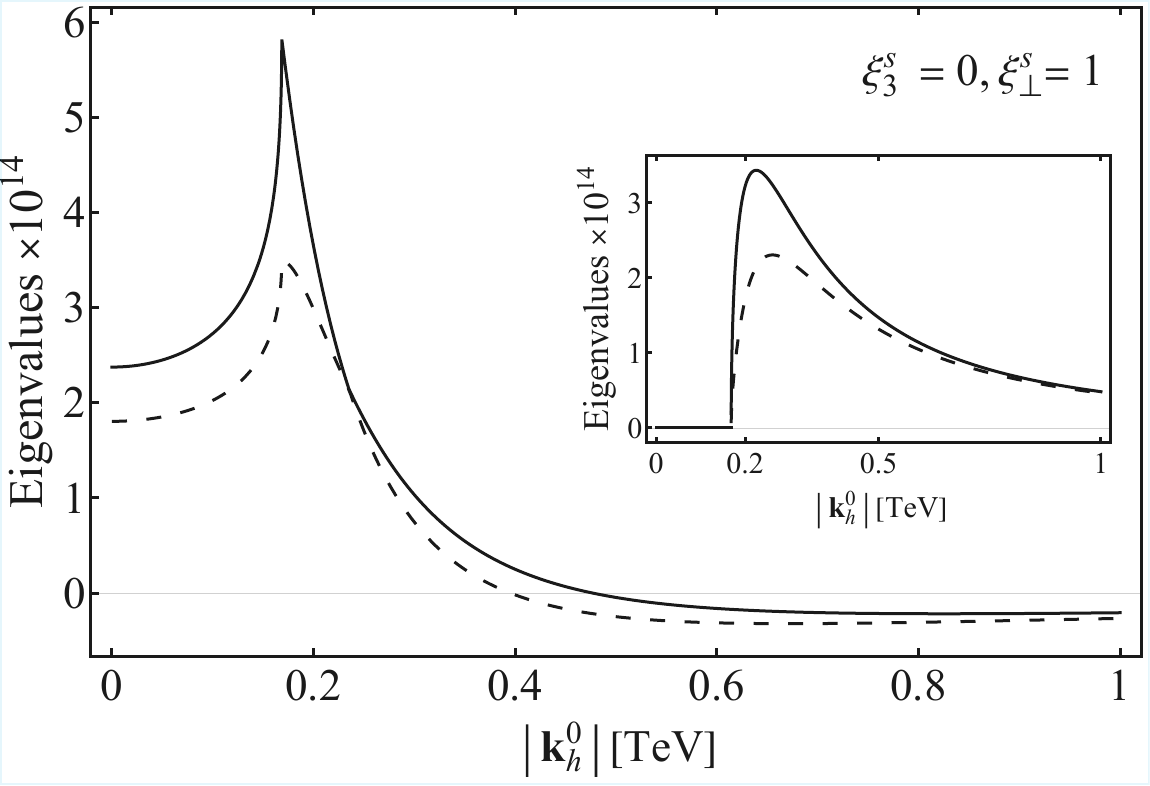}
\caption{{\footnotesize The eigenvalues \eqref{chi_eigenval} of the the tensors $\chi'_{ij}$ (main plot) and $\chi''_{ij}$ (inset) as functions of the probe photon energy for the different polarizations of the photon gas. The mean values of momenta: $\spk_s^0=(0,0,1.55)$ eV and $\spk_h^0=(0,0,-|\spk^0_h|)$ eV. The undulator strength parameter $K_u=1$. The solid line is for $\la_+$ and the dashed line corresponds to $\la_-$.}}
\label{Plot_Suscept}
\end{figure}

%%%%%%%%%% figure here %%%%%%%%%%%%%

To give a clearer physical interpretation to \eqref{chi_app_1}, we split the susceptibility tensor into Hermitian and anti-Hermitian parts,
\begin{equation}
    \chi_{ij}=\chi'_{ij} +i\chi''_{ij},
\end{equation}
where $\chi'_{ij}$ has the form \eqref{chi_app_1} with functions $f_s$, $f_a$, and $g$ replaced by their real parts, whereas $\chi''_{ij}$ has also the form \eqref{chi_app_1} but where imaginary parts of the functions $f_s$, $f_a$, and $g$ are taken. The tensor $\chi''_{ij}$ is responsible for dissipation of the electromagnetic field in the photon gas. In the system of coordinates with the $z$ axis directed along the momentum $\spk_h$, the matrices $\chi_{ij}$, $\chi'_{ij}$, and $\chi''_{ij}$ have the only nonzero $2\times2$ block. The nonzero eigenvalues of $\chi'_{ij}$ and $\chi''_{ij}$ are
\begin{equation}\label{chi_eigenval}
\begin{split}
    \la_\pm&=4\al^2(\mathbf{n}_h-\mathbf{n}_s^0)^4\frac{\rho_s(x)|\spk_s^0|}{s_0^2} \Big[f_s \pm\sqrt{(\xi^s_3)^2f_a^2+(\xi_\perp^s)^2g^2/16 }\Big]\approx\\
    &\approx \frac{2\al}{\pi} K_u^2\frac{m^2 }{\spk_h^2} \Big[f_s \pm\sqrt{(\xi^s_3)^2f_a^2+(\xi_\perp^s)^2g^2/16 }\Big],
\end{split}
\end{equation}
where we have used the definition of the undulator strength parameter \eqref{K_u_defn} and the functions $f_s$, $f_a$, and $g$ must be replaced by their real and imaginary parts for the tensors $\chi'_{ij}$ and $\chi''_{ij}$, respectively. For example, in the high-energy limit, $s\gg 4m^2$, using the asymptotics \eqref{fsfag_asymp_HE}, we obtain
\begin{equation}\label{lambda_HE_asympt}
    \la_\pm'\approx - \frac{2\al^2\rho_s(x)}{|\spk_s^0| \spk_h^2} \Big[5-\frac{\pi^2}{2} +\ln^2\frac{s_0}{em^2} \mp\frac12\sqrt{\pi^4(\xi^s_3)^2+(\xi_\perp^s)^2} \Big],\qquad \la_\pm''\approx \frac{2\pi\al^2\rho_s(x)}{|\spk_s^0| \spk_h^2} \Big( \ln\frac{s_0}{em^2}\pm|\xi^s_3|\ln\frac{s_0}{e^3m^2}\Big).
\end{equation}
Thus, except for a slow logarithmic scaling of the susceptibility tensor, a photon gas and a single photon behave as a birefringent plasma in coherent scattering of a high-energy probe photon. The factor at $-1/\spk^2_h$ in $\la'_\pm$ can be interpreted as the plasma frequency squared of a photon gas.

Keeping only the leading contributions in the low energy limit, $s\ll4m^2$, the dielectric susceptibility tensor \eqref{chi_app_1} becomes
\begin{equation}
    \chi^{ij}(x,k_{h})=\frac{11\al^2}{90}\frac{\rho_s(x)|\spk^0_s|}{m^4} (\spn_h-\spn_s^0)^4 \Big[ \big(1+\frac{3}{22}\xi^s_\perp\big) \de_\perp^{ij}
    -\frac{4 i}{77}\frac{|\spk_0^s|}{m^2}(\spn_h-\spn_s^0)^2 \xi^s_3\e_{ijk} k_h^k -\frac{3}{11}\xi^s_\perp f^i_{-\psi_s} f^j_{-\psi_s} \Big].
\end{equation}
As we see, in this limit, the dielectric susceptibility is proportional to the energy density, $w$, of photons in the target at a given point and by the order of magnitude\footnote{Formula (56) of \cite{KazSol2023} for the estimate of the magnitude of the susceptibility tensor contains a superfluous factor $\al$.}
\begin{equation}\label{chi_app_1_LE}
    \chi^{ij}_{LE}(x,k_{h})\sim \frac{11\al^2}{90}\frac{w}{m^4}\approx 4.6\times 10^{-28} \frac{w}{[\text{J}/\text{cm}^3]}.
\end{equation}
For example, the energy density of order $w\approx 10^{10}$ J$/$cm${}^3$ is achieved in laser beams at the laser facilities \cite{LiLasRev2025} for the photons with energy $k^0_s=1.55$ eV.

For a fixed energy density of a photon gas and the energies of these photons near $|\spk_s^0|$, the ratio of the dielectric susceptibility for high energy probe photon \eqref{chi_app_1} to the dielectric susceptibility for low energy probe photon \eqref{chi_app_1_LE} is of order
\begin{equation}
    \frac{\chi^{ij}}{\chi^{ij}_{LE}}\sim \frac{360}{11} \frac{m^4}{s_0^2}|f_s(s_0)|.
\end{equation}
This ratio is maximal for the probe photon energies near the electron-positron pair creation threshold, i.e., for $s_0\approx 4 m^2$. In this case, $f_s\sim1$ and so $\chi^{ij}/\chi^{ij}_{LE}\sim2$ (see formula (49) of \cite{KazSol2023} and Fig. \ref{Plot_Suscept}). Thus, by the order of magnitude, for the energies in the range where $s\lesssim 4m^2$, the dielectric susceptibility of a photon gas weakly depends on the energy of a probe photon \cite{KazSol2023}. For higher energies of a probe photon, the dielectric susceptibility of a photon gas drops as $1/\spk_h^2$. So long as the optical path of the probe photon in the photon gas is proportional to $\chi^{ij}|\spk_h|$, the effect of light-by-light scattering manifests stronger for high energy photons with the energy near the electron-positron pair creation threshold where $s_0\approx 4m^2$. For example, as it follows from \eqref{s_0_variable}, for the energies of target photons near $k^0_s=1.55$ eV, the optimum energy of the probe photon in head-on scattering is $k^0_h=168.5$ GeV, while for the energy of a probe photon $k^0_h=7$ GeV \cite{Hajima2016} the optimum energies of target photons are near $k^0_s=37.3$ eV. The plots of the eigenvalues \eqref{chi_eigenval} of the dielectric susceptibility tensor \eqref{chi_app_1} are presented in Fig. \ref{Plot_Suscept}.

The optical path of the probe photon and, consequently, the magnitude of the effect increase when the probe photon propagates along the target beam. This situation is realized for almost head-on scattering. The growth of the magnitude of the effect in such a case can be easily understood if one considers this process in the other inertial frame. Let the momenta of photons in the target be concentrated near $k^\mu_s=(k^0_s,0,0,k^0_s)$ and the probe photon momentum be $k^\mu_h=(k^0_h,0,k^h_\perp,-k^h_\parallel)$. Performing the boost along the $z$ axis, we pass to the reference frame where $k^\mu_h\rightarrow p^\mu_h=(p^0_h,0,k_\perp^h,0)$ and $p^0_h=|k_\perp^h|$. It is clear that
\begin{equation}
    k^0_h=\gamma p^0_h,\qquad k^h_\parallel=\ga p^0_h\be,\qquad\gamma=\frac{k^0_h}{|k^h_\perp|},\qquad \be=\frac{k^h_\parallel}{k^0_h}.
\end{equation}
In this reference frame, the momenta of photons in the target, $k_s^\mu$, become
\begin{equation}
    p_s^\mu=\ga(1+\be) k^0_s(1,0,0,1)\approx 2\ga k^0_s(1,0,0,1),
\end{equation}
where it is assumed in the approximate equality that $\ga\gg1$ and $\be>0$. For example, for $k^h_\perp=1$ eV and $k^0_h=1$ GeV, $\ga=10^9$ that leads to a $2\ga^2$ fold growth of the energy density of the target photon gas. As long as $\chi^{ij}/\chi^{ij}_{LE}\sim2$, estimate \eqref{chi_app_1_LE} implies in this reference frame
\begin{equation}\label{chi_est}
    \chi^{ij}(x,k_{h})\sim  4.6\times 10^{-28} \frac{2w\ga^2}{[\text{J}/\text{cm}^3]}.
\end{equation}
Therefore, we have $\chi^{ij}\sim 10$ for $w\approx 10^{10}$ J$/$cm${}^3$ and $\ga=10^9$. The probe photon with energy $1$ eV falls on the target beam along the $y$ axis in this reference frame, the energies of photons in the target beam being of order $3\times 10^9$ GeV. In the case of scattering by a lattice considered in Sec. \ref{Scatt_by_Lattice}, resonant scattering occurs when the lattice period along the $y$ axis is a multiple of the probe photon wavelength.

\section{Conclusion}

Let us sum up the results. We have found the photon hologram \eqref{inclus_probab_3}, \eqref{rho_out_0} of a photon gas and of a single photon prepared in an arbitrary quantum state in the leading nontrivial order of perturbation theory including the case where the energy of a probe photon is higher than the electron-positron pair creation threshold. We have also obtained the general expression \eqref{chi_gen} for the dielectric susceptibility tensor of this gas on the mass-shell of a probe photon. This tensor depends nontrivially on the transferred momentum and describes an anisotropic medium with spatial and frequency dispersions. The hologram and the susceptibility tensor are shown to be determined by the one-particle density matrix of the target and can be constructed even for a single photon. It turns out that the photon gas possesses linear and circular birefringence and it becomes absorbing for the energy of a probe photon above the electron-positron pair creation threshold (see Fig. \ref{Plot_Suscept}). In the small quantum recoil limit, the susceptibility tensor of a photon gas is expressed through the forward scattering amplitude convolved in a certain way with the one-particle density matrix of the photon gas. If additionally the one-particle density matrices of the photon gas and of the probe photon are narrow in the momentum space then the photon hologram of a photon gas is determined by a convolution \eqref{1_ave_as_convol}, \eqref{I_integral_as_convol} of the probe photon density matrix with the Fourier transforms of the particle number and spin densities of the target photon gas in the coordinate space.

We have derived equations \eqref{de_rho_de_xi} describing the evolution of the Stokes parameters of the probe photon density matrix in coherent scattering. These parameters describe the photon hologram of the quantum state of target photons. In the particular case, with appropriate identifications, the evolution equations derived agrees with the equations obtained in \cite{KotSerb97} for head-on scattering. In the case when the photon gas is circularly polarized and possesses a small dispersion of photon momenta, they also are in conformity with the analogous evolution equations for the photon hologram of a spin polarized neutron gas. In this case, the photon gas appears to be a purely gyrotropic medium without linear birefringence just as the spin polarized neutron gas. The photon gas in such a state has a natural optical activity. As for the target photon gas prepared in a linearly polarized quantum state with small dispersion of momenta, the circular birefringence is absent in it. It is also absent in the low energy approximation in the domain of applicability of the Heisenberg-Euler Lagrangian.

We have obtained the estimates \eqref{estimate_rel_mag_1}, \eqref{effect_rel_val} for the magnitude of the effect of coherent light-by-light scattering and shown that it can be measured with existing experimental facilities. As for the dielectric susceptibility tensor of a photon gas, we have established in \eqref{chi_app_1} that it is proportional to the energy density of the photon gas when its one-particle density matrix is concentrated near a certain momentum. This has allowed us to provide a simple physical explanation of a large magnitude of coherent scattering of a hard probe photon hitting target photons in almost head-on configuration. In the high-energy limit, $\sqrt{s}\gg2m$, the dielectric susceptibility tensor of a photon gas and of a single photon behaves as the dielectric susceptibility tensor of a birefringent plasma \eqref{lambda_HE_asympt} save for a slow logarithmic scaling.

We have elaborated several examples of coherent scattering of probe photons by a photon gas prepared in various quantum states. We have obtained the hologram of the general Gaussian one-particle density state of a photon gas including the case where this Gaussian is time shifted by some $t_0$ from the density matrix of a probe photon in the interaction picture. The holograms of lattices made of photons  have also been investigated. The two cases have been considered: the lattice is an coherent sum of Gaussians and the lattice is a sum of Gaussians with random relative phases. We call these lattices the coherent and incoherent, respectively. The explicit expressions \eqref{1_averaged_1_E_0}, \eqref{1_averaged_incoh} for the parameters specifying a variations of the Stokes parameters of the density matrix of a probe photon in coherent scattering have been derived. It turns out that the holograms of coherent and incoherent lattices differs drastically (cf. Figs. \ref{Plot_Coh_Latt_1}, \ref{Plot_Coh_Latt} and Fig. \ref{Plot_Incoh_Latt}) despite the fact that the parameters of these lattices have been chosen so that the photon number densities coincide in the coordinate space at the instant of time $t=0$. In average, the relative contrast of the holograms of these lattices is larger by a factor of $N_a$ than the same hologram of a Gaussian state without periodic structure and with the same average photon number density. Here $N_a$ is the number of Gaussians in the lattice along the direction of propagation of a probe photon. As in the case of photon scattering by crystals, there are the resonant cones in the momentum space of a scattered photon for both coherent and incoherent lattices. These cones are determined by condition \eqref{resonance_cond} for coherent lattices and by condition \eqref{resonance_cond_incoh} for incoherent ones and these conditions are different. In the case of constructive interference, there is an increase of the relative contrast of a hologram at the intersection points of resonant cones.

In the present paper, we studied the dielectric susceptibility tensor of a photon gas on the mass-shell of a probe photon. It would be interesting to find the off-shell photon polarization operator in the presence of other photons (or a photon) along the same lines as it was done in the papers \cite{AKS2025,KazSokNeut} for the photon polarization operator in the presence of electrons or neutrons. This will allow one to investigate quasiparticles and plasmon-polaritons in a photon gas or a single photon. Another possible direction of research would be to trace the influence of a quantum measurement on a hologram employing the formalism developed in \cite{radet}. The quantum measurement changes the state of target photons what should be imprinted in its hologram.

%\paragraph{Acknowledgments.}

%This research was supported by the TPU development program Priority 2030.

%The reported study was supported by the Ministry of Science and Higher Education of the Russian Federation, the contract FSWM-2025-0007.

\appendix
\section{Polarization vectors}\label{Polarization_Vectors_App}

Let us introduce the vectors of linear polarization of photons as in the paper \cite{KazSokNeut}. We choose some timelike $4$-vector $N^\mu$, $N^2=1$, defining the reference frame. We also take the spacelike $4$-vector $d^\mu$, $(Nd)=0$, $d^2=-1$, specifying the direction in space with respect to which the polarization vectors will be defined. We construct the $4$-vector $k_\perp^\mu$ from the $4$-vector of the photon momentum $k^\mu$ as
\begin{equation}
    k_\perp^\mu:=k^\mu-(kN)N^\mu +(kd)d^\mu, \qquad k_\perp:=\sqrt{-k_\perp^2}=\sqrt{(kN)^2-(kd)^2}.
\end{equation}
This vector is orthogonal to $N^\mu$ and $d^\mu$. Then we introduce the vectors of linear polarization of a photon with momentum $k^\mu$,
\begin{equation}\label{polarization_vects}
	e_{1}^{\mu}(\spk)=\frac{N^{\mu}(kd)-d^\mu(kN)-k^\mu (kd)/(kN)}{k_\perp},\qquad
    e_{2}^{\mu}(\spk)=\frac{\e^{\mu\nu\rho\la}N_\nu k_\rho d_{\la}}{k_\perp}.
\end{equation}
These vectors constitute an orthonormal right-handed triple with the vector $k^\mu/(kN)-N^\mu$ and possess the properties
\begin{equation}
	k_\mu e_{l}^\mu(\spk)=0,\qquad (e_{l}(\spk) e_{l'}^*(\spk))=-\delta_{ll'},\qquad N_\mu e_{l}^\mu(\spk)=0,\qquad l=\overline{1,2}.
\end{equation}

In the papers \cite{KarpNeu50,KarpNeu51,Tollis64,Tollis65,ConTollPist71}, the polarization vectors \eqref{polarization_vects} were used with
\begin{equation}
    N^\mu=\frac{k_1^\mu+k_2^\mu}{\sqrt{2(k_1k_2)}},\qquad d^\mu=\frac{\e^{\mu\nu\rho\s} k^2_\nu k^1_\rho k^3_\s}{\sqrt{2(k_1k_2)(k_2k_3)(k_3k_1)}}.
\end{equation}
The vector $N^\mu$ singles out the center-of-mass reference frame and the vector $d^\mu$ is orthogonal to the reaction plane. The polarization vectors for the photons with momenta $k_n^\mu$, $n=\overline{1,4}$, are constructed with the aid of formulas \eqref{polarization_vects}, where one should put $k^\mu=k_n^\mu$. It turns out in this case that
\begin{equation}\label{polarization_vects_KN}
\begin{gathered}
    e^\mu_1(\spk_n)=-d^\mu=-\frac{\e^{\mu\nu\rho\s} k^2_\nu k^1_\rho k^3_\s}{\sqrt{2(k_1k_2)(k_2k_3)(k_3k_1)}},\quad n=\overline{1,4},\\
    e^\mu_2(\spk_1)=\frac{k_4^\mu(k_1k_2) -k_1^\mu(k_2k_4) -k_2^\mu(k_4k_1)}{\sqrt{2(k_1k_2)(k_2k_3)(k_3k_1)}}=-e^\mu_2(\spk_2),\\ e^\mu_2(\spk_4)=\frac{k_2^\mu(k_3k_4) -k_3^\mu(k_4k_2) -k_4^\mu(k_2k_3)}{\sqrt{2(k_1k_2)(k_2k_3)(k_3k_1)}}=-e^\mu_2(\spk_3).
\end{gathered}
\end{equation}
In the small recoil limit in the leading order in the vector $q$ introduced in \eqref{q_mu}, we have
\begin{equation}
    e^\mu_1(\spk_n)\approx-\frac{\e^{\mu\nu\rho\s}k_{24\nu} q_\rho k_{13\s}}{(k_{24}k_{13})\sqrt{-q^2}} \qquad e^\mu_2(\spk_1)\approx -e^\mu_2(\spk_4)\approx\frac{q^\mu}{\sqrt{-q^2}},
\end{equation}
The vectors of circular polarization are defined as \cite{KarpNeu50,KarpNeu51,Tollis64,Tollis65,ConTollPist71}
\begin{equation}\label{polarization_vects_circ}
    e_{(\la)}^\mu(\spk_n)=\big[e_{1}^\mu(\spk_n)+i\la e_{2}^\mu(\spk_n)\big]/\sqrt{2},
\end{equation}
where $\la=\pm1$.

Let us find the connection between the polarization vectors \eqref{polarization_vects_KN} and the polarization vectors \eqref{polarization_vects} where $N^\mu$ defines the laboratory frame and $d^\mu$ is some fixed spacelike $4$-vector. We denote these last polarization vectors as $f^\mu_l(\spk_n)$ for the linear polarization and as $f^\mu_{(\la)}(\spk_n)$ for the circular one. Introduce the transition matrices
\begin{equation}\label{U_n_O_n_matrices}
    U^n_{\la'\la}=-(e^*_{(\la')}(\spk_n)f_{(\la)}(\spk_n)),\qquad O^n_{l'l}=-(e_{l'}(\spk_n)f_{l}(\spk_n)).
\end{equation}
It is clear that the matrices $U^n$ are unitary and the matrices $O^n$ are orthogonal. Moreover,
\begin{equation}
    U^n_{\la'\la}=e^{i\la\vf_n}\de_{\la'\la},\qquad
    O^n=
    \left[
      \begin{array}{cc}
        a_n & b_n \\
        -b_n & a_n \\
      \end{array}
    \right],
\end{equation}
where $a_n=\cos\vf_n$ and $b_n=\sin\vf_n$. The elements of the matrices $O^n$ are
\begin{subequations}
\begin{align}
    a_1&=\frac{(dNk_1k_2)(k_1k_3)-(dNk_1k_3)(k_1k_2)}{k_{1\perp}\sqrt{2(k_1k_2)(k_2k_3)(k_3k_1)}},&\qquad b_1&=\frac{N_{[\mu} d_{\nu]}  [k_1^\mu k_3^\nu(k_2k_1) +k_2^\mu k_1^\nu(k_1k_3) ]}{k_{1\perp}\sqrt{2(k_1k_2)(k_2k_3)(k_3k_1)}},\\
    a_2&=\frac{(dNk_2k_3)(k_2k_1)-(dNk_2k_1)(k_2k_3)}{k_{2\perp}\sqrt{2(k_1k_2)(k_2k_3)(k_3k_1)}},&\qquad b_2&=\frac{N_{[\mu} d_{\nu]} [k_2^\mu k_1^\nu(k_3k_2) +k_3^\mu k_2^\nu(k_2k_1) ]}{k_{2\perp}\sqrt{2(k_1k_2)(k_2k_3)(k_3k_1)}},\\
    a_3&=\frac{(dNk_2k_3)(k_1k_3)-(dNk_1k_3)(k_2k_3)}{k_{3\perp}\sqrt{2(k_1k_2)(k_2k_3)(k_3k_1)}},&\qquad b_3&=\frac{N_{[\mu} d_{\nu]} [k_3^\mu k_2^\nu(k_1k_3) +k_1^\mu k_3^\nu(k_3k_2) ]}{k_{3\perp}\sqrt{2(k_1k_2)(k_2k_3)(k_3k_1)}},\\
    a_4&=\frac{(dNk_1k_4)(k_2k_4)-(dNk_2k_4)(k_1k_4)}{k_{4\perp}\sqrt{2(k_1k_2)(k_2k_3)(k_3k_1)}},&\qquad b_4&=\frac{N_{[\mu} d_{\nu]} [k_4^\mu k_1^\nu(k_2k_4) +k_2^\mu k_4^\nu(k_4k_1) ]}{k_{4\perp}\sqrt{2(k_1k_2)(k_2k_3)(k_3k_1)}},
\end{align}
\end{subequations}
where the notation has been introduced
\begin{equation}
    (p_1 p_2 p_3 p_4):=\e_{\mu\nu\rho\s}p^\mu_1 p^\nu_2 p^\rho_3 p^\s_4.
\end{equation}
In the small recoil limit in the leading order in $q$, we obtain
\begin{equation}\label{a_n_b_n_appr}
\begin{aligned}
    a_1\approx a_3\approx a_s&= \frac{(dNk_sq)}{k_{s\perp}\sqrt{-q^2}},&\qquad a_2\approx a_4\approx a_h &=-\frac{(dNk_hq)}{k_{h\perp}\sqrt{-q^2}}  ,\\
    b_1\approx b_3\approx b_s& =-\frac{N_{[\mu} d_{\nu]}k_s^\mu q^\nu}{k_{s\perp}\sqrt{-q^2}},&\qquad b_2\approx b_4\approx b_h &=\frac{N_{[\mu} d_{\nu]}k_h^\mu q^\nu}{k_{h\perp}\sqrt{-q^2}} ,
\end{aligned}
\end{equation}
where the notation \eqref{k_13_k_24} has been used. It is clear that in this limit
\begin{equation}\label{phi_s_phi_h}
    \vf_1\approx\vf_3\approx\vf_s,\qquad \vf_2\approx\vf_4\approx\vf_h.
\end{equation}

\section{Evaluation of Gaussian integrals}\label{Evaluation_Gaauss_Int_App}

The Gaussian integrals are performed by using the standard formula
\begin{equation}\label{Gaussian_int_gen}
    \int d^nk e^{-S_0 -S^ik_i -\frac12 k_i S^{ij} k_j}=\frac{(2\pi)^{n/2}}{\det^{1/2}(S^{ij})} e^{-S_0 +\frac12 S^iS^{-1}_{ij} S^j}.
\end{equation}
In evaluating the integral \eqref{I_int_Gauss} over $\spq_\perp$, we need to perform the integral of the form
\begin{equation}
    I_q=\int d\spq_\perp e^{-\spq_\perp \mathbf{v}_\perp-\frac12 \spq_\perp g_\perp \spq_\perp},
\end{equation}
where
\begin{equation}
    g_\perp =\pr g \pr,\qquad \pr_{ij}=\de_{ij}-\tau_i\tau_j,\qquad (\spq_\perp \bs\tau)=0,\qquad \bs\tau^2=1.
\end{equation}
In order to apply formula \eqref{Gaussian_int_gen}, we have to find $\det\nolimits'(g_\perp)$ and $g_\perp^{-1}$ in the subspace distinguished by the projector $\pr$. Introduce the antisymmetric matrix
\begin{equation}
    \tau_{ij}:=\e_{ijk}\tau_k.
\end{equation}
Then for an arbitrary $3\times3$ matrix $A$, we have
\begin{equation}
    \det\nolimits'(\pr A\pr)=-\frac12\Sp(\tau A\tau A^T),
\end{equation}
and
\begin{equation}
    (\pr A\pr)^\vee =\frac{2}{\Sp(\tau A\tau A^T)} \tau A^T\tau,
\end{equation}
where $X^\vee$ is the pseudoinverse matrix to $X$. Hence, employing formula \eqref{Gaussian_int_gen}, we get
\begin{equation}\label{I_q_int}
    I_q=\frac{2\sqrt{2}\pi}{\sqrt{-\Sp(\tau g\tau g)}} \exp\Big(\frac{\mathbf{v}_\perp \tau g\tau \mathbf{v}_\perp}{\Sp(\tau g\tau g)} \Big).
\end{equation}

In Sec. \ref{Scatt_by_Lattice}, in approximate evaluating the integral \eqref{F_int_cell}, the Gaussian integral \eqref{F_int_cell_appr} of the form \eqref{Gaussian_int_gen} arises with
\begin{equation}
    S_0=0,\qquad S^\al=0,\qquad S^{\al\be}=
    \left[
      \begin{array}{cc}
        b_{ij} +\e^{-2}\ups_i\ups_j & -\e^{-2}\ups_i\ups'_j \\
        -\e^{-2}\ups'_i\ups_j & b_{ij} +\e^{-2}\ups'_i\ups'_j \\
      \end{array}
    \right],
\end{equation}
where, for conciseness, we omit the indices $rr'$ at $\ups_i$ and $\ups'_i$. Besides,
\begin{equation}
    b_{ij}=\frac{2}{\pi} \sum_{a=1}^3 N_a(N_a+1) b^{(a)}_i b^{(a)}_j.
\end{equation}
The matrix $S^{\al\be}$ takes the form
\begin{equation}\label{S_albe}
    S=B+\frac1{\e^2}V\otimes V,\qquad B=
    \left[
      \begin{array}{cc}
        b & 0 \\
        0 & b \\
      \end{array}
    \right],\qquad
    V=\left[
        \begin{array}{c}
          \bs\ups \\
          -\bs\ups' \\
        \end{array}
      \right].
\end{equation}
Therefore,
\begin{equation}\label{S_det}
    \det S=\big(1+\frac1{\e^2}VB^{-1}V\big) \det B\approx \frac1{\e^2} (\bs\ups b^{-1}\bs\ups +\bs\ups' b^{-1}\bs\ups') \det\nolimits^2b,
\end{equation}
where we have taken into account that $\e\rightarrow0$ in the approximate equality. It is not difficult to see that
\begin{equation}\label{det_b_b_inv}
    \det b= \Big(\frac{2}{\pi}\Big)^3  \det\nolimits^2(b^{(a)}_i) \prod_{a=1}^3 N_a(N_a+1),\qquad b^{-1}_{ij}= \sum_{a=1}^3\frac{w^{(a)}_i w^{(a)}_j}{8\pi N_a(N_a+1)}.
\end{equation}
Substituting these expressions into the general formula \eqref{Gaussian_int_gen}, we deduce \eqref{F_int_cell_appr1} for $N_a\gg1$ in the limit $\e\rightarrow0$.

\section{Relative time shift of Gaussian states}\label{Time_Shift_App}

Let us generalize expression \eqref{I_int_Gauss_2} derived in Sec. \ref{Scat_by_Gaussian} to the case where the density matrix of a probe photon has the form \eqref{one_part_dens_matr} at the instant of time $t=t_0$. Then at the instant of time $t=0$, this density matrix reads
\begin{equation}\label{dens_matr_probe_t0}
    \rho^{h}(\spk_h,\spk'_h)=c_he^{-\frac14\de \spk_h g_h \de \spk_h -\frac14\de \spk'_h g_h \de \spk'_h}e^{-i(\spk_h-\spk'_h)\mathbf{b}_h +it_0(|\spk_h| -|\spk'_h|)}.
\end{equation}
We suppose that it is narrow in the momentum space. Then
\begin{equation}\label{t_0_contrib}
    t_0(|\spk_h| -|\spk'_h|)=t_0(|\spk_4-\spq|-|\spk_4|)\approx -t_0\big[(\spq\spn_4) -\frac{1}{2|\spk_4|} (\spq^2-(\spq\spn_4)^2) \big],
\end{equation}
where it is assumed that the condition,
\begin{equation}\label{t_0_appr_applic1}
    \frac{|t_0||\spq|^3}{|\spk_4|^2}\ll1,
\end{equation}
is fulfilled. Solving perturbatively the equation, $E=0$, expressing the energy conservation law (see \eqref{energy_cons_law_app}) with respect to $q_\parallel$ and substituting its solution into \eqref{t_0_contrib}, we obtain
\begin{equation}\label{t_0_contrib_appr}
    t_0(|\spk_4-\spq|-|\spk_4|)\approx -t_0\big[(\spq_\perp\spn_4) -\frac{1}{4|\spk_4|} (\spq_\perp^2-(\spq_\perp\spn_4)^2) \big],
\end{equation}
where it is implied that
\begin{equation}\label{t_0_appr_applic2}
    \frac{|t_0||\spq|^3}{|\spk_s|^2}\ll1.
\end{equation}
The terms in the integrand of \eqref{inclus_int} depending on $q_\parallel$ that do not enter into the rapidly varying exponents are taken in the leading order in $q$ as it has been done in Secs. \ref{Scat_by_Gaussian} and \ref{Scatt_by_Lattice}. Substituting expression \eqref{t_0_contrib_appr} into \eqref{dens_matr_probe_t0}, where $\spk_h=\spk_4-\spq$ and $\spk_h'=\spk_4$, and comparing the resulting expression with the corresponding contributions to \eqref{S_intI}, we see that the nonzero $t_0$ can be taken into account in the integral \eqref{I_int_Gauss} by the replacement \eqref{t_0_b_g_replacement}. The applicability conditions for this approximation, \eqref{t_0_appr_applic1} and \eqref{t_0_appr_applic2}, give rise to the condition \eqref{t_0_cond}.

In Sec. \ref{Scatt_by_Lattice_Coh}, coherent scattering of a probe photon by target photons has been considered where the initial one-particle density matrix of the photon gas is a coherent sum of Gaussians with centers located at the sites of some lattice. Let us consider how the hologram of such a target changes when the density matrix of a probe photon takes the form \eqref{dens_matr_probe_t0}. To simplify the calculations, we assume that a stronger condition than \eqref{t_0_appr_applic1} holds,
\begin{equation}\label{t_0_appr_applic3}
    \frac{|t_0||\spq|^2}{|\spk_4|}\ll1.
\end{equation}
All the below formulas are generalized to the case where the weaker condition \eqref{t_0_appr_applic1} is satisfied. However, the resulting expression becomes cumbersome in this case.

Under the fulfillment of condition \eqref{t_0_appr_applic3}, one can discard the second term in the square brackets in \eqref{t_0_contrib_appr}. Then the main integral \eqref{F_int_cell} is written as
\begin{equation}
    F_{rr'}(\spk_4)=\int_{C_{rr'}} d\spk d\spk' \de \big(E(\spk,\spk';\spk_4)\big) \prod_{a=1}^3 \Big[\frac{\sin\big(\frac{\spk\spb_{(a)}}{2}(2N_a+1)\big)}{\sin\frac{\spk\spb_{(a)}}{2}} \frac{\sin\big(\frac{\spk'\spb_{(a)}}{2}(2N_a+1)\big)}{\sin\frac{\spk'\spb_{(a)}}{2}}\Big] e^{-it_0\spn_4(\de\spk_r-\de\spk'_{r'})}.
\end{equation}
This integral is evaluated analogously to the integral \eqref{F_int_cell} that has been performed in Appendix \ref{Evaluation_Gaauss_Int_App}. Replacing the sine ratios by the Gaussian exponents and the delta function by the delta shaped sequence as in \eqref{F_int_cell_appr}, we come to the Gaussian integral \eqref{Gaussian_int_gen} with
\begin{equation}
    S_0=0,\qquad S^\al=it_0
    \left[
      \begin{array}{c}
        \spn_4 \\
        -\spn_4 \\
      \end{array}
    \right]
    ,\qquad S^{\al\be}=
    \left[
      \begin{array}{cc}
        b_{ij} +\e^{-2}\ups_i\ups_j & -\e^{-2}\ups_i\ups'_j \\
        -\e^{-2}\ups'_i\ups_j & b_{ij} +\e^{-2}\ups'_i\ups'_j \\
      \end{array}
    \right].
\end{equation}
The matrix $S^{\al\be}$ is of the form \eqref{S_albe} and so
\begin{equation}\label{S_inv}
    S^{-1}=B^{-1}-\frac1{\e^2} \frac{(B^{-1}V)\otimes(B^{-1}V)}{1+VB^{-1}V/\e^2} \approx B^{-1} - \frac{(B^{-1}V)\otimes(B^{-1}V)}{VB^{-1}V}.
\end{equation}
If the weaker condition \eqref{t_0_appr_applic1} is satisfied instead of \eqref{t_0_appr_applic3}, then the additional terms appear in $S^{\al\be}$ that depend on $t_0$. Evaluating the resulting integral by using formula \eqref{Gaussian_int_gen}, we deduce that for $N_a\gg1$ the main integral \eqref{F_int_cell} is reduced to
\begin{equation}
    F_{rr'}(\spk_4)\approx \frac{\det\nolimits^2(w^{(a)}_i)}{\sqrt{2\pi}\sqrt{\bs\ups_{rr'} b^{-1}\bs\ups_{rr'} +\bs\ups'_{rr'} b^{-1}\bs\ups'_{rr'}}} \exp\Big\{-t_0^2\Big[\spn_4b^{-1}\spn_4 -\frac{(\spn_4 b^{-1}\bs\ups_{rr'}+\spn_4 b^{-1}\bs\ups'_{rr'})^2}{2(\bs\ups_{rr'} b^{-1}\bs\ups_{rr'} +\bs\ups'_{rr'} b^{-1}\bs\ups'_{rr'})}\Big] \Big\},
\end{equation}
i.e., it differs from \eqref{F_int_cell_appr1} by the exponential factor. As a result, expression \eqref{1_averaged_1} becomes
\begin{equation}\label{1_averaged_1_t_0}
\begin{split}
    \lan 1\ran \approx& -\frac{\det\nolimits^2(w^{(a)}_i)}{8\pi^2|\spk_s^0| |\spk_4| }\Big[ \frac{ \rho^h(\spk_4,\spk_4)}{\sqrt{4\pi\De\spn_0 b^{-1}\De\spn_0}} \exp\Big\{-t_0^2\Big[\spn_4b^{-1}\spn_4 -\frac{(\spn_4 b^{-1}\De\spn)^2}{\De\spn b^{-1}\De\spn}\Big] \Big\} \sum_{\{r_a\}=-\infty}^\infty \rho^{s(1)}_0(\spk_r,\spk_{r}) +\\
    &+\sideset{}{'}\sum_{\{r_a\},\{r'_a\}=-\infty}^\infty \frac{\rho^{s(1)}_0(\spk_r,\spk_{r'}) \rho^h(\spk_4-\spk_r+\spk_{r'},\spk_4)}{\sqrt{4\pi\De\spn_0 b^{-1}\De\spn_0}}  \exp\Big\{-t_0^2\Big[\spn_4b^{-1}\spn_4 -\frac{(\spn_4 b^{-1}\bs\ups_{rr'}+\spn_4 b^{-1}\bs\ups'_{rr'})^2}{2(\bs\ups_{rr'} b^{-1}\bs\ups_{rr'} +\bs\ups'_{rr'} b^{-1}\bs\ups'_{rr'})}\Big] \Big\} \Big].
\end{split}
\end{equation}
On substituting into formula \eqref{eta_narrow_wp} for the parameters $\eta$ and $\eta_a$, this expression determines a variation of the Stokes parameters of the density matrix of a probe photon.

Now we consider a change of the hologram of a photon gas with the one-particle density matrix in the form of an incoherent sum of Gaussians \eqref{one_part_dens_matr_incoh} when the density matrix of a probe photon has the form \eqref{dens_matr_probe_t0}. In order to find the parameters $\eta$ and $\eta_a$, it is necessary to evaluate \eqref{1_averaged_incoh} where only the functions $G^0(\spk_4)$ and $G_r(\spk_4)$ are modified. Under the fulfillment of condition \eqref{t_0_appr_applic3}, these function are written as
\begin{equation}
\begin{split}
    G^0(\spk_4)&=c_s\int d\spk_s e^{-\frac12\de \spk_sg_s\de \spk_s} \int_{C_0} d \spq e^{-it_0\spn_4\spq}\de\big(E(\spk_s+\spq/2,\spk_s-\spq/2;\spk_4) \big) \prod_{a=1}^3 \frac{\sin\big(\frac{\spq\spb_{(a)}}{2}(2N_a+1)\big)}{\sin\frac{\spq\spb_{(a)}}{2}},\\
    G_{r}(\spk_4)&= c_s\int d\spk_s e^{-\frac12\de \spk_sg_s\de \spk_s} \de\big(E(\spk_s+\spq_r/2,\spk_s-\spq_r/2;\spk_4) \big)\int_{C_r}d\spq e^{-it_0\spn_4(\spq-\spq_r)} \prod_{a=1}^3 \frac{\sin\big(\frac{\spq\spb_{(a)}}{2}(2N_a+1)\big)}{\sin\frac{\spq\spb_{(a)}}{2}}.
\end{split}
\end{equation}
These integrals are performed in the same way as it has been done for the integrals \eqref{G_r_ints} in Sec. \ref{Scatt_by_Lattice_Incoh} by reducing them to the Gaussian integrals of the form \eqref{Gaussian_int_gen} with
\begin{equation}
    S_0=0,\qquad S^i=it_0n_4^i,\qquad S^{ij}= b_{ij} +\frac{1}{\e^2}\De n_i\De n_j.
\end{equation}
Then for $N_a\gg1$ we arrive at
\begin{equation}\label{G_r_ints_t0}
\begin{split}
    G^0(\spk_4)&\approx c_s\int d\spk_s e^{-\frac12\de \spk_sg_s\de \spk_s} \frac{\det(w^{(a)}_i)}{\sqrt{2\pi\De \spn b^{-1}\De \spn}} \exp\Big\{-\frac{t_0^2}{2}\Big[\spn_4b^{-1}\spn_4 -\frac{(\spn_4 b^{-1}\De\spn)^2}{\De\spn b^{-1}\De\spn}\Big] \Big\},\\
    G_{r}(\spk_4)&=\det(w^{(a)}_i)e^{-\frac{t_0^2}{2}\spn_4 b^{-1}\spn_4} c_s\int d\spk_s e^{-\frac12\de \spk_sg_s\de \spk_s} \de\big(E(\spk_s+\spq_r/2,\spk_s-\spq_r/2;\spk_4) \big).
\end{split}
\end{equation}
The expression for $G_r(\spk_4)$ differs from that is given in \eqref{G_r_ints} only by the exponential factor. Therefore, it can be evaluated as it has been described in Sec. \ref{Scatt_by_Lattice_Incoh}. As for the expression for $G^0(\spk_4)$, it can be easily calculated numerically. For a small $|t_0|$ and a sufficiently narrow one-particle density matrix of target photons, one can neglect a variation of $\De\spn$ in the exponent in the integrand. In that case,
\begin{equation}
    G^0(\spk_4)\approx \frac{N_s\det(w^{(a)}_i)}{\sqrt{2\pi\De \spn_0b^{-1}\De \spn_0}} \exp\Big\{-\frac{t_0^2}{2}\Big[\spn_4b^{-1}\spn_4 -\frac{(\spn_4 b^{-1}\De\spn_0)^2}{\De\spn_0 b^{-1}\De\spn_0}\Big] \Big\},
\end{equation}
i.e., this function differs from \eqref{G_0_int} only by the exponential factor.

\end{document}